\documentclass[preprint,aps,prd]{revtex4-1}
\usepackage{t1enc}
\usepackage{aas_macros}
\usepackage[utf8]{inputenc}
\usepackage[english]{babel}
\usepackage{subcaption}
\usepackage{graphicx, float}
\usepackage[toc,page]{appendix}
\usepackage{verbatim}
\usepackage{url}
\usepackage{subfiles}
\usepackage{xcolor}
\usepackage{amsmath}
\usepackage{amsfonts}
\usepackage{amssymb}
\usepackage{mathrsfs}
\usepackage{mathtools}
\usepackage{hyperref}
\usepackage{booktabs}
\usepackage[noabbrev]{cleveref}
\usepackage{listings} 
\usepackage{xcolor}   
\usepackage{cleveref}
\usepackage{booktabs}
\usepackage{multirow}
\usepackage{soul}
\usepackage[mathlines]{lineno}

\begin{document}

\title{Neural network approach to preferred event selection for low-latency gravitational-wave alerts}

\author{Pratyusava Baral}
\email{pbaral@uwm.edu}
\affiliation{University of Wisconsin-Milwaukee, Milwaukee, Wisconsin 53201, USA}

\author{Cody Messick}
\affiliation{University of Wisconsin-Milwaukee, Milwaukee, Wisconsin 53201, USA}

\author{Patrick Brady}
\affiliation{University of Wisconsin-Milwaukee, Milwaukee, Wisconsin 53201, USA}

\begin{abstract}
    The LIGO-Virgo-KAGRA Collaboration uses multiple independent search pipelines to detect gravitational waves, often resulting in multiple triggers (so-called g-events) for a single astrophysical event. These triggers are grouped into superevents, raising a critical question for multimessenger astronomy: which g-event provides the most accurate sky localization in low latency for neutrino and electromagnetic follow-up? Currently, the g-event with the highest signal-to-noise ratio (SNR) is selected, under the assumption that it should provide the best estimators of the source's parameters, including its location on the sky. Analysis of simulated signals reveals that the highest-SNR event often does not yield the smallest searched area---the metric we use to quantify localization accuracy. We present a neural-network-based selector trained on simulated signals comprising binary black holes, binary neutron stars, and neutron star--black hole mergers to identify the g-event with the minimum searched area. The network uses information such as detector SNRs, false-alarm rate (FAR), and chirp mass from all triggers associated with each astrophysical event and is designed to be pipeline agnostic. Across all events recovered below a FAR threshold of one per five months, the network reduces the mean searched area by $\sim$1.4\% relative to the SNR-based selector. The two selectors agree on $\sim$41\% of superevents; in the remaining cases, the network outperforms SNR by more than 100~deg$^2$ for $\sim$12\% of superevents (68\% of these contain a neutron star), while SNR outperforms the network by a comparable margin in another $\sim$11\% (67\% of these contain a neutron star). The net mean improvement is therefore modest. Unlike traditional selectors, the neural network preserves the underlying distribution of pipeline contributions, avoiding systematic biases toward specific pipelines. The network can be trained in approximately one minute on a few thousand events, making retraining straightforward as detector noise and pipeline configurations evolve between observing runs.
    Selection itself is instantaneous, making it suitable for low-latency applications.
    These results demonstrate that machine learning can enhance multimessenger astronomy capabilities while maintaining fairness across detection pipelines. We recommend exploring this approach for future observing runs.
\end{abstract}

\maketitle

\section{Introduction}
\label{sec:intro}

The LIGO-Virgo-KAGRA (LVK) Collaboration has detected gravitational waves (GWs) \cite{LIGOScientific:2014pky, VIRGO:2014yos, KAGRA:2020tym, 2025arXiv250818082T} from the merger of binary black holes (BBHs) \cite{AbEA2016b}, binary neutron stars (BNSs) \citep{AbEA2017b}, and neutron star--black hole (NSBH) mergers \cite{AbEA2021}.
The fourth observing run (O4) ended on November 18, 2025 \footnote{\url{https://observing.docs.ligo.org/plan}}.
Building upon the $\sim$100 confident detections from the first three observing runs (O1–O3), the fourth observing run yielded $\sim$250 additional gravitational-wave candidates.
The first BNS merger GW170817 \citep{AbEA2017b}, was detected in low latency during the second observing run in coincidence with  GRB170817A~\citep{GoVe2017,SaFe2017,AbEA2017e} and the optical counterpart AT2017gfo \citep{2017Sci...358.1556C,SmCh2017,AbEA2017f}.
Such coincident multimessenger observations enable measurements of the neutron-star equation-of-state (EOS) \citep{BaJu2017, MaMe2017, CoDi2018b, CoDi2018, CoDi2019b, AnEe2018, MoWe2018,RaPe2018,Lai2019,DiCo2020,Huth:2021bsp}, the Hubble constant \citep{CoDi2019,CoAn2020,2017Natur.551...85A,HoNa2018,DiCo2020}, and $r$-process nucleosynthesis \citep{ChBe2017,2017Sci...358.1556C, CoBe2017,RoFe2017,SmCh2017,WaHa2019,KaKa2019}.

There are two types of low-latency search pipelines in the current observing run \cite{2025arXiv250818080T, 2025arXiv250818081T, 2025arXiv250818082T}, namely compact-binary-coalescence (CBC) searches that use matched filtering based on precomputed waveforms and weakly modeled burst searches.
Data from the first part of the fourth observing run were analyzed by four CBC search pipelines: Multi-Band Template Analysis (MBTA) \cite{Adams:2015ulm, Aubin:2020goo, Allene:2025saz}, GStreamer LIGO Algorithm Library (GstLAL) \cite{Messick:2016aqy, Sachdev:2019vvd, Hanna:2019ezx, Cannon:2020qnf, 2024PhRvD.109d4066S, Ewing:2023qqe, Tsukada:2023edh, Ray:2023nhx, Joshi:2025nty, Joshi:2025zdu}, the Python-based CBC search software (PyCBC) \citep{Allen:2004gu, Usman_2016, Nitz:2017lco, DalCanton:2020vpm, DalCanton:2020vpm, Davies:2020tsx}, and Summed Parallel Infinite Impulse Response (SPIIR) \cite{s_hooper_2012,j_luan_2012}.
The CBC searches \cite{AbEA2021d} target primarily BNS, NSBH, and BBH mergers.
Dedicated searches targeting subsolar mass binaries and other extreme regions of parameter space have also been deployed in O4 but are not considered in this work.
Weakly modeled burst searches \citep{AbEA2021b} like Coherent WaveBurst (cWB) \cite{klimenko2008coherent, Klimenko:2015ypf, drago2021coherent} and Omicron LALInferenceBurst (oLIB) \cite{Lynch_2017}, look for consistent signal morphologies from various detectors and are capable of detecting a variety of astrophysical sources like core-collapse supernova and magnetic starquakes, in addition to standard CBC sources \cite{AbEA2021b}.
Multiple independent search pipelines process strain data to identify GW triggers.
Each pipeline employs distinct methods for trigger generation, background estimation, and significance estimation, resulting in slightly different sensitivities \cite{Messick:2016aqy, Aubin:2020goo, s_hooper_2012,j_luan_2012, Usman_2016, DalCanton:2020vpm, piotrzkowski2022searching} across the parameter space.
This multipipeline approach maximizes detection efficiency \cite{2025arXiv250818082T}, enables cross-validation of results, and provides operational redundancy.
For this paper, we only consider triggers from CBC pipelines, although the method could be extended.

The triggers from low-latency pipelines are fed into an event orchestrator called GWCelery \footnote{\url{https://git.ligo.org/emfollow/gwcelery}} \footnote{\url{https://rtd.igwn.org/projects/gwcelery/en/latest/}}, which annotates the triggers with data-quality information, computes a number of astrophysical attributes, and issues public alerts when predefined conditions are satisfied.
The Gravitational-Wave Candidate Event (Grace) Database \footnote{\url{https://gracedb.ligo.org/}} stores information about the candidate events and provides a web interface to view them.
The low-latency alert infrastructure is described in detail in \cite{doi:10.1073/pnas.2316474121}, and the LVK Alert User Guide \footnote{\url{https://emfollow.docs.ligo.org/userguide/}} is used to communicate changes to this system to the broader astrophysical community.

Each GW trigger is assigned a Grace ID and is referred to as a g-event.
We often see multiple g-events corresponding to the same candidate event.
This occurs both because we employ multiple search pipelines and because a single pipeline may upload several g-events for a given astrophysical event.
These g-events are grouped when they are believed to originate from the same astrophysical source, forming what is known as a superevent (e.g., S230306ch, an event discussed as a case study in Appendix~\ref{app:case_studies}).
A natural question that arises is, among all the g-events in a superevent, which one most closely represents the properties of the astrophysical source?
This work proposes a neural-network- (NN) based selector trained to select the event with the \textit{best} sky map in low latency.
For training purposes, the best sky map for a simulated signal is chosen using a metric called searched area, which is described in the following section.

The paper is organized as follows. In Sec.~\ref{sec:background}, we describe the method we use currently for selecting the preferred event. Section~\ref{sec:data} describes the data used for the paper. Section~\ref{sec:methods} describes the details of the neural-network-based selector developed in this paper.
Section~\ref{sec:results} presents our key results, and Sec.~\ref{sec:discussion} is reserved for discussions.

\section{Background and Motivation}
\label{sec:background}
A key motivation for low-latency GW detection is multimessenger astronomy.
Each CBC g-event in a superevent has a sky map generated in low latency by an algorithm called Bayesian triangulation and rapid localization (BAYESTAR) \citep{2016PhRvD..93b4013S}.
The sky map is a two-dimensional map of the posterior probabilities of right ascension and declination, or a three-dimensional map that also includes the luminosity distance to the source.
Three-dimensional sky maps generated using BAYESTAR within seconds of trigger upload are used for public alerts.
Our goal is to identify the event with the \textit{best} sky map for multimessenger follow-up.
A key parameter associated with CBC triggers is the signal-to-noise ratio, which represents the ratio of the power contained in the signal to the noise.
Under the assumption of a stationary Gaussian noise model, the highest-SNR event is expected to be the best match to the observed data.
It also yields the most constrained sky map, with the smallest 90\% credible region, making it a natural choice for multimessenger follow-up.
Unfortunately, real detector noise is not perfectly Gaussian, and these desirable properties are not guaranteed to hold.
This leads us to ask the question: \textit{How can we select the best sky map for multimessenger follow-up in low latency?}

The primary constraint for electromagnetic follow-up observations is accurate localization.
In simulations, the searched area, defined as the sky area encompassed by pixels in the localization sky map with posterior probability densities exceeding those of the true source location, can be used as a metric to identify the best sky map.
Better localization results in decreased telescope time allocation and enhanced probability of transient detection.
Under the assumption of stationary Gaussian noise, the trigger with maximum SNR is also the maximum likelihood estimator of the signal \cite{jolien} and is expected to yield the minimum searched area.
We evaluate this hypothesis by injecting simulated GW signals into real detector data from the third observing run, which captures the full non-Gaussian noise environment of operating interferometers rather than a synthetic noise model.
Given our knowledge of the injected source positions, we compute the corresponding searched areas and compare the performance of events ranked by minimum searched area versus maximum SNR.
Our analysis reveals a systematic deviation from expectations that may be due to non-Gaussianities in detector noise.

Selecting the g-event with the lowest FAR has been proposed as an alternative to selecting the highest-SNR g-event.
A key complication arises from the fact that each pipeline defines a FAR in a self-consistent way, and as a result FAR values are not directly comparable across pipelines \cite{2025arXiv250818082T}.
Selecting on FAR, therefore, tends to favor pipelines whose FAR estimates are systematically lower for a given astrophysical signal, biasing the selection toward those pipelines.
We note, however, that SNR systematics across pipelines are also non-negligible \cite{2025arXiv250818082T}, so SNR-based selection is not free of pipeline bias either.

\section{Data}
\label{sec:data}
Prior to the start of the fourth observing run, the LVK Collaboration prepared a dataset comprising 50,000 simulated CBC waveforms, or injections.
Each injection, simulated using IMRPhenomPv2\_NRTidalv2 \cite{NRPv2} waveform family, has an optimal network SNR greater than 4, ensuring that signals are realistically detectable by search pipelines and avoiding those that would be impossible to recover.
The SLy EOS \cite{ChBo1998} is used to map from component masses to the tidal deformability parameters.
The maximum neutron-star (NS) mass allowed by the SLy EOS is $\sim 2.05 M_\odot$ \cite{ChBo1998}, and all objects above this mass are classified as black holes with the tidal parameters set to 0.
The injection set is comprised of 40.9\% BNS, 35.8\% NSBH, and 23.3\% BBH mergers.
These simulated events are distributed uniformly in comoving volume, assuming a flat $\Lambda$CDM cosmology based on Planck 2018 results \cite{Planck18}.
The rate of injections is artificially high and not astrophysical.

These injections were added to a stretch of data collected by the LVK instruments between January 05, 2020 and February 14, 2020 as a part of the O3 run \cite{doi:10.1073/pnas.2316474121}.
The simulated strain is projected onto the detectors, time shifted to fit within the 40-day window, and streamed in 1-sec segments to enable low-latency analysis by search and other annotation pipelines.
This exercise is repeated in several cycles, even during the observation run, to benchmark analysis, continuously track improvements in the alert infrastructure, and provide avenues for pipelines to test their changes.
It serves as an end-to-end test bed for data ingestion, execution of online search pipelines, event annotation, and the issuance of alerts to the astrophysics community.

This study uses the data products from the 11th iteration of this exercise, referred to as mock data challenge 11 (MDC11), conducted by the LVK Collaboration prior to O4.
The MDC11 data were analyzed by all four CBC search pipelines: MBTA, GstLAL, PyCBC, and SPIIR.
In addition to the CBC searches, two burst searches---cWB \cite{klimenko2008coherent, Klimenko:2015ypf, drago2021coherent} and oLIB \cite{Lynch_2017}---and an external coincidence search, rapid on-source VOEvent coincident monitor (RAVEN) \cite{piotrzkowski2022searching, urban2016, cho2019low}, were also operational during the challenge.
However, we focus exclusively on the CBC search pipeline uploads in this work.

\subsection{Data preparation}
A g-event from a CBC pipeline is labeled as significant if it is associated with a FAR below 3.9 $\times$ $10^{-7}$ Hz (one per month).
At the time of the MDC, four distinct CBC search pipelines, plus RAVEN \cite{piotrzkowski2022searching, urban2016, cho2019low}, operated independently in low latency.
While these pipelines are not strictly independent (as they are all searching for the same underlying signals), they often respond differently to detector noise.
To account for the use of multiple pipelines, we take the number of trials to equal the number of pipelines.
This likely leads to an overestimate of the false-alarm rate, but we adopt this approach to avoid overestimating significance.
Accounting for the trials factor, we mark g-events from CBC searches as significant when $\textrm{FAR} < 7.7 \times 10^{-8}$ Hz (one per 5 months).

For this study, g-events were classified into two distinct categories based on the FAR.
Events with a $\textrm{FAR} < 7.7 \times 10^{-8}$ Hz are referred to as significant g-events.
In a low-latency search, such events would correspond to astrophysical events \footnote{\url{https://emfollow.docs.ligo.org/userguide/}}.
A superevent which has at least one significant g-event is referred to as a significant event henceforth.
The other category mimics the set of events made public during O4 and is comprises the set of superevents that have at least a g-event recovered with $\textrm{FAR} < 2.3 \times 10^{-5}$ Hz (approximately two false alarms per day).
This category is referred to as all events henceforth.
This dataset typically has triggers of both astrophysical and nonastrophysical origin.
However, for this study, we retain only the injected astrophysical triggers and exclude all nonastrophysical triggers from consideration because we need to compute the searched area.

Each g-event is characterized by five observational features: the SNRs from the H1, L1, and V1 detectors, the logarithm of the FAR, and the chirp mass.
We include the chirp mass because it is the best-measured intrinsic parameter for CBC signals and provides a stable identifier for ordering g-events.
We impose a maximum limit of 22 g-events per superevent.
This corresponds to the maximum number of g-events corresponding to a superevent in MDC11.
While 22 serves as a practical threshold in this study, larger values can be adopted in future analyses without loss of generality.
Superevents containing fewer than 22 constituent g-events are zero-padded to maintain uniform input dimensionality.
Consequently, each superevent is encoded as a fixed-dimension matrix of 22 $\times$ 5 floating-point values.
The g-events within each superevent are ordered by ascending chirp mass and indexed from 0 to 21, with index 0 corresponding to the g-event with the lowest chirp mass.
This sorting procedure ensures that the neural network predictions remain invariant to the temporal sequence of g-event detection.
Alternative sorting parameters may be substituted.
The sorting does not change the outcome of the neural-network-based selector as shown in Appendix~\ref{app:random_sort}.

To assess model performance and evaluate robustness against data partitioning, the complete superevent dataset is split into training and testing sets using a 9:1 ratio.
This split is a standard choice in supervised learning when the training set is modest in size, balancing sufficient data for parameter learning against retention of an independent test set. \cite{Goodfellow}.
Identical superevent allocations are maintained across both high- and low-significance filtered datasets, ensuring consistency.
This partitioning procedure is iterated 10 times, yielding ten training-testing pairs.
The ten test pairs do not overlap and are independent datasets.
This approach enables us to assess the stability and predictive uncertainty of the neural network across different data configurations, helping ensure that performance is not overly sensitive to a particular split.

\section{Methods}
\label{sec:methods}
We present a neural-network-based approach to identify the preferred event based solely on the outputs of the search pipelines, without relying on any knowledge of the true sky location of the injected signal.
The goal of this work is to assess whether a neural network can be developed that takes as input the collective outputs from multiple low-latency search pipelines and successfully selects the event with the \textit{best} sky map for multimessenger follow-up.
For training purposes, we use the sky map with the lowest searched area as the \textit{best} sky map.
Our network is designed to be pipeline agnostic, a deliberate choice to mitigate systematic biases toward any specific search algorithm.
By treating all pipelines equally and learning only from the aggregate features they report, we aim to achieve robust generalization across diverse pipeline-specific characteristics.

\subsection{Network architecture}
We implement a fully connected feedforward architecture using PyTorch, comprising an input layer, two hidden layers, and an output layer.
\begin{itemize}
\item[(i)] Input layer: A linear transformation from 110 (flattened from the 22 $\times$ 5 matrix representing pipeline outputs) input features to 256 hidden units. The inputs of the layer are normalized using one-dimensional batch normalization \cite{ioffe2015batchnormalizationacceleratingdeep}, followed by a rectified linear unit (ReLU) activation function to introduce nonlinearity.
\item[(ii)] Hidden layer: Two successive hidden layers progressively reduce dimensionality from 256 to 128, then to 64 units. Each transformation incorporates batch normalization and ReLU activation to maintain stable training dynamics.
\item[(iii)] Output layer: The final layer projects the hidden representation consisting of 64 units to 22 output units, corresponding to the 22 possible g-event indices within each superevent.
These outputs represent raw scores (logits).
Padding slots corresponding to zero-padded g-events are masked to $-\infty$ before computing the loss, ensuring they cannot be selected.
We use a categorical cross-entropy loss function \footnote{\url{https://docs.pytorch.org/docs/stable/generated/torch.nn.CrossEntropyLoss.html}}, with a soft-ranking loss: the soft loss is a Kullback--Leibler (KL) divergence \cite{KLD} between the predicted log-probabilities and target weights inversely proportional to the searched area for each g-event, encouraging the network to assign higher scores to better-localized events.
The two loss terms are combined with equal weight (0.5 each).
\end{itemize}

To prevent overfitting and ensure robust generalization to unseen GW events, we implemented several regularization techniques throughout the network architecture.
Batch normalization, introduced in 2015 \cite{ioffe2015batchnormalizationacceleratingdeep}, provides fast and stable training by scaling and recentering the inputs to each layer.
Dropout regularization \cite{hinton2012improvingneuralnetworkspreventing} with a rate of 10\% was employed between hidden layers to prevent coadaptation of neurons and improve model robustness.
This essentially randomly ignores some nodes, which helps to improve performance on unseen data by reducing overfitting.
L2 weight decay with a coefficient of $1 \times 10^{-4}$ was applied during optimization to keep the weights small, which prevents the gradients from exploding during training.
Network weights were initialized using Xavier uniform initialization \cite{pmlr-v9-glorot10a} to ensure stable gradient flow during early training phases.
These comprehensive regularization strategies were chosen due to the limited size of the dataset relative to the model complexity.

To optimize the network, we use the Adam optimizer \cite{kingma2017adammethodstochasticoptimization} with an initial learning rate of 0.001.
The composite loss described above is well suited for the multiclass selection task, with the cross-entropy term anchoring the hard classification target and the soft-ranking term providing graded feedback across all candidate g-events.
Training is performed with minibatches of 64 samples to balance computational efficiency with stable gradient estimation.

We train using the entire training dataset and use the test dataset for validation.
Validation loss is computed starting from epoch 20 to allow the network to stabilize before early stopping is applied.
Early stopping is implemented with a patience of 50 epochs based on validation loss to prevent overfitting, complemented by learning rate reduction on plateau scheduling (factor $=$ 0.5, patience $=$ 20 epochs) to fine-tune convergence.
Training is allowed to proceed for a maximum of 1000 epochs, though early stopping typically terminates training significantly earlier.

All computations are performed on Mac M3 Pro hardware utilizing metal performance shaders (MPS) for GPU acceleration, with models operating in float32 precision to ensure MPS compatibility.
Training the neural network is computationally inexpensive, taking around a minute on an Apple M3 Pro processor (with MPS).
After training, the step involving identifying the preferred event can be performed instantaneously using the models stored on disk.
This efficiency makes the neural-network-based preferred event selector particularly well suited for low-latency applications. We also expect the neural network to have two additional properties:

\begin{enumerate}
    \item The event selected by the neural network should not be worse than the event selected using the present SNR-based algorithm on average.
    \item The per pipeline fraction of events selected by the neural network should closely follow the distribution of events with the lowest searched area per pipeline.
\end{enumerate}

In a realistic deployment scenario, we envision training the neural network on a set of simulated signals injected into real detector data, ideally from a stretch of time that closely mimics the noise and conditions of the observational run.
The neural network, trained to recognize which events tend to have the lowest searched area, could then be applied to low-latency data during actual observations to select the most promising candidate for electromagnetic follow-up.

However, for the scope of this study, we evaluate the neural network on the MDC11 dataset, which comprises simulated CBC injections into archival detector data.
As outlined in earlier sections, we split the dataset into training and testing subsets using a 9:1 ratio, and repeat this procedure across ten different random seeds to quantify the variability and robustness of the model’s predictions.
Each 9:1 split is referred to as an iteration in this work.
This enables a thorough assessment of the model’s accuracy, generalization, and sensitivity to data splits, establishing a strong foundation for future deployment on real observational data.

\section{Results}
\label{sec:results}
We train ten models on the ten iterations of the significant dataset and ten other models on ten iterations of the datasets containing all triggers recovered at $\textrm{FAR} < 2$/day (referred to as all triggers henceforth).
Out of 3248 superevents in the significant dataset, 61\% of the significant events have at least one g-event with a neutron star (component of one of the masses $< 2.05 M_\odot$), 26\% have at least one g-event with a 90\% sky area less than 200~deg$^2$, and 16\% satisfy both criteria.
These subsets motivate the stratified analyses presented below.
Including all triggers increases the size of the training set by roughly 47\%.
The performance of each model is then evaluated on both datasets.
Training models on both the significant-only and all-event datasets enable a direct comparison of their performance.
The neural network takes $\sim 70$ iterations to learn sufficiently with a training loss $\leq 0.5$ ($\leq 0.55$) and a validation loss of $\leq 0.9$ ($\leq 1.0$) when trained on significant (all) events.

We do not use the fraction of correctly identified preferred events as our primary performance metric.
Many superevents contain only a single g-event, for which any selector trivially makes the correct choice.
Including these in a counting metric inflates apparent accuracy without reflecting the selector's behavior on the multi-g-event superevents that actually present a selection problem.
Instead, we evaluate selectors by the searched area distribution of their chosen events.

\begin{figure}[h!]
    \centering
    \includegraphics[width=\textwidth]{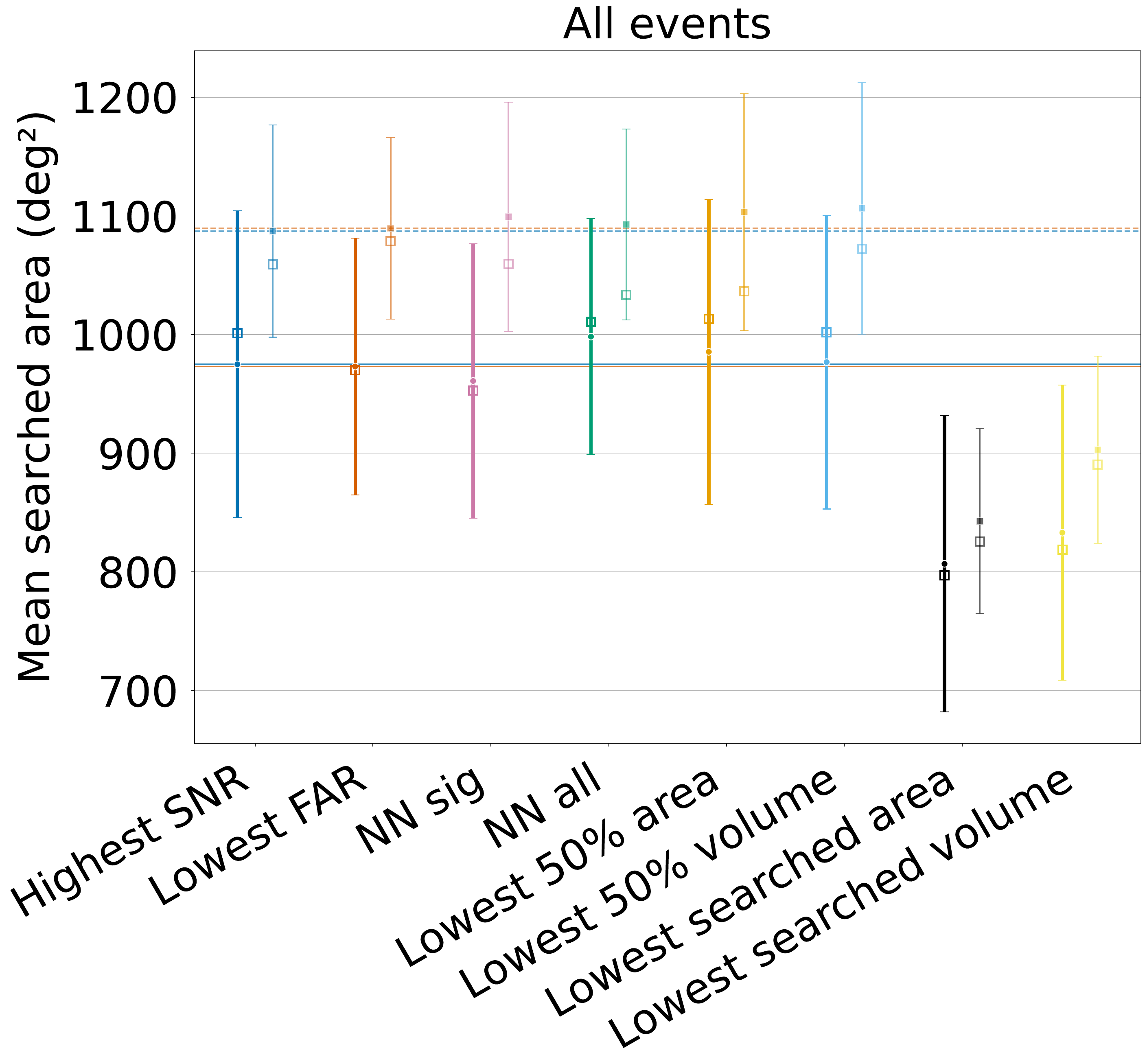}
    \caption{Mean searched areas of the sky maps chosen for significant (darker and thicker lines) and all (lighter and thinner lines) events by different selection algorithms. NN sig denotes the neural network trained on significant events, and NN all denotes the neural network trained on all events. The spread ($1\sigma$) in the mean searched area is due to the ten test sets. The solid (dashed) line denotes the mean searched areas for the sky maps of significant (all) events chosen by highest SNR (blue) and lowest FAR (orange).The boxes denote the mode of the distribution of means of the ten test sets.}
    \label{fig:summary}
\end{figure}

\begin{figure}[ht]
    \centering
    \begin{subfigure}[t]{0.475\textwidth}
        \centering

        \includegraphics[width=\textwidth]{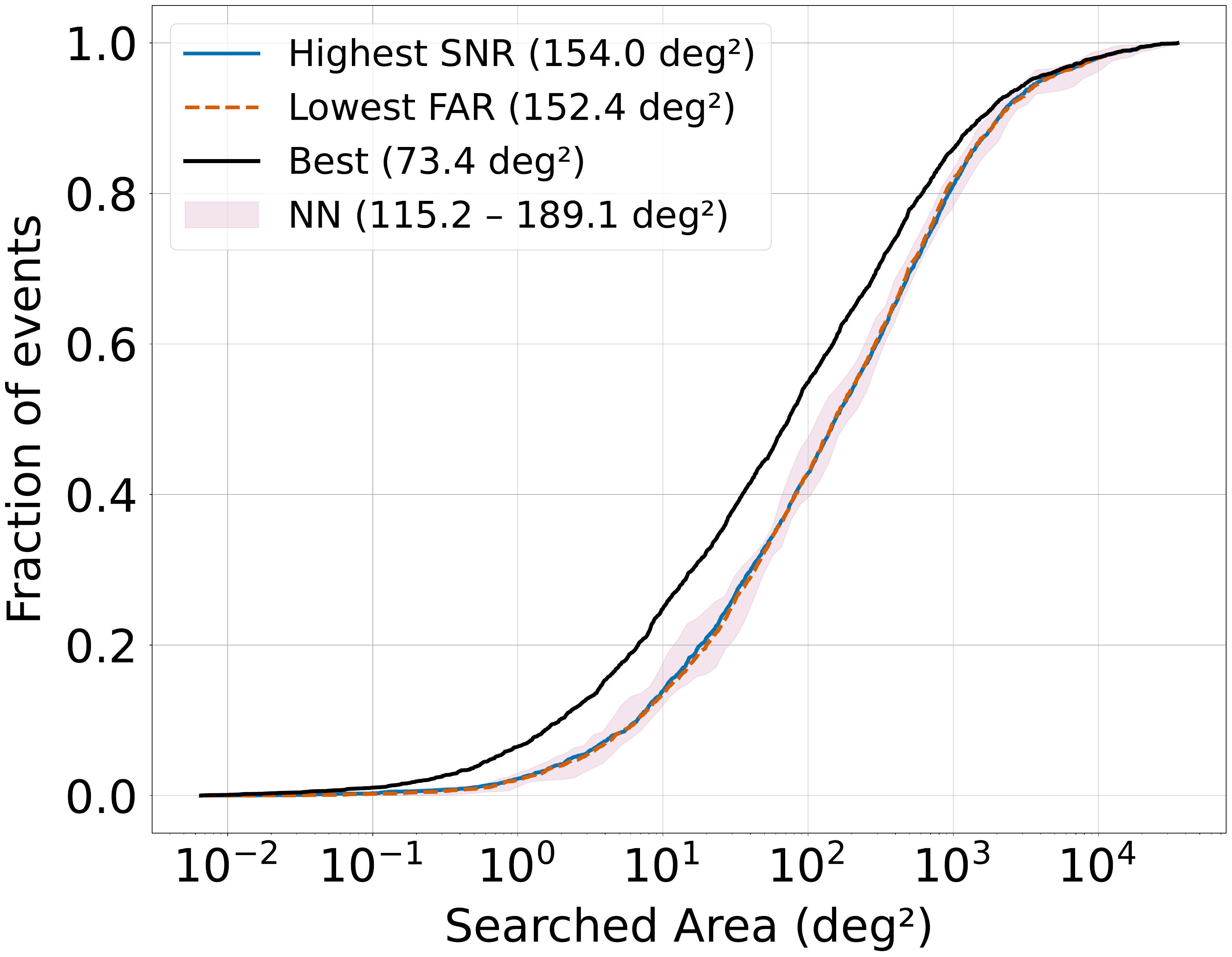}
        \caption{Significant events (FAR < 1/5 months)}
    \end{subfigure}
    \hfill
    \begin{subfigure}[t]{0.475\textwidth}
        \centering
        \includegraphics[width=\textwidth]{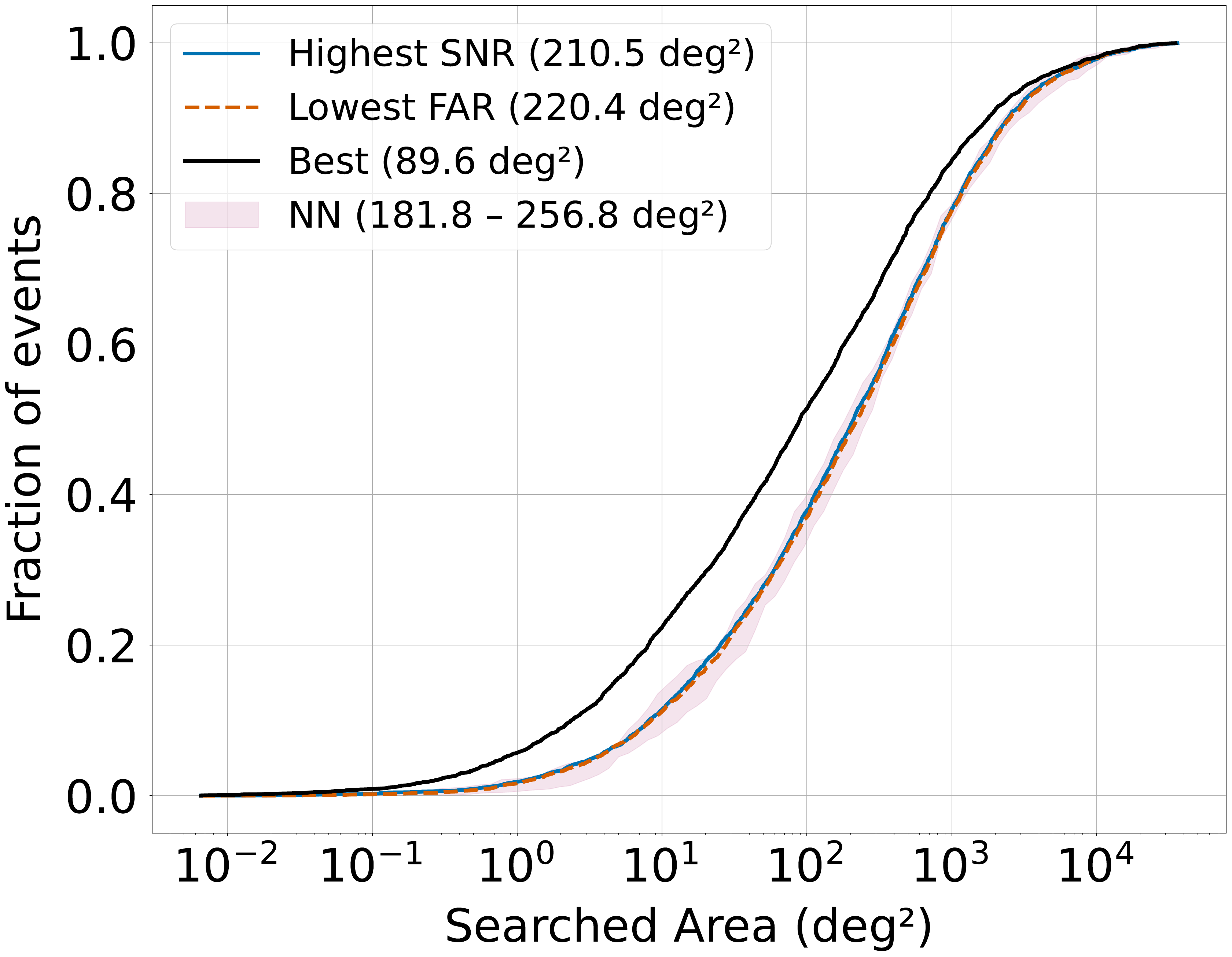}
        \caption{All events recovered with FAR < 2/day}
    \end{subfigure}
\caption{The cumulative distribution of searched areas for g-event sky maps comprising (a) significant events and (b) all events selected by lowest FAR, by highest SNR, by lowest searched area (denoted by best), and the NN trained on (a) significant and (b) all events. The median searched areas have been reported in parentheses. The bands have been obtained by evaluating different preferred event selectors on the ten test datasets.}
\label{fig:SAFrac_all}
\end{figure}

To assess the performance of various selection schemes, we examine the mean searched area for events obtained by various selectors (see Fig.~\ref{fig:summary}).
Based on the means, we find that the SNR-based selector (mean searched area of 975~deg$^2$ for significant events and 1088~deg$^2$ for all events) performs comparable to the FAR-based selector (mean searched area of 973~deg$^2$ for significant events and 1090~deg$^2$ for all events).
Selectors based on metrics such as the lowest 50\% sky area and the lowest 50\% sky volume perform poorly as seen in Fig.~\ref{fig:summary}.
The error bars have been computed from the standard deviation of the means obtained from the ten smaller datasets used to test the neural network's performance.
As described earlier, we train two neural networks: one only on significant events and the other on all events.
The neural network trained on significant events achieves a mean searched area of 961~deg$^2$ on the significant test set, marginally below the means of both traditional selectors.
The neural network trained on the all-events dataset achieves comparable performance to the SNR selector when evaluated on its matched all-events test set, but transfers poorly to the significant test set as seen in Fig.~\ref{fig:summary}.
The fraction of events versus the searched area for all events is shown in Fig.~\ref{fig:SAFrac_all}.

\begin{figure}[ht]
    \centering
    \begin{subfigure}[t]{0.475\textwidth}
        \centering

        \includegraphics[width=\textwidth]{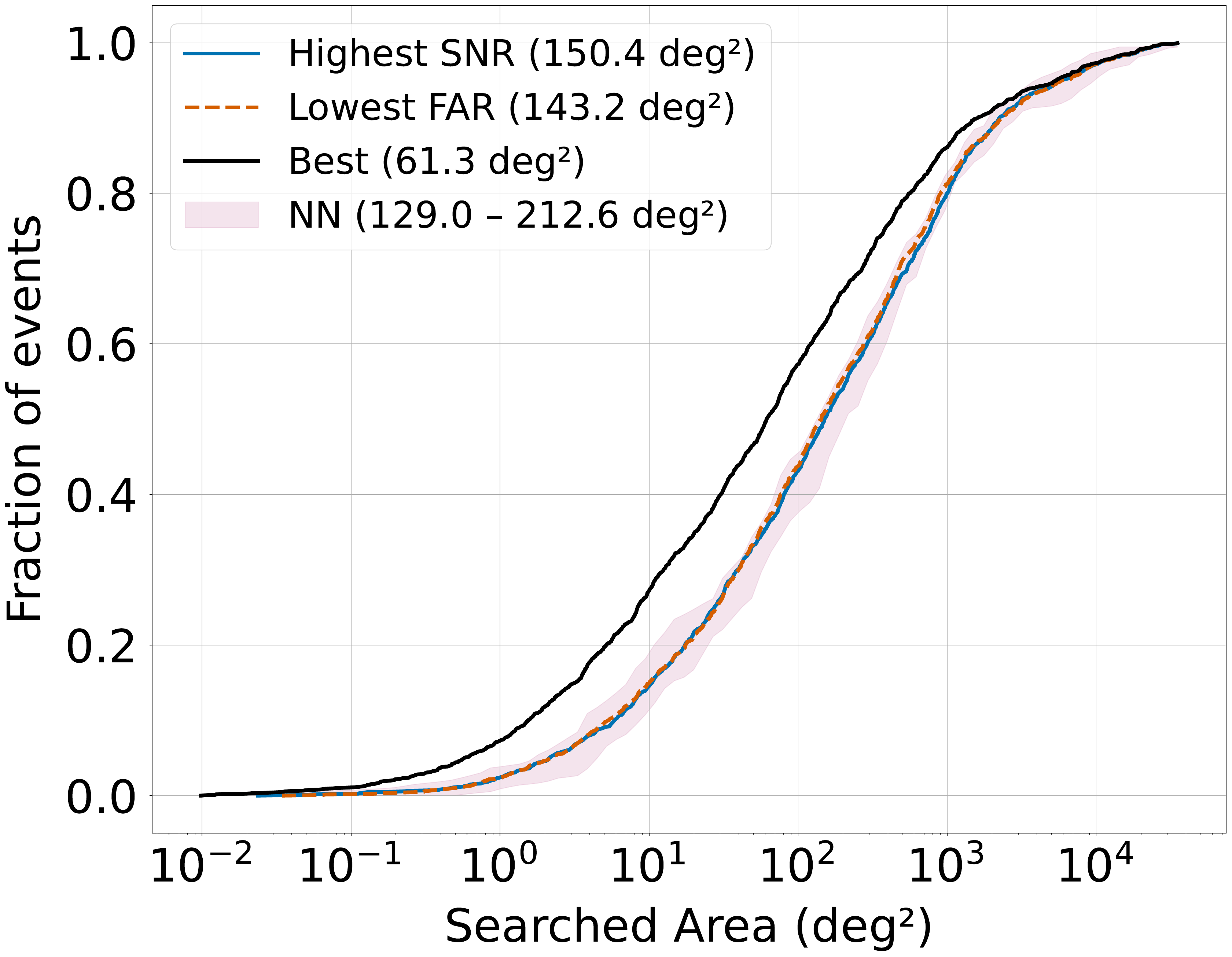}
        \caption{Significant events with neutron stars}
    \end{subfigure}
    \hfill
    \begin{subfigure}[t]{0.475\textwidth}
        \centering
        \includegraphics[width=\textwidth]{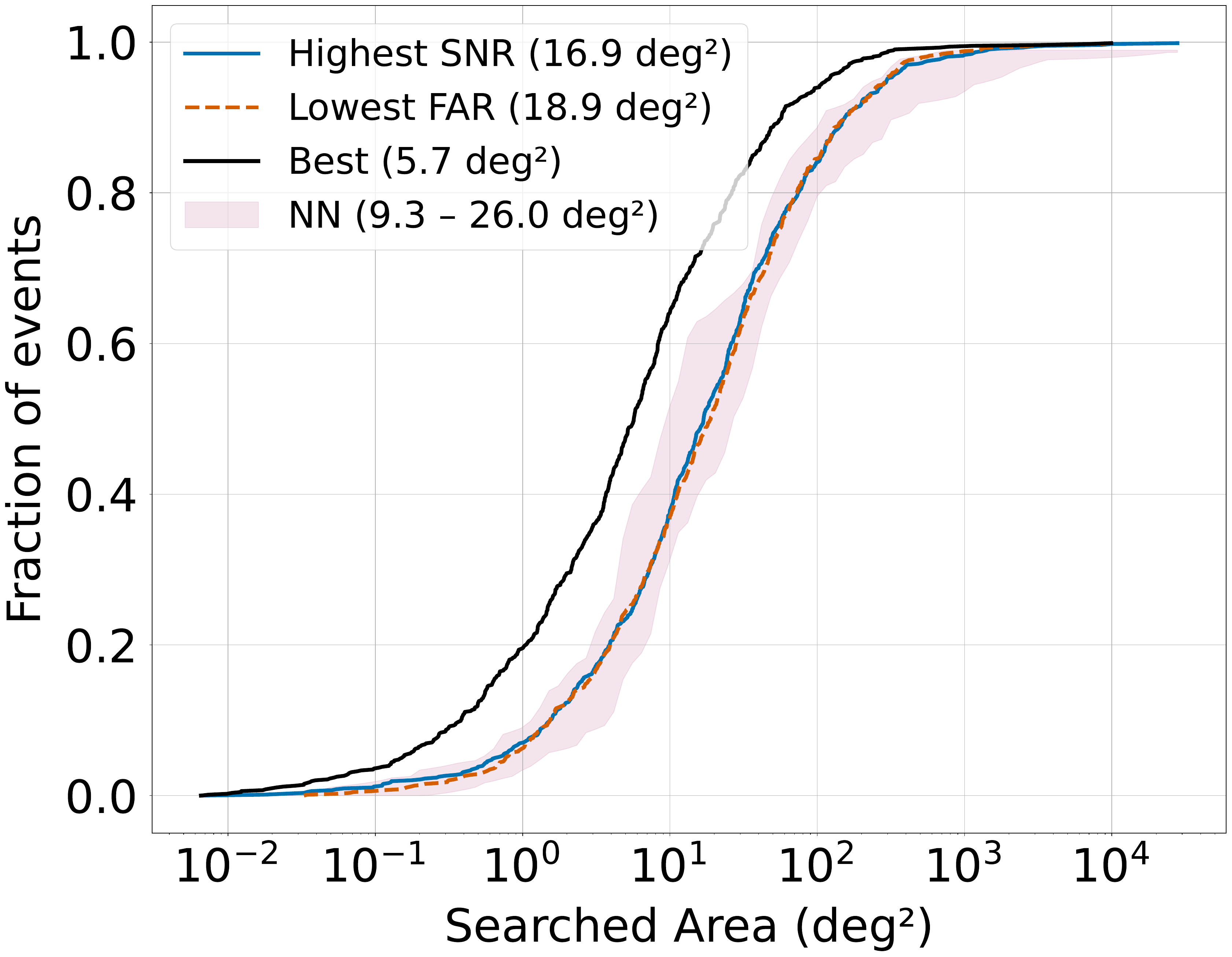}
        \caption{Significant events with 90\% sky area $< 200$~deg$^2$}
    \end{subfigure}
\caption{The cumulative distribution of searched areas for g-event sky maps comprising significant events (a) with neutron stars and (b) with good localization selected by lowest FAR, by highest SNR, by lowest searched area (denoted by best), and the NN trained on significant events. The median searched areas have been reported in parentheses. The bands have been obtained by evaluating different preferred event selectors on the ten test datasets.}
\label{fig:SAFrac_subset}
\end{figure}

We note an important caveat about how to interpret the summary statistics.
The searched-area distributions for all selectors (Fig.~\ref{fig:SAFrac_all}) span more than 4 orders of magnitude and are strongly non-Gaussian, with heavy tails extending to $\sim 10^4$~deg$^2$.
In this regime, the mean is highly sensitive to a small number of poorly localized events and can shift substantially between dataset splits.
At the same time, the median and mode characterize the bulk of the distribution and respond more weakly to outliers.
The cumulative distributions in Figs.~\ref{fig:SAFrac_all} and~\ref{fig:SAFrac_subset} show that the NN, SNR, and FAR selectors produce distributions that overlap nearly everywhere across this dynamic range and they differ only in subtle, distribution-level shifts rather than in the bulk shape.
We therefore report the means primarily because they are the conventional summary statistic, but we caution against overinterpreting small differences in mean values.
For this reason, the bands shown in Figs.~\ref{fig:SAFrac_all}, \ref{fig:SAFrac_subset}, \ref{fig:RCA}, and \ref{fig:CCA}, which capture the spread of cumulative distributions across the ten train/test splits, are the most informative summary of selector behavior.

\begin{figure}[ht]
    \centering
      \begin{subfigure}[t]{0.45\textwidth}
      \centering
      \includegraphics[width=\textwidth]{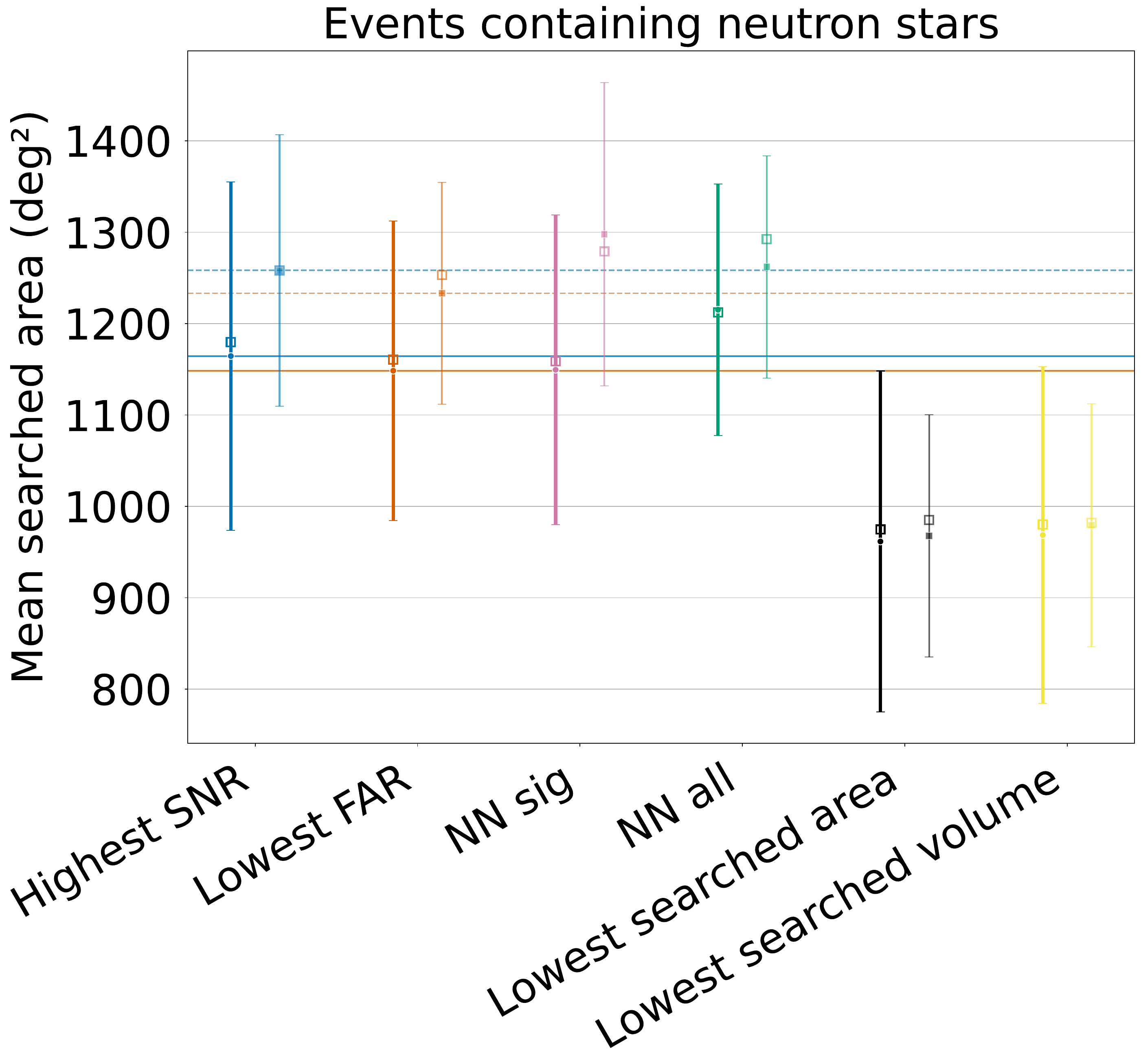}
      \caption{Events containing neutron stars}
      \label{fig:summary_has_NS}
      \end{subfigure}
    \vspace{1em}
            \begin{subfigure}[t]{0.45\textwidth}
            \centering
            \includegraphics[width=\textwidth]{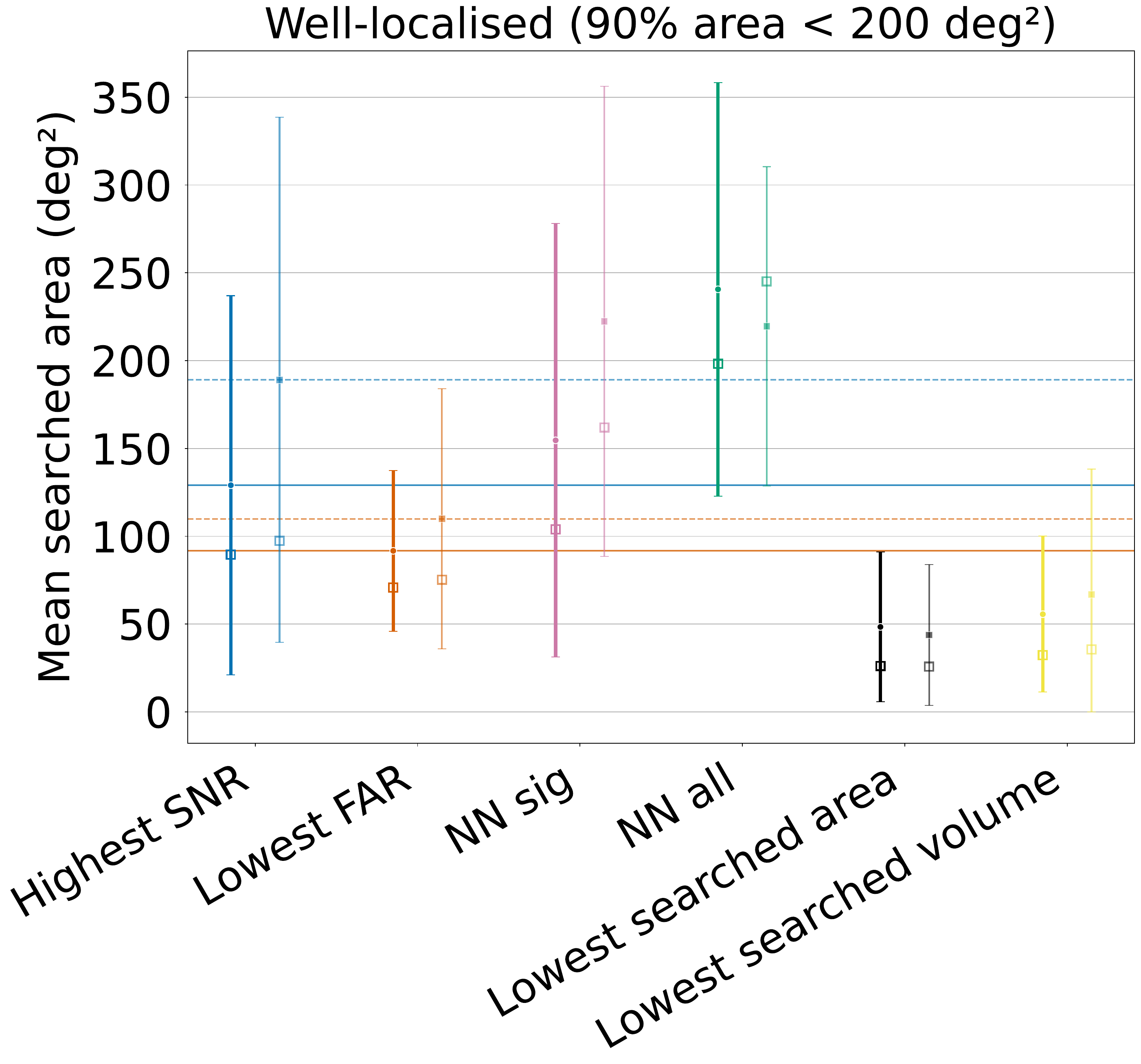}
          \caption{Well localized events (90\% credible area $< 200$\,deg$^2$)}
          \label{fig:summary_good_loc}
          \end{subfigure}
    \caption{Mean searched areas of the sky maps chosen for significant (darker and thicker lines) and all (lighter and thinner lines) events for (a) superevents with at least one g-event containing a neutron star and (b) superevents with at least one g-event having 90\% credible sky area $< 200$~deg$^2$. NN sig denotes the NN trained on significant events, and NN all denotes the neural network trained on all events. The spread ($1\sigma$) in the mean searched area is due to the ten test sets. The solid (dashed) line denotes the mean searched areas for the sky maps of significant (all) events chosen by highest SNR (blue) and lowest FAR (orange). The boxes denote the mode of the distribution of means of the ten test sets.}
      \label{fig:summary_subsets}
  \end{figure}

We further investigate two subsets of superevents: (a) those with at least one g-event with a component below 2.05 $M_\odot$, i.e., containing a neutron star, and (b) those containing at least a g-event with a 90\% credible sky area smaller than 200~deg$^2$ (the well-localized subset).
As shown in Fig.~\ref{fig:summary_has_NS}, the trends for neutron-star-bearing events are similar to the overall population.
For the well-localized subset [Fig.~\ref{fig:summary_good_loc}], however, the neural-network-based selector performs worse than the highest-SNR selector and the lowest-FAR selector on average.

Table~\ref{tab:area_regimes} provides a finer stratification based on superevents with at least a g-event with 90\% credible sky area below various thresholds.
For well-localized sources with a 90\% credible region below 200~deg$^2$, the FAR-based selector outperforms both the SNR and the neural-network-based selectors as seen in Fig.~\ref{fig:summary_good_loc}.
For very well-localized sources (90\% area $< 10$~deg$^2$), the mean searched area of the SNR and neural network selectors is elevated by spurious pipeline uploads.
Some pipelines produce G-events with spuriously high SNR and very tight sky localizations that are pathological.
For example, S230219ar contains two g-events with 90\% credible areas of $\sim$1.2~deg$^2$ and combined SNR $\sim$42, yet their sky maps place negligible probability at the true source location, resulting in searched areas exceeding 27,000~deg$^2$.
The neural network is similarly misled for such outliers.
The neural-network-based selector performs well for events with 90\% area $> 200$~deg$^2$.
This is due to the fact that well-localized events are a minority in the training dataset.
Training a separate network on a well-localized subset without any pathologies is expected to improve performance.
However, this requires the construction of a training set with a large number of well-localized sources.
We leave this for the future.

\begin{table}[ht]
  \caption{Mean searched area for the highest-SNR, lowest-FAR, and neural-network-based
  selectors, stratified by whether the superevent contains at least one g-event whose
  90\% credible region area falls below the regime threshold.
  Each superevent is assigned to the first (smallest) regime for which such a g-event exists.}
  \label{tab:area_regimes}
\begin{tabular}{lrrrr}
    \hline\hline
    Regime (min 90\% & & Highest SNR & Lowest FAR & NN \\
    CR area, deg$^2$) & Count & \multicolumn{3}{c}{Mean searched area (deg$^2$)} \\
    \hline
    $< 10$        &   83 &  360.5 &   36.0 &  412.3 \\
    $10$--$200$   &  754 &  103.6 &   97.8 &  126.3 \\
    $200$--$1000$ & 1236 &  370.0 &  359.2 &  347.1 \\
    $> 1000$      & 1175 & 2213.8 & 2246.8 & 2181.0 \\
    \hline\hline
  \end{tabular}
\end{table}

To characterize the operational impact of the NN selector, we examine the distribution of differences in selected-event searched area between the NN and the highest-SNR selector across all significant superevents, which we denote $\Delta$.
In 41\% of cases, the two selectors choose the same g-event ($\Delta = 0$), so the NN introduces no change for nearly half of all events.
In 12\% of superevents ($\Delta < -100$~deg$^2$), the NN reduces the mean selected area by more than a factor of 2, from 1407 to 644~deg$^2$.
These are predominantly events containing at least one neutron star (68\%), for which the best possible low-latency sky map is most valuable for multimessenger follow-up.
Within this $\Delta < -100$ subset, the NN brings 50\% of events below the 200~deg$^2$ follow-up threshold, compared to only 10\% for the SNR-based selector.
The picture is not uniformly positive: in 11\% of cases, the SNR selector outperforms the NN by more than 100~deg$^2$ ($\Delta > +100$~deg$^2$), and these events have a similar neutron-star fraction (67\%).
The symmetric appearance of neutron-star-bearing events in both tails indicates that the network is not selectively biased toward neutron-star sources.
It uses information complementary to SNR that helps in some events and hurts in others, with a net positive effect on the population mean.

\begin{figure}[ht]
    \centering

    \begin{subfigure}[t]{0.45\textwidth}
        \centering
        \includegraphics[width=\textwidth]{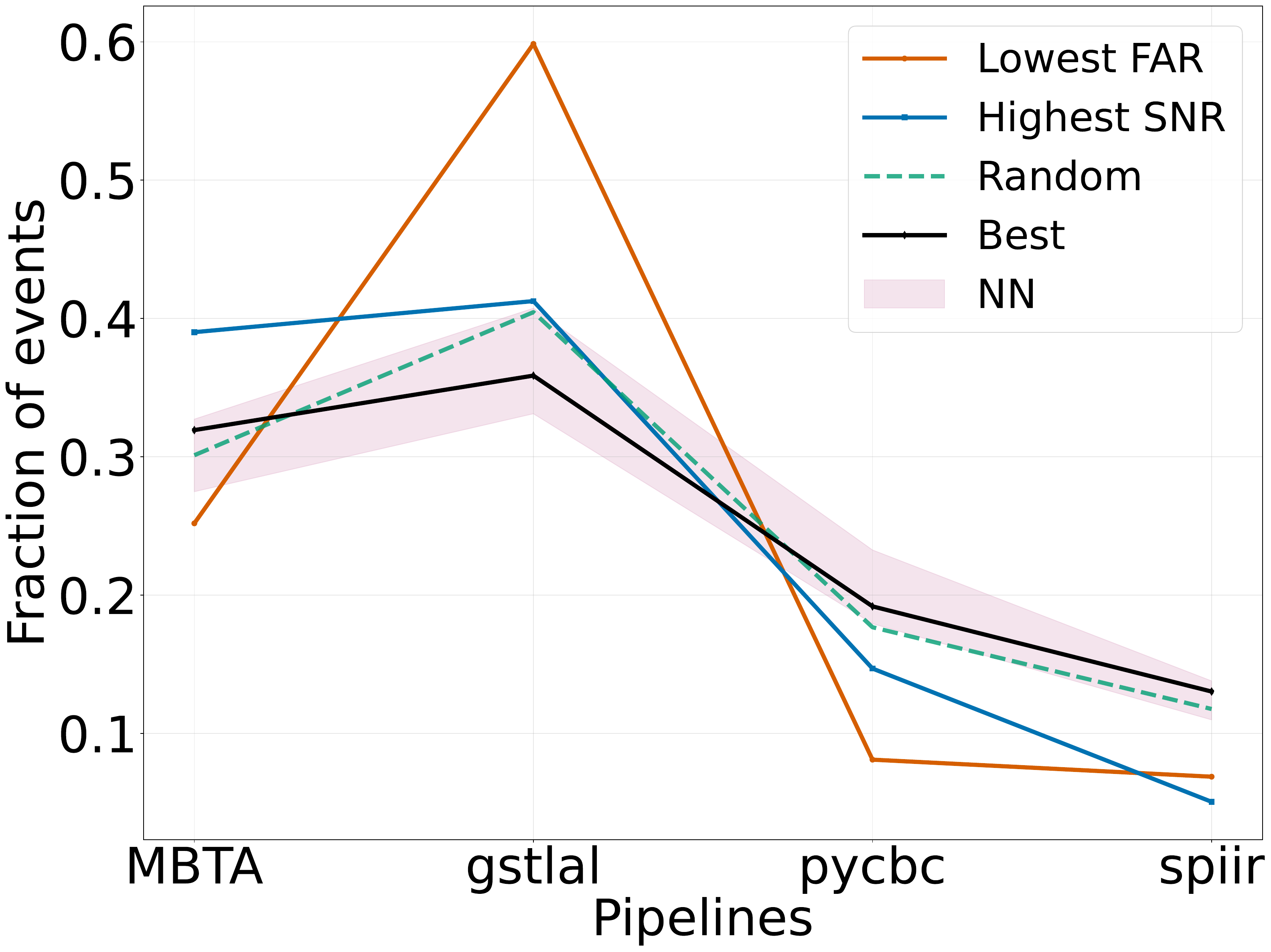}
        \caption{All significant events}
    \end{subfigure}
    \hfill
    \begin{subfigure}[t]{0.45\textwidth}
        \centering
        \includegraphics[width=\textwidth]{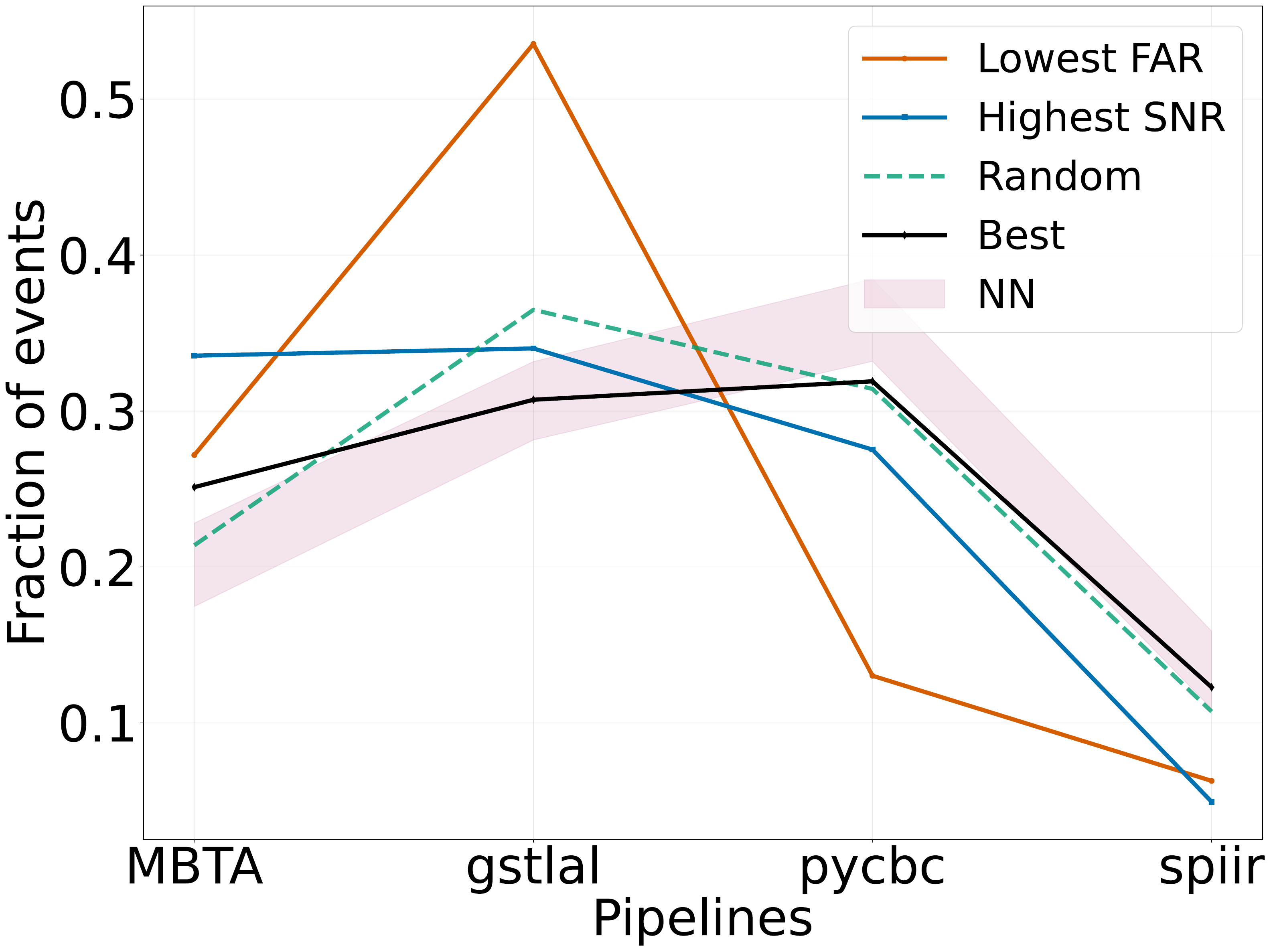}
        \caption{All events recovered with FAR < 2/day}
    \end{subfigure}

    \vspace{8pt}

    \begin{subfigure}[t]{0.45\textwidth}
        \centering
        \includegraphics[width=\textwidth]{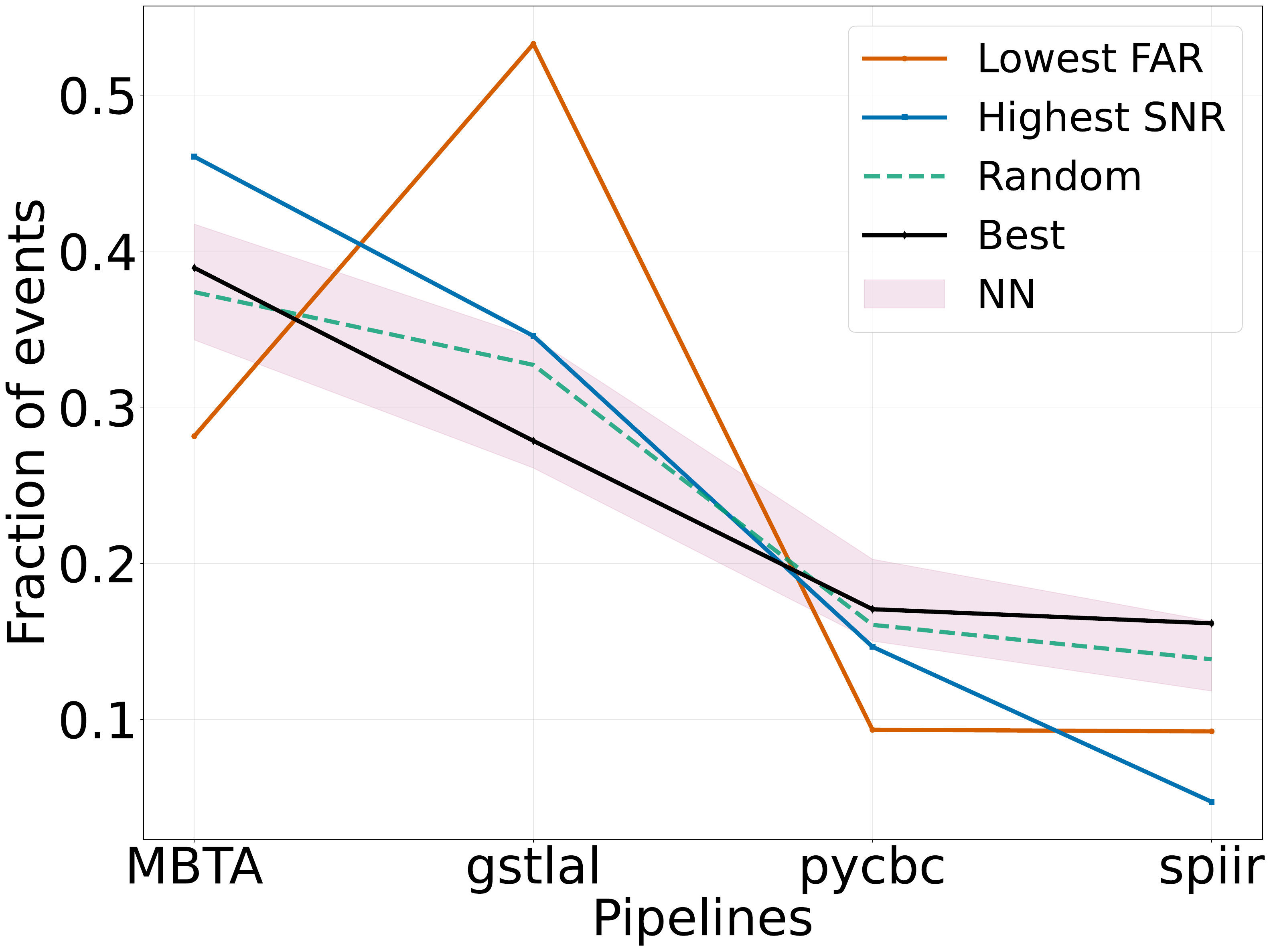}
        \caption{Significant events containing a neutron star}
    \end{subfigure}
    \hfill
    \begin{subfigure}[t]{0.45\textwidth}
        \centering
        \includegraphics[width=\textwidth]{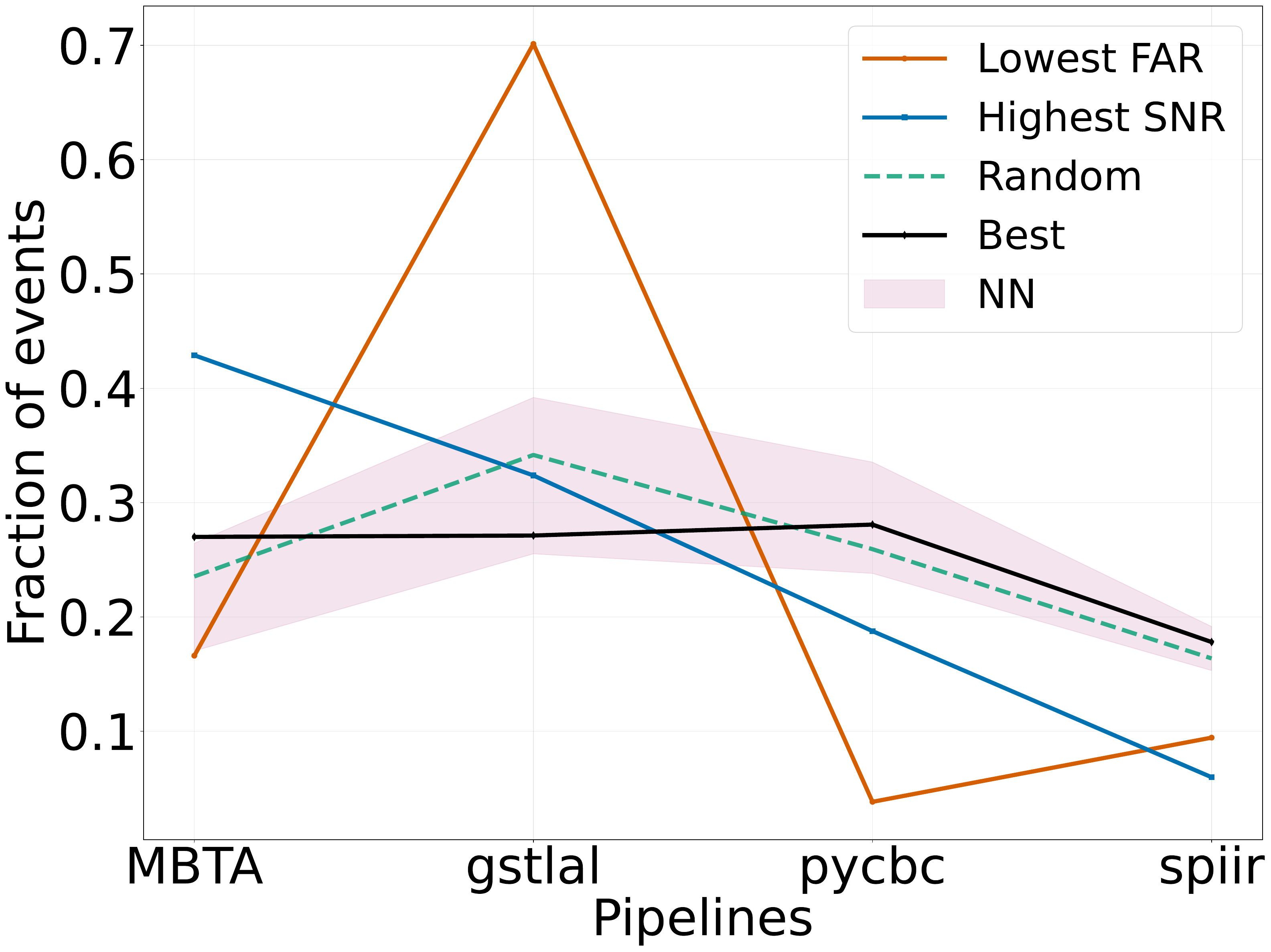}
        \caption{Significant events with 90\% credible region $< 200$~deg$^2$}
    \end{subfigure}

    \caption{Fraction of selected g-events per pipeline for different selection choices, namely (a) significant events, (b) all events, (c) significant events with an NS, and (d) well-localized significant events. The best line corresponds to the g-event with the lowest searched area in each superevent (the ground truth). The dashed line shows the average fraction of g-events per superevent recovered by each pipeline. This is equivalent to picking g-events at random from a superevent.}
    \label{fig:pipeline_fraction}
\end{figure}

The developed neural-network-based selector, however, stands out in its representation of pipelines across the selected events as shown in Fig.~\ref{fig:pipeline_fraction}.
Since MDC11 injections have a known ground truth, we can quantify how often each pipeline recovers the lowest-searched-area event.
Both the highest-SNR and lowest-FAR selectors exhibit strong biases toward specific pipelines, deviating from the true underlying distribution.
The neural-network-based selector, which is pipeline agnostic by design, in contrast, closely follows the underlying distribution while selecting events.
This is crucial in realistic applications to ensure equitable treatment of all low-latency pipelines.

To summarize, using MDC11 injections, we find that the lowest-FAR and highest-SNR selectors achieve comparable mean searched areas, with the FAR-based selector marginally better.
Our neural-network-based selector achieves searched-area performance similar to both traditional selectors but, unlike them, preserves the underlying pipeline distribution of ground-truth best-localized events.
The network is computationally inexpensive to train ($\sim$1 min on consumer hardware) and selects events instantaneously, making it suitable for low-latency deployment.

\section{Discussions}
\label{sec:discussion}
This work shows that a neural-network-based preferred event selector can serve as an effective alternative to the current strategies employed in low-latency gravitational-wave follow-up.
Our selector achieves searched-area performance comparable to the highest-SNR and lowest-FAR methods on average, while uniquely preserving the distribution of pipeline contributions seen in ground-truth injected events.
Once trained, the neural network can identify the preferred event almost instantaneously, making it highly suitable for low-latency applications.
Given these benefits, we conclude that a neural-network-based approach is worthy of serious consideration for the next observing run (O5).

The conclusions in our study are entirely based on MDC11, which was conducted before the start of the current observing run using data from the third observing run.
The majority of the events have 90\% credible intervals greater than 200~deg$^2$, and so the neural network performs the best in this regime.
The poor performance of the neural network for well-localized sources is a reflection of the training dataset and not the methodology used.
In principle, different training sets can be constructed to select g-events for different problems.
Fortunately, the lightweight nature of our neural network model makes retraining on a more representative dataset both straightforward and computationally inexpensive.

The LIGO detectors have become more sensitive, while Virgo has a similar BNS range.
Moreover, all pipelines have evolved significantly, with new features such as the SNR optimizers \cite{Joshi:2025nty} being introduced and additional pipelines like Aframe \cite{Aframe} and cwb-bbh \cite{2025arXiv250818081T} joining the search.
Our results remain relevant, as the goal is not to compare pipeline performance but to demonstrate, as a proof of concept, that a neural-network-based event selector can be deployed in low latency.
Nonetheless, substantial work is still required before deployment in an observing run.
To enable realistic deployment, it is essential to train the neural network on data that more accurately reflect the state of the interferometers and the low-latency pipelines during an observing run.

An additional unexplored dimension is the dependence of selector performance on sky position.
The sensitivity of a particular detector network varies across the sky, and the NN may behave differently in directions where BAYESTAR localization is systematically better or worse.
We defer this analysis to future work with larger injection sets.

Moreover, there is scope to enhance the model’s performance by incorporating additional features.
For sources observed in multiple detectors, incorporating additional information like the time delay and phase difference between detectors could potentially enhance the selection of the event with a lower searched area.
These features capture aspects of the SNR time series that are sensitive to sky localization and may help the network more reliably identify the preferred event.
Exploring such physically motivated features is left for future work.

A second extension concerns the choice of training target.
We train the network to identify the g-event with the lowest searched area, which is a two-dimensional metric that ignores distance.
The g-event with the lowest searched area is generally not the same as the one with the lowest searched volume [see Figs.~\ref{fig:summary}, \ref{fig:summary_has_NS}, and \ref{fig:summary_good_loc}], indicating that searched-area-optimal selections may carry suboptimal distance estimates.
For science cases where accurate distance is critical in low latency, like host-galaxy identification or volume-limited follow-up strategies, the same network architecture could be retrained against the searched volume as the objective and is a natural avenue for future work.

Our neural network architecture arranges the g-events within each superevent in ascending order based on chirp mass.
This is done to ensure that the network is invariant to the order in which g-events arrive.
This setup ensures that the preferred event is determined only by detection pipeline outputs (SNR, FAR, and chirp mass).
Nevertheless, this restriction is not required.
We observe that key statistical properties of the selection, including searched-area performance and pipeline representation, remain stable even when the g-events are ordered randomly.
However, we would like our network to be permutation invariant for reasons of reproducibility.
This design choice reduces the effect of extraneous factors such as data transmission latencies, computational resource allocation, or cluster processing schedules.
We also evaluated a fundamentally different architecture, a PointNet-style permutation-invariant network, which achieves comparable searched-area performance but does not preserve the pipeline distribution as well as the feedforward neural network (see Appendix~\ref{app:cnn}).
This indicates that pipeline fairness is sensitive to the inductive biases of the chosen architecture, not just to the pipeline-agnostic input design.

Our neural network is pipeline agnostic by input design.
We intentionally omit the identity of the pipeline that produced each g-event.
Although one might expect that including such information could enhance performance, given the inherent differences in FAR and SNR distributions across pipelines, our experiments found no significant performance gains from incorporating pipeline identity.
This might be partly because FAR is sufficiently informative to identify which pipeline sent the trigger.
We emphasize, however, that pipeline-agnostic inputs do not automatically yield pipeline-fair behavior.
As seen in Appendix~\ref{app:cnn}, an alternative pipeline-agnostic architecture (a PointNet-style network) does not preserve the ground-truth pipeline distribution as well as the feedforward neural network.
The pipeline-fairness property of our main result, therefore, reflects the specific combination of architecture and loss design we have adopted.


\begin{acknowledgments}
We thank the anonymous referee for constructive feedback that helped in improving the overall quality of the manuscript. The authors would like to thank Deep Chatterjee for providing valuable feedback during the internal LIGO review.
P.~B. would like to thank Gaurav Waratkar, Andrew Toivonen, and Sushant Sharma-Chaudhary for their assistance in calculating the searched area for sky maps corresponding to each g-event in MDC11.
P.~B. is also grateful to Soichiro Morisaki, Leo Singer, Leo Tsukada, Carl Haster, Jolien Creighton, Lucy Thomas, and Adrian Cornell for their valuable comments and suggestions, which helped improve this work.
This material is based upon work supported by NSF's LIGO Laboratory which is a major facility fully funded by the National Science Foundation.
This work was supported partially by the National Science Foundation (NSF) Awards No.~NSF PHY-2513124, No.~PHY-2207728, No.~PHY-2110594, and No.~PHY-2513358 and partially by Wisconsin Space Grant Consortium Award No.~RFP25\_9-0. This work carries LIGO Document No.~LIGO-P2500665.
\end{acknowledgments}

\section*{Data Availability}
This work reuses the MDC11 data, which is described and available at \cite{doi:10.1073/pnas.2316474121}. The codes for this project are hosted at \cite{preferred_event_ml}.

\appendix
\section{Results without sorting by chirp mass}
\label{app:random_sort}
\begin{figure}[ht]
    \centering
    \begin{subfigure}[t]{0.475\textwidth}
        \centering

        \includegraphics[width=\textwidth]{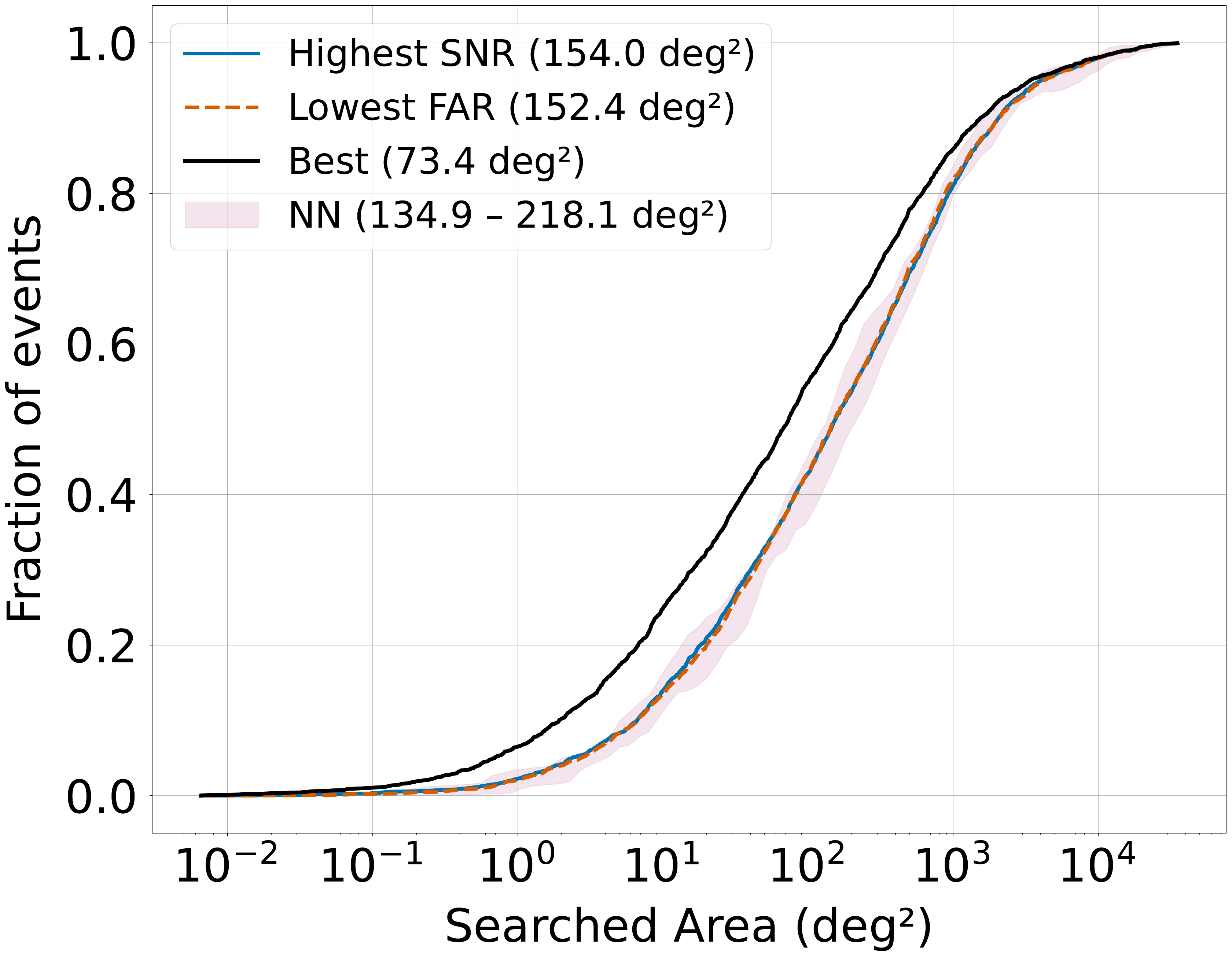}
        \caption{Significant events}
    \end{subfigure}
    \hfill
    \begin{subfigure}[t]{0.475\textwidth}
        \centering
        \includegraphics[width=\textwidth]{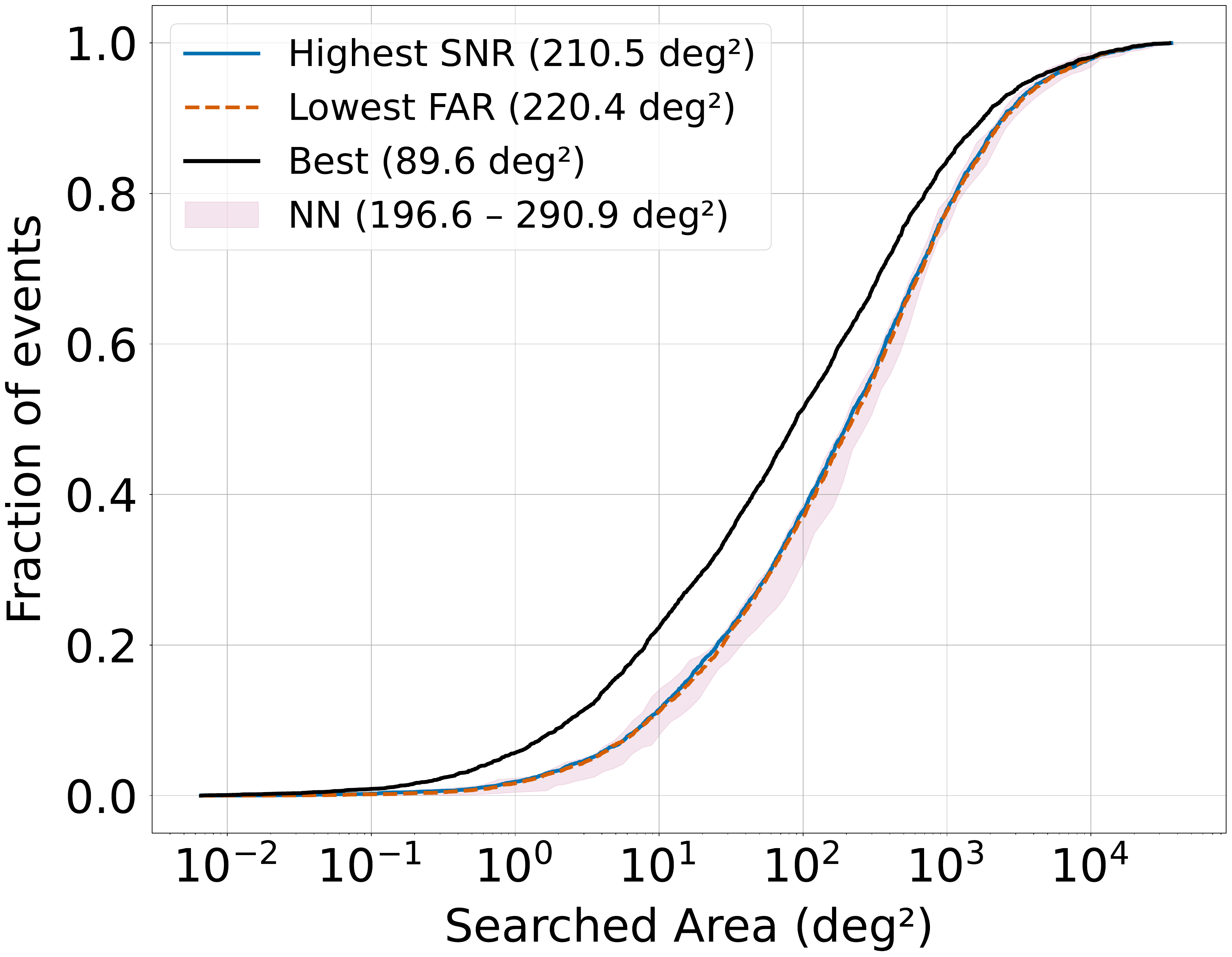}
        \caption{All events recovered with FAR < 2/day}
    \end{subfigure}
\caption{Cumulative distribution of searched areas for g-event sky maps selected by lowest FAR, highest SNR, lowest searched area (denoted by best), and the NN trained on inputs that are not sorted by chirp mass, for (a) significant events and (b) all events. Median searched areas are reported in parentheses. Bands are obtained by evaluating the selectors on the ten test datasets. The distributions are consistent with those obtained when inputs are sorted by chirp mass (Fig.~\ref{fig:SAFrac_all}), confirming that the chirp-mass ordering serves only to enforce permutation invariance and does not affect the scientific outcomes.}
\label{fig:RCA}
\end{figure}

\begin{figure}[ht]
    \centering

    \begin{subfigure}[t]{0.45\textwidth}
        \centering
        \includegraphics[width=\textwidth]{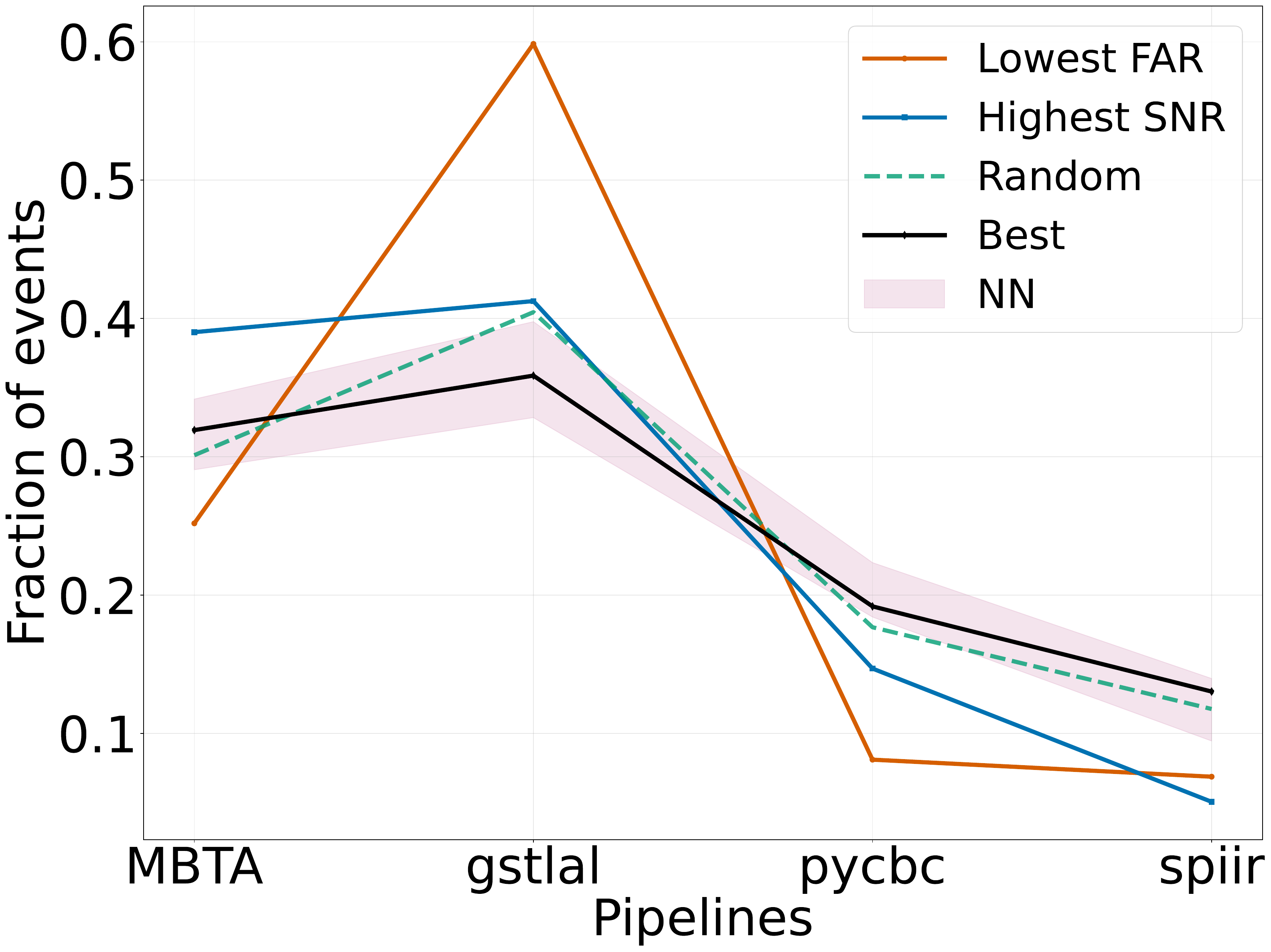}
        \caption{Significant events}
    \end{subfigure}
    \hfill
    \begin{subfigure}[t]{0.45\textwidth}
        \centering
        \includegraphics[width=\textwidth]{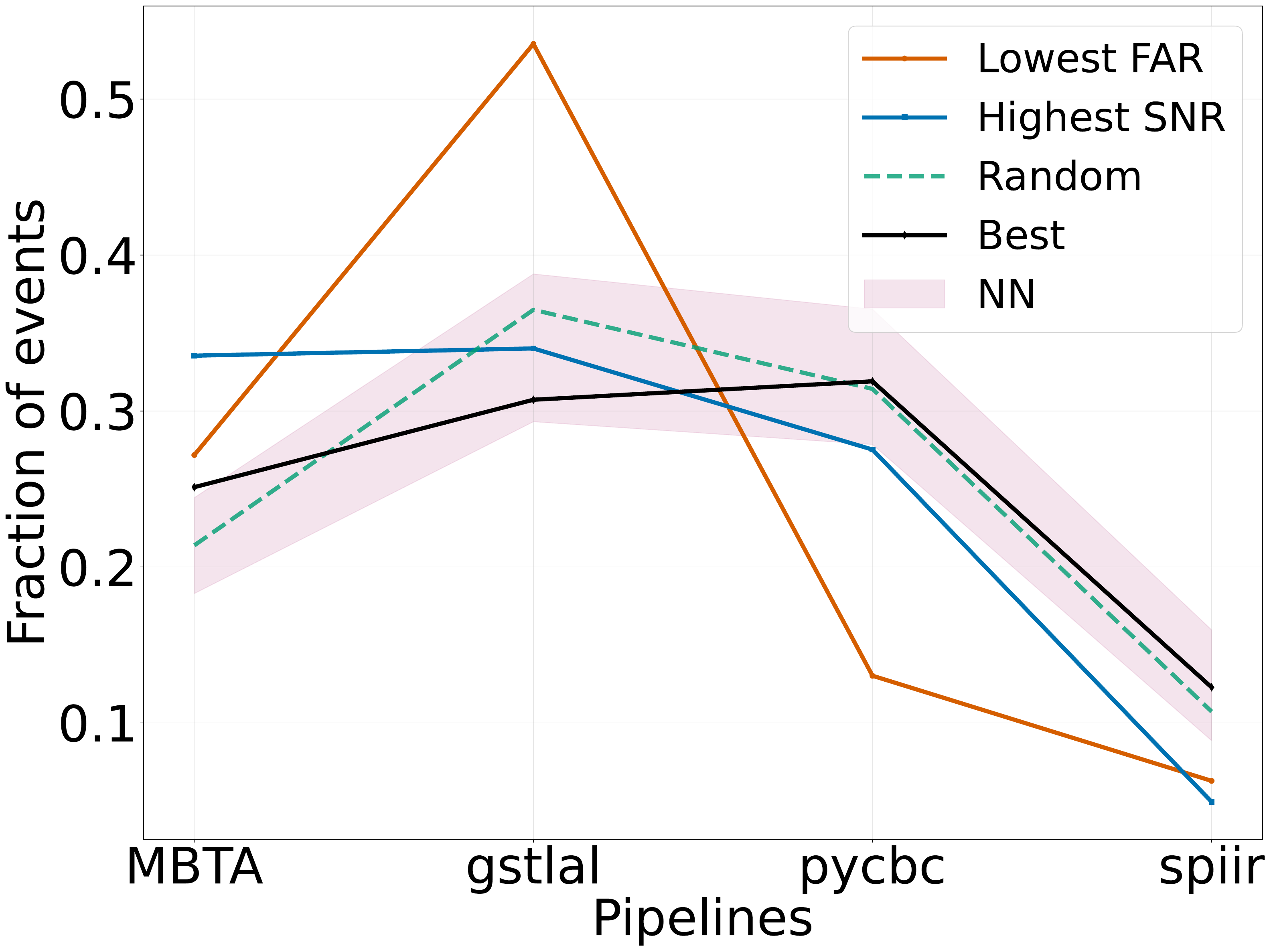}
        \caption{All events recovered with FAR < 2/day}
    \end{subfigure}
    \caption{Fraction of selected g-events per pipeline for the neural network trained without chirp-mass sorting of inputs, for (a) significant events and (b) all events. Conventions follow Fig.~\ref{fig:pipeline_fraction}: the best line corresponds to the g-event with the lowest searched area in each superevent (the ground truth), and the dashed ``random'' line shows the average fraction of g-events per superevent recovered by each pipeline. The pipeline distributions remain close to the ground truth and are consistent with those of the chirp-mass-sorted neural network shown in Fig.~\ref{fig:pipeline_fraction}, confirming that pipeline fairness is preserved regardless of input ordering.}
    \label{fig:Rpipeline_fraction}
\end{figure}
The chirp-mass ordering introduced in Sec.~\ref{sec:methods} is a design choice, not a requirement.
It provides a deterministic, reproducible input ordering that decouples the network's behavior from the order in which g-events happen to arrive in low latency.
One concern is that this specific ordering might introduce a bias.
For example, the network might learn to exploit the fact that the first or last position in the sorted sequence tends to correspond to a certain type of g-event, rather than learning from the feature values themselves.
If so, the scientific results would depend on the choice of sorting parameter, which would be undesirable.

To test this, we train and evaluate the same network architecture on inputs in which g-events within each superevent are placed in a random order rather than sorted by chirp mass.
If the sorting is merely a convention, we expect the performance to be similar.

The searched-area cumulative distributions (Fig.~\ref{fig:RCA}) and per-pipeline selection fractions (Fig.~\ref{fig:Rpipeline_fraction}) of the randomly ordered network match those of the chirp-mass-sorted network in the main text (Figs.~\ref{fig:SAFrac_all} and~\ref{fig:pipeline_fraction}) within the spread of the ten test sets.
This confirms that the chirp-mass ordering does not introduce a detectable bias.
The network learns from the feature values themselves, not from positional cues.
We therefore conclude that the chirp-mass ordering in the main analysis only ensures that the preferred event selection is invariant to transmission latency and cluster scheduling artifacts.

\section{Results using a permutation-invariant architecture}
\label{app:cnn}

\begin{figure}[ht]
    \centering
    \begin{subfigure}[t]{0.475\textwidth}
        \centering

        \includegraphics[width=\textwidth]{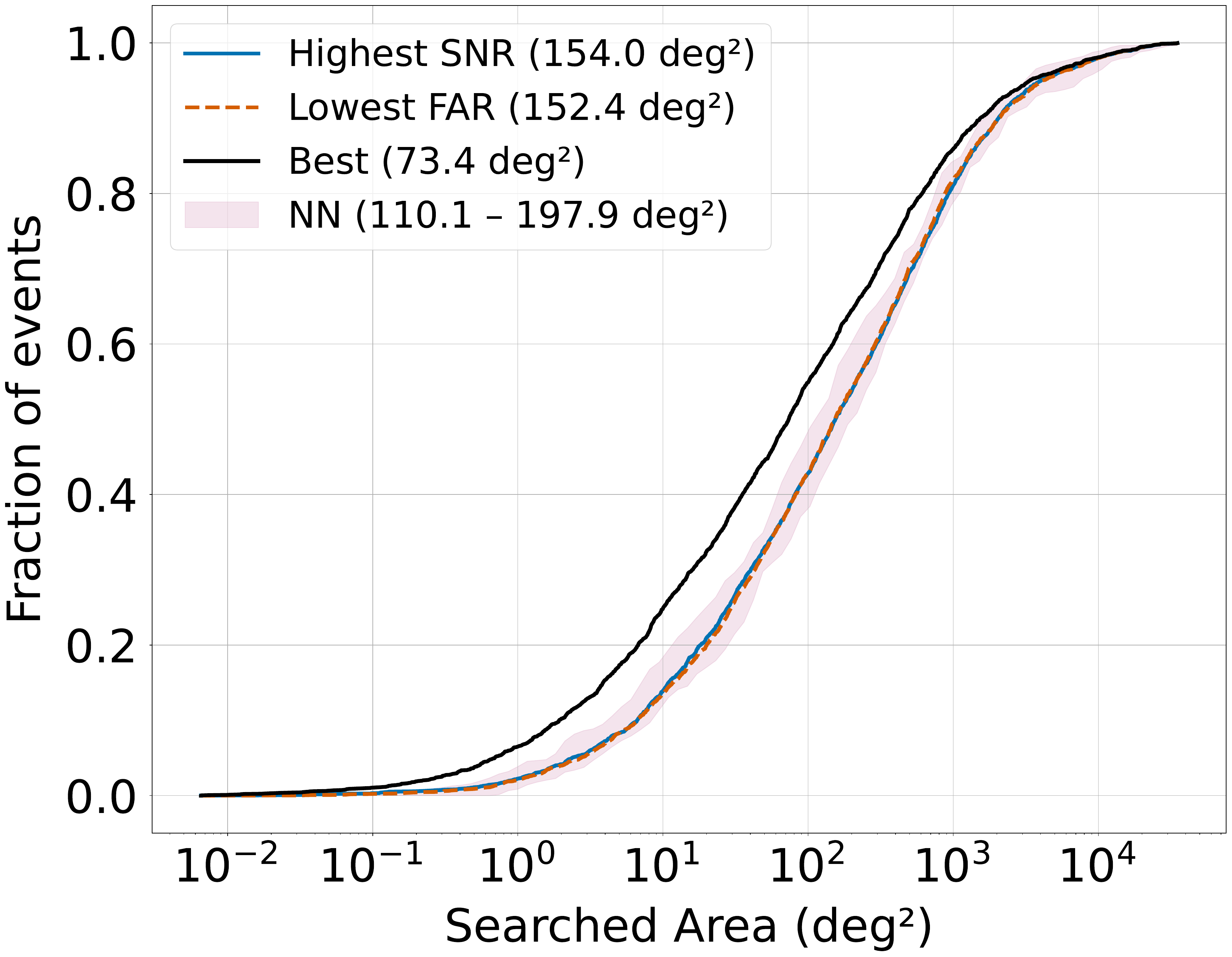}
        \caption{Significant events}
    \end{subfigure}
    \hfill
    \begin{subfigure}[t]{0.475\textwidth}
        \centering
        \includegraphics[width=\textwidth]{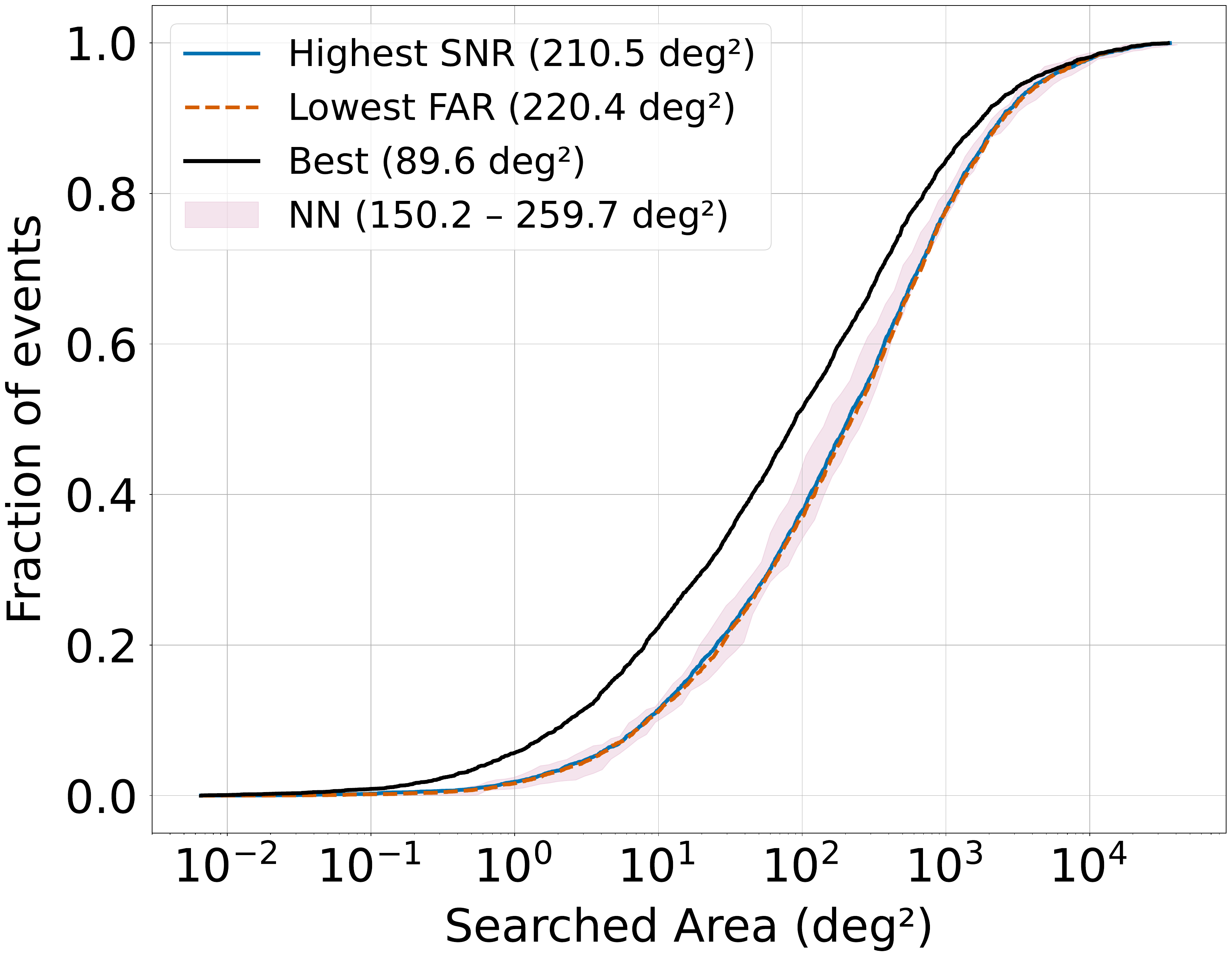}
        \caption{All events}
    \end{subfigure}
\caption{Cumulative distribution of searched areas for g-event sky maps selected by lowest FAR, highest SNR, lowest searched area (denoted by best), and the PointNet-style permutation-invariant neural network described in Appendix~\ref{app:cnn}, trained on (a) significant events and (b) all events. Median searched areas are reported in parentheses. Bands are obtained by evaluating the selectors on the ten test datasets. The searched-area performance is consistent with that of the feedforward network in the main text (Fig.~\ref{fig:SAFrac_all}).}
\label{fig:CCA}
\end{figure}

\begin{figure}[ht]
    \centering
    \begin{subfigure}[t]{0.475\textwidth}
        \centering
        \includegraphics[width=\textwidth]{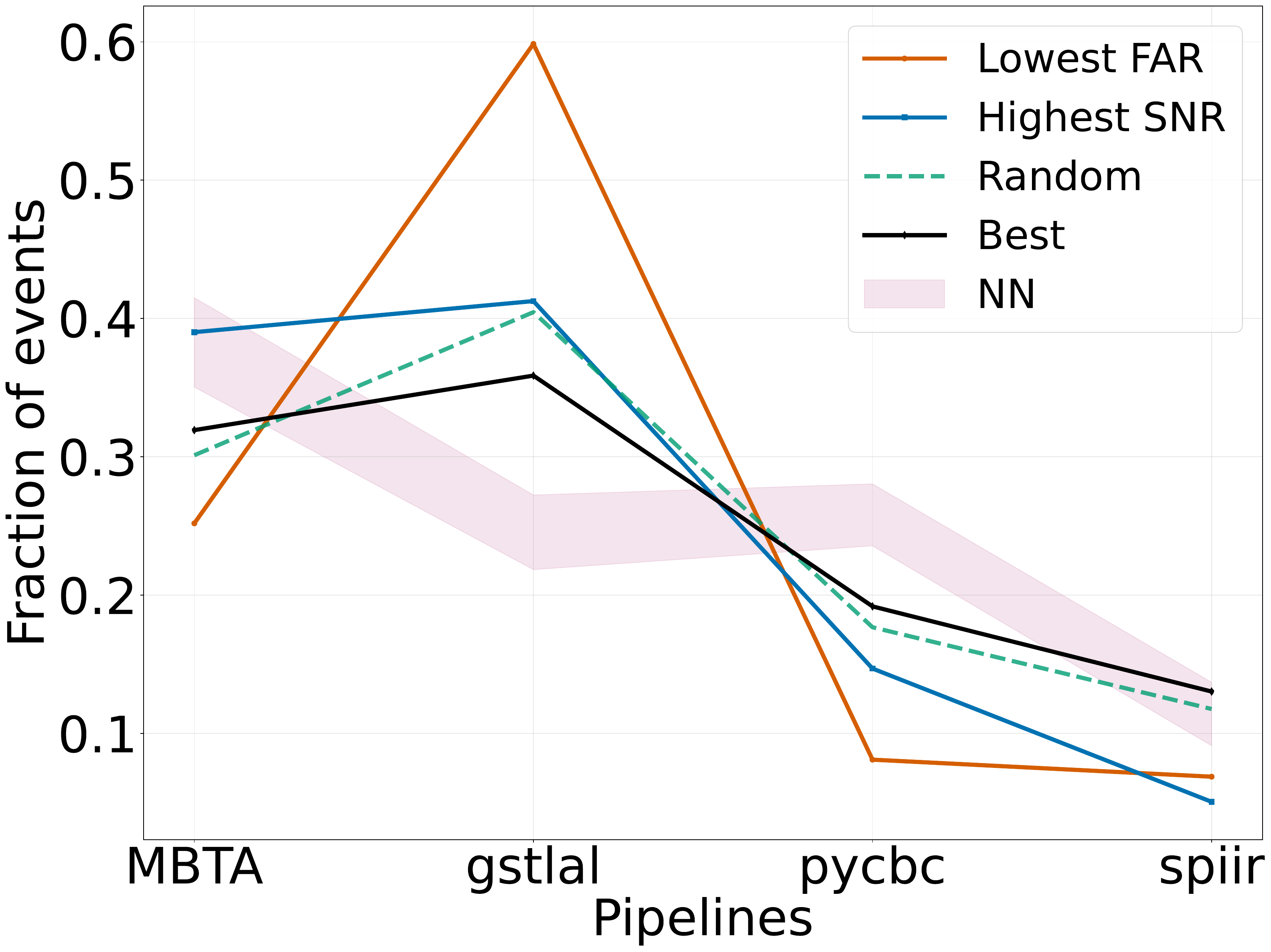}
        \caption{Significant events}
    \end{subfigure}
    \hfill
    \begin{subfigure}[t]{0.475\textwidth}
        \centering
        \includegraphics[width=\textwidth]{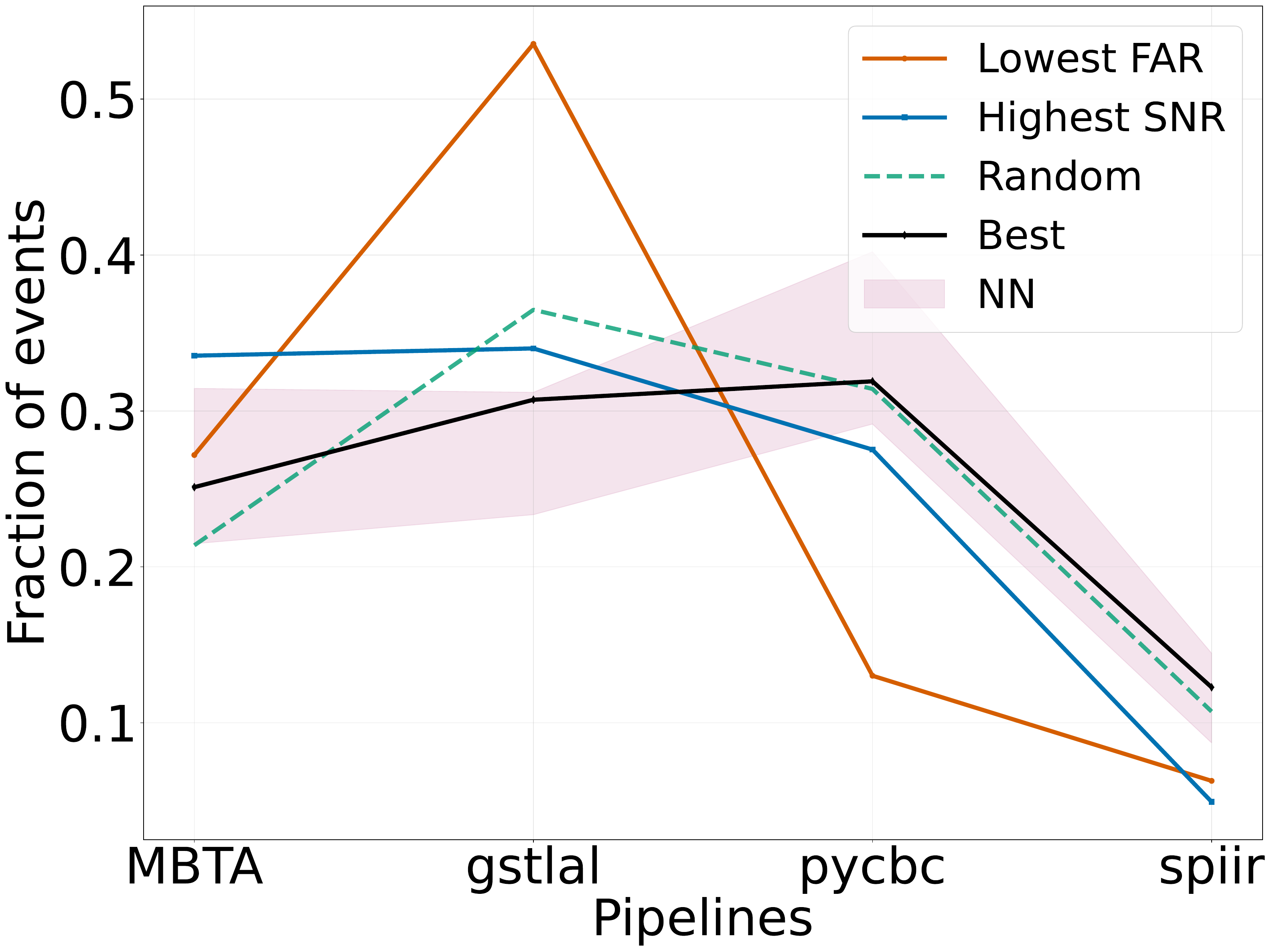}
        \caption{All events}

    \end{subfigure}
    \caption{Fraction of selected g-events per pipeline for the PointNet-style permutation-invariant architecture described in Appendix~\ref{app:cnn}, for (a) significant events and (b) all events. Conventions follow Fig.~\ref{fig:pipeline_fraction}: the best line corresponds to the g-event with the lowest searched area in each superevent (the ground truth), and the dashed random line shows the average fraction of g-events per superevent recovered by each pipeline. Unlike the feedforward neural network (Fig.~\ref{fig:pipeline_fraction}), the PointNet-style architecture deviates more strongly from the ground-truth pipeline distribution, indicating that pipeline fairness is sensitive to architectural choices and not solely to the pipeline-agnostic input design.}
\label{fig:Cpipeline_fraction}
\end{figure}

The main-text neural network achieves permutation invariance indirectly, through chirp-mass sorting, which converts the unordered set of g-events into a canonical sequence before flattening.
As a complementary test, we evaluate a PointNet-style architecture that is permutation-invariant by construction and therefore requires no sorting at all.

The PointNet-style network processes each g-event independently through a shared subnetwork before aggregating information across the full superevent.
Concretely, each g-event's five-dimensional feature vector (per-detector SNRs, log-FAR, chirp mass) is passed through a shared two-layer multilayer perceptron with batch normalization and ReLU activations, producing a per-event embedding of fixed dimension.
An elementwise max-pooling operation then aggregates these embeddings across all nonpadded g-events in the superevent, yielding a single global feature vector that summarizes the collective pipeline outputs.
This aggregation step is the key source of permutation invariance because max pooling is symmetric with respect to the order of its inputs, the global feature is independent of how the g-events happen to be ordered.
A final fully connected layer maps the global feature to 22 output logits, one per possible g-event index.
As in the main-text network, padding slots are masked to $-\infty$ before the loss is computed.
The model uses the same input features, the same composite loss (cross-entropy plus KL-divergence soft ranking, equally weighted), and the same ten-iteration train/test split protocol as the main-text network.

The searched-area cumulative distributions (Fig.~\ref{fig:CCA}) are consistent with those of the main-text neural network (Fig.~\ref{fig:SAFrac_all}), confirming that the PointNet-style architecture achieves comparable localization performance.
The network is therefore a viable alternative from a searched-area standpoint.

The per-pipeline selection fractions (Fig.~\ref{fig:Cpipeline_fraction}) tell a different story.
Unlike the main-text neural network, the PointNet-style network deviates more noticeably from the ground-truth pipeline distribution.
This shows pipeline fairness is therefore not guaranteed by pipeline-agnostic inputs alone and it also depends on how those inputs are processed.
We leave a systematic investigation of the relationship between architecture and pipeline fairness for future work.

\section{Case studies of individual superevents}
\label{app:case_studies}

To complement the population-level statistics presented in the main text, we examine three individual MDC11 superevents that illustrate distinct regimes of selector behavior.
Figures~\ref{fig:case_good}--\ref{fig:case_all_struggle} show the 90\% credible sky area region of the BAYESTAR sky map selected by each algorithm, overlaid on the full sky in equatorial coordinates.
The red cross marks the true injected source location, and an inset provides an enlarged view of the region around it.

\subsection{NN outperforms both traditional selectors}

\begin{figure}[h!]
    \centering
    \includegraphics[width=\textwidth]{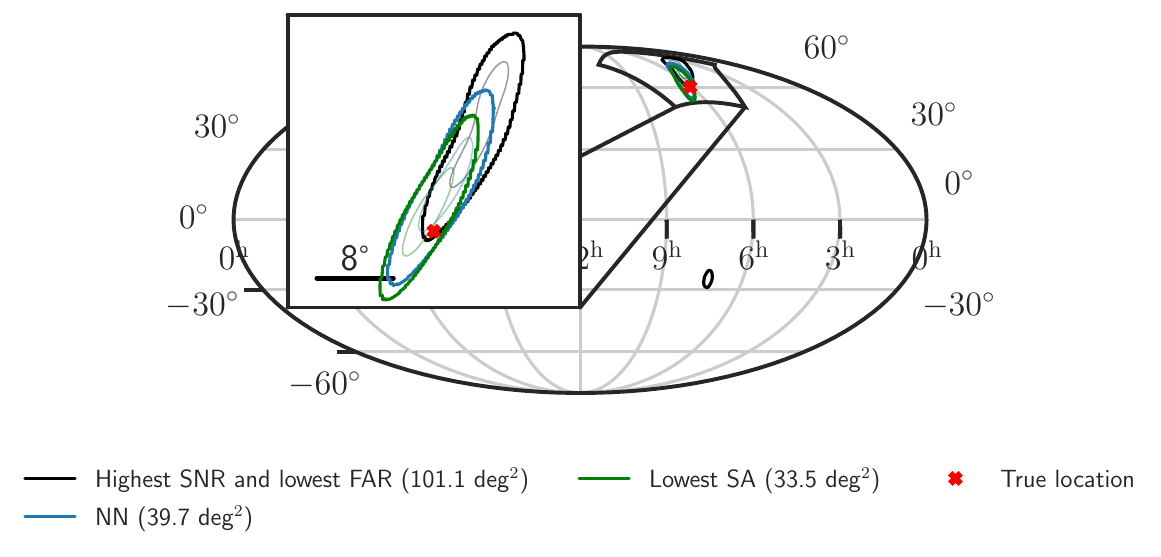}
    \caption{Sky localization for MDC11 superevent S230306ch, an example where the neural network substantially outperforms both traditional selectors. In this superevent, the highest-SNR and lowest-FAR criteria select the same g-event (black contour, 101.1~deg$^2$). The neural network selects a different g-event (blue contour, 39.7~deg$^2$) that closely approaches the optimal localization (green contour, 33.5~deg$^2$), reducing the searched area by a factor of $\sim$2.5 relative to the SNR/FAR selection. All three contours enclose the true source location (red cross), but the NN-selected sky map is considerably tighter and would require significantly less telescope time to search. This event is representative of the $\sim$11.8\% of significant superevents where the NN reduces the searched area by more than 100~deg$^2$ relative to the highest-SNR selector. SA stands for searched area.}
    \label{fig:case_good}
\end{figure}

Superevent S230306ch (as shown in Fig.~\ref{fig:case_good}) illustrates the regime where the neural network performs well.
The highest-SNR and lowest-FAR selectors agree, both ranking the same g-event as preferred with a searched area of 101.1~deg$^2$.
The neural network selects a g-event with a searched area of 39.7~deg$^2$, close to the true optimum of 33.5~deg$^2$.
The SNR of the g-event selected by the NN is lower than that of the SNR-preferred event, but its BAYESTAR sky map is better constrained.
This case exemplifies how non-Gaussian noise can inflate the SNR of a trigger without improving its sky localization.

\subsection{NN underperforms in the well-localized regime}

\begin{figure}[h!]
    \centering
    \includegraphics[width=\textwidth]{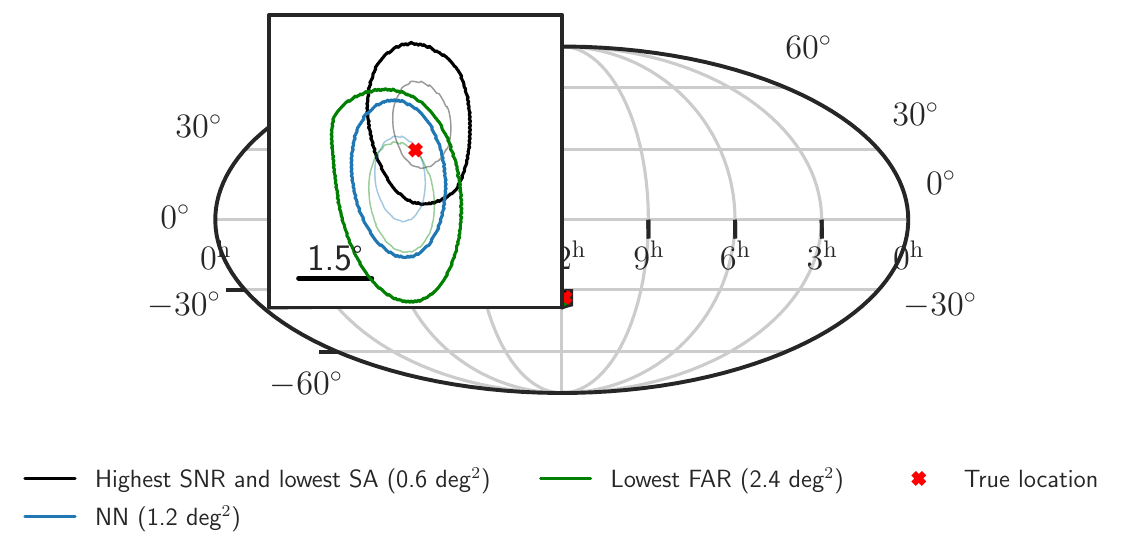}
    \caption{Sky localization for MDC11 superevent S230315em, an example where the highest-SNR selector identifies the optimal g-event. The SNR-selected event coincides with the lowest-searched-area event (black contour, 0.6~deg$^2$). The neural network selects a g-event with a searched area of 1.2~deg$^2$ (blue contour), and the lowest-FAR selector yields 2.4~deg$^2$ (green contour). Although the NN underperforms the SNR selector by a factor of 2 in searched area, all three localization regions are sub-3~deg$^2$, making the absolute difference operationally negligible for any realistic multimessenger follow-up campaign. The true source location (red cross) lies well within all three credible regions.}
    \label{fig:case_snr_wins_well_loc}
\end{figure}

Superevent S230315em (as shown in Fig.~\ref{fig:case_snr_wins_well_loc}) illustrates the well-localized regime, where the SNR-based selector happens to identify the optimal g-event.
The SNR-preferred and lowest-searched-area events coincide (0.6~deg$^2$), while the neural network selects a g-event with a searched area of 1.2~deg$^2$ and the FAR-based selector yields 2.4~deg$^2$.
While the NN's selected area is a factor of 2 larger than the SNR-based selection, the absolute difference of $\sim$0.6~deg$^2$ is negligible in practice.

\subsection{NN underperforms when all selectors struggle}

\begin{figure}[h!]
    \centering
    \includegraphics[width=\textwidth]{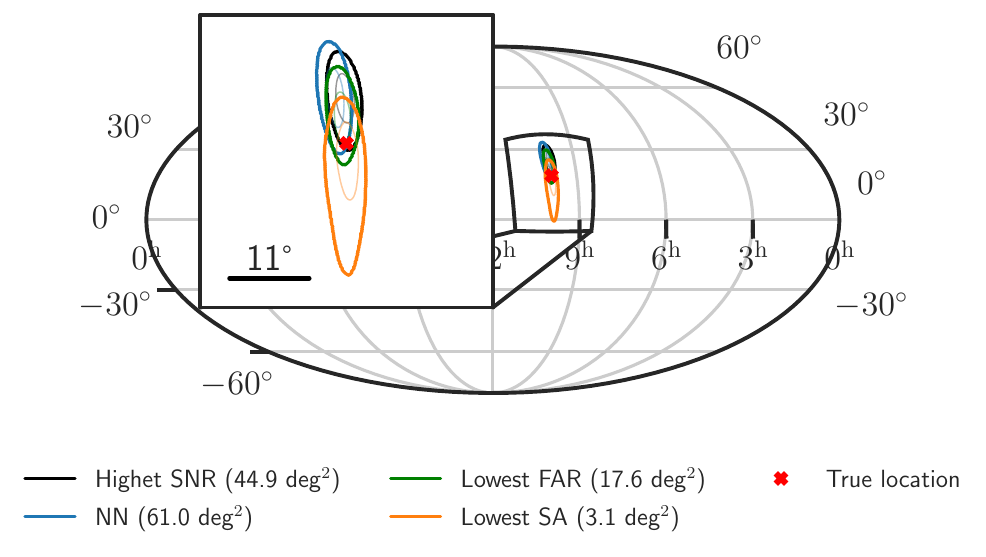}
    \caption{Sky localization for MDC11 superevent S230311ey, an example where the neural network performs poorly and all selectors deviate substantially from the optimal localization. The four selectors choose four distinct g-events. The lowest-FAR selector (green contour, 17.6~deg$^2$) most closely approaches the optimal localization (orange contour, 3.1~deg$^2$), while the highest-SNR selector (black contour, 44.9~deg$^2$) and the neural network (blue contour, 61.0~deg$^2$) are both considerably worse. All four contours enclose the true source location (red cross). This event illustrates a general limitation shared by all three strategies: when the pipeline outputs do not strongly discriminate the best-localized event from other candidates in a superevent, all selectors can fail to identify the optimal g-event.}
    \label{fig:case_all_struggle}
\end{figure}

Superevent S230311ey (as shown in Fig.~\ref{fig:case_all_struggle}) illustrates a case where the neural network performs worst among the three selectors and where none of the selectors recovers the optimal g-event.
The lowest searched area is 3.1~deg$^2$.
The lowest-FAR selector comes closest (17.6~deg$^2$), followed by the highest-SNR selector (44.9~deg$^2$), with the neural network performing worst (61.0~deg$^2$).
This reflects a configuration in which the per-pipeline statistics---SNR, FAR, and chirp mass---do not carry sufficient information to identify the best-localized trigger.
A potential mitigation would be to include additional physically motivated features such as time delays and phase differences between detectors that carry localization-specific information.
\bibliographystyle{apsrev4-2}
\bibliography{references}

\begin{thebibliography}{85}%
\makeatletter
\providecommand \@ifxundefined [1]{%
 \@ifx{#1\undefined}
}%
\providecommand \@ifnum [1]{%
 \ifnum #1\expandafter \@firstoftwo
 \else \expandafter \@secondoftwo
 \fi
}%
\providecommand \@ifx [1]{%
 \ifx #1\expandafter \@firstoftwo
 \else \expandafter \@secondoftwo
 \fi
}%
\providecommand \natexlab [1]{#1}%
\providecommand \enquote  [1]{``#1''}%
\providecommand \bibnamefont  [1]{#1}%
\providecommand \bibfnamefont [1]{#1}%
\providecommand \citenamefont [1]{#1}%
\providecommand \href@noop [0]{\@secondoftwo}%
\providecommand \href [0]{\begingroup \@sanitize@url \@href}%
\providecommand \@href[1]{\@@startlink{#1}\@@href}%
\providecommand \@@href[1]{\endgroup#1\@@endlink}%
\providecommand \@sanitize@url [0]{\catcode `\\12\catcode `\$12\catcode
  `\&12\catcode `\#12\catcode `\^12\catcode `\_12\catcode `\%12\relax}%
\providecommand \@@startlink[1]{}%
\providecommand \@@endlink[0]{}%
\providecommand \url  [0]{\begingroup\@sanitize@url \@url }%
\providecommand \@url [1]{\endgroup\@href {#1}{\urlprefix }}%
\providecommand \urlprefix  [0]{URL }%
\providecommand \Eprint [0]{\href }%
\providecommand \doibase [0]{https://doi.org/}%
\providecommand \selectlanguage [0]{\@gobble}%
\providecommand \bibinfo  [0]{\@secondoftwo}%
\providecommand \bibfield  [0]{\@secondoftwo}%
\providecommand \translation [1]{[#1]}%
\providecommand \BibitemOpen [0]{}%
\providecommand \bibitemStop [0]{}%
\providecommand \bibitemNoStop [0]{.\EOS\space}%
\providecommand \EOS [0]{\spacefactor3000\relax}%
\providecommand \BibitemShut  [1]{\csname bibitem#1\endcsname}%
\let\auto@bib@innerbib\@empty
\bibitem [{\citenamefont {{LIGO Scientific
  Collaboration}}(2015)}]{LIGOScientific:2014pky}%
  \BibitemOpen
  \bibfield  {author} {\bibinfo {author} {\bibnamefont {{LIGO Scientific
  Collaboration}}},\ }\href {https://doi.org/10.1088/0264-9381/32/7/074001}
  {\bibfield  {journal} {\bibinfo  {journal} {Classical and Quantum Gravity}\
  }\textbf {\bibinfo {volume} {32}},\ \bibinfo {eid} {074001} (\bibinfo {year}
  {2015})},\ \Eprint {https://arxiv.org/abs/1411.4547} {arXiv:1411.4547
  [gr-qc]} \BibitemShut {NoStop}%
\bibitem [{\citenamefont {{Acernese}}\ \emph {et~al.}(2015)\citenamefont
  {{Acernese}}, \citenamefont {{Agathos}}, \citenamefont {{Agatsuma}} \emph
  {et~al.}}]{VIRGO:2014yos}%
  \BibitemOpen
  \bibfield  {author} {\bibinfo {author} {\bibfnamefont {F.}~\bibnamefont
  {{Acernese}}}, \bibinfo {author} {\bibfnamefont {M.}~\bibnamefont
  {{Agathos}}}, \bibinfo {author} {\bibfnamefont {K.}~\bibnamefont
  {{Agatsuma}}}, \emph {et~al.},\ }\href
  {https://doi.org/10.1088/0264-9381/32/2/024001} {\bibfield  {journal}
  {\bibinfo  {journal} {Classical and Quantum Gravity}\ }\textbf {\bibinfo
  {volume} {32}},\ \bibinfo {eid} {024001} (\bibinfo {year} {2015})},\ \Eprint
  {https://arxiv.org/abs/1408.3978} {arXiv:1408.3978 [gr-qc]} \BibitemShut
  {NoStop}%
\bibitem [{\citenamefont {{Akutsu}}\ \emph {et~al.}(2021)\citenamefont
  {{Akutsu}}, \citenamefont {{Ando}}, \citenamefont {{Arai}} \emph
  {et~al.}}]{KAGRA:2020tym}%
  \BibitemOpen
  \bibfield  {author} {\bibinfo {author} {\bibfnamefont {T.}~\bibnamefont
  {{Akutsu}}}, \bibinfo {author} {\bibfnamefont {M.}~\bibnamefont {{Ando}}},
  \bibinfo {author} {\bibfnamefont {K.}~\bibnamefont {{Arai}}}, \emph
  {et~al.},\ }\href {https://doi.org/10.1093/ptep/ptaa125} {\bibfield
  {journal} {\bibinfo  {journal} {Progress of Theoretical and Experimental
  Physics}\ }\textbf {\bibinfo {volume} {2021}},\ \bibinfo {eid} {05A101}
  (\bibinfo {year} {2021})},\ \Eprint {https://arxiv.org/abs/2005.05574}
  {arXiv:2005.05574 [physics.ins-det]} \BibitemShut {NoStop}%
\bibitem [{\citenamefont {{The LVK
  Collaboration}}(2026{\natexlab{a}})}]{2025arXiv250818082T}%
  \BibitemOpen
  \bibfield  {author} {\bibinfo {author} {\bibnamefont {{The LVK
  Collaboration}}},\ }\href@noop {} {\bibfield  {journal} {\bibinfo  {journal}
  {\apjl}\ }\textbf {\bibinfo {volume} {1004}},\ \bibinfo {pages} {L22}
  (\bibinfo {year} {2026}{\natexlab{a}})},\ \Eprint
  {https://arxiv.org/abs/2508.18082} {arXiv:2508.18082 [gr-qc]} \BibitemShut
  {NoStop}%
\bibitem [{\citenamefont {{Abbott}}\ \emph {et~al.}(2016)\citenamefont
  {{Abbott}}, \citenamefont {{Abbott}}, \citenamefont {{Abbott}} \emph
  {et~al.}}]{AbEA2016b}%
  \BibitemOpen
  \bibfield  {author} {\bibinfo {author} {\bibfnamefont {B.~P.}\ \bibnamefont
  {{Abbott}}}, \bibinfo {author} {\bibfnamefont {R.}~\bibnamefont {{Abbott}}},
  \bibinfo {author} {\bibfnamefont {T.~D.}\ \bibnamefont {{Abbott}}}, \emph
  {et~al.},\ }\href {https://doi.org/10.3847/2041-8205/826/1/L13} {\bibfield
  {journal} {\bibinfo  {journal} {\apjl}\ }\textbf {\bibinfo {volume} {826}},\
  \bibinfo {eid} {L13} (\bibinfo {year} {2016})},\ \Eprint
  {https://arxiv.org/abs/1602.08492} {arXiv:1602.08492 [astro-ph.HE]}
  \BibitemShut {NoStop}%
\bibitem [{\citenamefont {{Abbott}}\ \emph
  {et~al.}(2017{\natexlab{a}})\citenamefont {{Abbott}}, \citenamefont
  {{Abbott}}, \citenamefont {{Abbott}} \emph {et~al.}}]{AbEA2017b}%
  \BibitemOpen
  \bibfield  {author} {\bibinfo {author} {\bibfnamefont {B.~P.}\ \bibnamefont
  {{Abbott}}}, \bibinfo {author} {\bibfnamefont {R.}~\bibnamefont {{Abbott}}},
  \bibinfo {author} {\bibfnamefont {T.~D.}\ \bibnamefont {{Abbott}}}, \emph
  {et~al.},\ }\href {https://doi.org/10.1103/PhysRevLett.119.161101} {\bibfield
   {journal} {\bibinfo  {journal} {\prl}\ }\textbf {\bibinfo {volume} {119}},\
  \bibinfo {eid} {161101} (\bibinfo {year} {2017}{\natexlab{a}})},\ \Eprint
  {https://arxiv.org/abs/1710.05832} {arXiv:1710.05832 [gr-qc]} \BibitemShut
  {NoStop}%
\bibitem [{\citenamefont {{Abbott}}\ \emph {et~al.}(2021)\citenamefont
  {{Abbott}}, \citenamefont {{Abbott}}, \citenamefont {{Abraham}} \emph
  {et~al.}}]{AbEA2021}%
  \BibitemOpen
  \bibfield  {author} {\bibinfo {author} {\bibfnamefont {R.}~\bibnamefont
  {{Abbott}}}, \bibinfo {author} {\bibfnamefont {T.~D.}\ \bibnamefont
  {{Abbott}}}, \bibinfo {author} {\bibfnamefont {S.}~\bibnamefont {{Abraham}}},
  \emph {et~al.},\ }\href {https://doi.org/10.3847/2041-8213/ac082e} {\bibfield
   {journal} {\bibinfo  {journal} {\apjl}\ }\textbf {\bibinfo {volume} {915}},\
  \bibinfo {eid} {L5} (\bibinfo {year} {2021})},\ \Eprint
  {https://arxiv.org/abs/2106.15163} {arXiv:2106.15163 [astro-ph.HE]}
  \BibitemShut {NoStop}%
\bibitem [{Note1()}]{Note1}%
  \BibitemOpen
  \bibinfo {note} {\protect \url
  {https://observing.docs.ligo.org/plan}}\BibitemShut {NoStop}%
\bibitem [{\citenamefont {{Goldstein}}\ \emph {et~al.}(2017)\citenamefont
  {{Goldstein}}, \citenamefont {{Veres}}, \citenamefont {{Burns}} \emph
  {et~al.}}]{GoVe2017}%
  \BibitemOpen
  \bibfield  {author} {\bibinfo {author} {\bibfnamefont {A.}~\bibnamefont
  {{Goldstein}}}, \bibinfo {author} {\bibfnamefont {P.}~\bibnamefont
  {{Veres}}}, \bibinfo {author} {\bibfnamefont {E.}~\bibnamefont {{Burns}}},
  \emph {et~al.},\ }\href {https://doi.org/10.3847/2041-8213/aa8f41} {\bibfield
   {journal} {\bibinfo  {journal} {\apjl}\ }\textbf {\bibinfo {volume} {848}},\
  \bibinfo {eid} {L14} (\bibinfo {year} {2017})},\ \Eprint
  {https://arxiv.org/abs/1710.05446} {arXiv:1710.05446 [astro-ph.HE]}
  \BibitemShut {NoStop}%
\bibitem [{\citenamefont {{Savchenko}}\ \emph {et~al.}(2017)\citenamefont
  {{Savchenko}}, \citenamefont {{Ferrigno}}, \citenamefont {{Kuulkers}} \emph
  {et~al.}}]{SaFe2017}%
  \BibitemOpen
  \bibfield  {author} {\bibinfo {author} {\bibfnamefont {V.}~\bibnamefont
  {{Savchenko}}}, \bibinfo {author} {\bibfnamefont {C.}~\bibnamefont
  {{Ferrigno}}}, \bibinfo {author} {\bibfnamefont {E.}~\bibnamefont
  {{Kuulkers}}}, \emph {et~al.},\ }\href
  {https://doi.org/10.3847/2041-8213/aa8f94} {\bibfield  {journal} {\bibinfo
  {journal} {\apjl}\ }\textbf {\bibinfo {volume} {848}},\ \bibinfo {eid} {L15}
  (\bibinfo {year} {2017})},\ \Eprint {https://arxiv.org/abs/1710.05449}
  {arXiv:1710.05449 [astro-ph.HE]} \BibitemShut {NoStop}%
\bibitem [{\citenamefont {{Abbott}}\ \emph
  {et~al.}(2017{\natexlab{b}})\citenamefont {{Abbott}}, \citenamefont
  {{Abbott}}, \citenamefont {{Abbott}} \emph {et~al.}}]{AbEA2017e}%
  \BibitemOpen
  \bibfield  {author} {\bibinfo {author} {\bibfnamefont {B.~P.}\ \bibnamefont
  {{Abbott}}}, \bibinfo {author} {\bibfnamefont {R.}~\bibnamefont {{Abbott}}},
  \bibinfo {author} {\bibfnamefont {T.~D.}\ \bibnamefont {{Abbott}}}, \emph
  {et~al.},\ }\href {https://doi.org/10.3847/2041-8213/aa920c} {\bibfield
  {journal} {\bibinfo  {journal} {\apjl}\ }\textbf {\bibinfo {volume} {848}},\
  \bibinfo {eid} {L13} (\bibinfo {year} {2017}{\natexlab{b}})},\ \Eprint
  {https://arxiv.org/abs/1710.05834} {arXiv:1710.05834 [astro-ph.HE]}
  \BibitemShut {NoStop}%
\bibitem [{\citenamefont {{Coulter}}\ \emph {et~al.}(2017)\citenamefont
  {{Coulter}}, \citenamefont {{Foley}}, \citenamefont {{Kilpatrick}} \emph
  {et~al.}}]{2017Sci...358.1556C}%
  \BibitemOpen
  \bibfield  {author} {\bibinfo {author} {\bibfnamefont {D.~A.}\ \bibnamefont
  {{Coulter}}}, \bibinfo {author} {\bibfnamefont {R.~J.}\ \bibnamefont
  {{Foley}}}, \bibinfo {author} {\bibfnamefont {C.~D.}\ \bibnamefont
  {{Kilpatrick}}}, \emph {et~al.},\ }\href
  {https://doi.org/10.1126/science.aap9811} {\bibfield  {journal} {\bibinfo
  {journal} {Science}\ }\textbf {\bibinfo {volume} {358}},\ \bibinfo {pages}
  {1556} (\bibinfo {year} {2017})},\ \Eprint {https://arxiv.org/abs/1710.05452}
  {arXiv:1710.05452 [astro-ph.HE]} \BibitemShut {NoStop}%
\bibitem [{\citenamefont {{Smartt}}\ \emph {et~al.}(2017)\citenamefont
  {{Smartt}}, \citenamefont {{Chen}}, \citenamefont {{Jerkstrand}} \emph
  {et~al.}}]{SmCh2017}%
  \BibitemOpen
  \bibfield  {author} {\bibinfo {author} {\bibfnamefont {S.~J.}\ \bibnamefont
  {{Smartt}}}, \bibinfo {author} {\bibfnamefont {T.-W.}\ \bibnamefont
  {{Chen}}}, \bibinfo {author} {\bibfnamefont {A.}~\bibnamefont
  {{Jerkstrand}}}, \emph {et~al.},\ }\href
  {https://doi.org/10.1038/nature24303} {\bibfield  {journal} {\bibinfo
  {journal} {\nat}\ }\textbf {\bibinfo {volume} {551}},\ \bibinfo {pages} {75}
  (\bibinfo {year} {2017})},\ \Eprint {https://arxiv.org/abs/1710.05841}
  {arXiv:1710.05841 [astro-ph.HE]} \BibitemShut {NoStop}%
\bibitem [{\citenamefont {{Abbott}}\ \emph
  {et~al.}(2017{\natexlab{c}})\citenamefont {{Abbott}}, \citenamefont
  {{Abbott}}, \citenamefont {{Abbott}} \emph {et~al.}}]{AbEA2017f}%
  \BibitemOpen
  \bibfield  {author} {\bibinfo {author} {\bibfnamefont {B.~P.}\ \bibnamefont
  {{Abbott}}}, \bibinfo {author} {\bibfnamefont {R.}~\bibnamefont {{Abbott}}},
  \bibinfo {author} {\bibfnamefont {T.~D.}\ \bibnamefont {{Abbott}}}, \emph
  {et~al.},\ }\href {https://doi.org/10.3847/2041-8213/aa91c9} {\bibfield
  {journal} {\bibinfo  {journal} {\apjl}\ }\textbf {\bibinfo {volume} {848}},\
  \bibinfo {eid} {L12} (\bibinfo {year} {2017}{\natexlab{c}})},\ \Eprint
  {https://arxiv.org/abs/1710.05833} {arXiv:1710.05833 [astro-ph.HE]}
  \BibitemShut {NoStop}%
\bibitem [{\citenamefont {{Bauswein}}\ \emph {et~al.}(2017)\citenamefont
  {{Bauswein}}, \citenamefont {{Just}}, \citenamefont {{Janka}},\ and\
  \citenamefont {{Stergioulas}}}]{BaJu2017}%
  \BibitemOpen
  \bibfield  {author} {\bibinfo {author} {\bibfnamefont {A.}~\bibnamefont
  {{Bauswein}}}, \bibinfo {author} {\bibfnamefont {O.}~\bibnamefont {{Just}}},
  \bibinfo {author} {\bibfnamefont {H.-T.}\ \bibnamefont {{Janka}}},\ and\
  \bibinfo {author} {\bibfnamefont {N.}~\bibnamefont {{Stergioulas}}},\ }\href
  {https://doi.org/10.3847/2041-8213/aa9994} {\bibfield  {journal} {\bibinfo
  {journal} {\apjl}\ }\textbf {\bibinfo {volume} {850}},\ \bibinfo {eid} {L34}
  (\bibinfo {year} {2017})},\ \Eprint {https://arxiv.org/abs/1710.06843}
  {arXiv:1710.06843 [astro-ph.HE]} \BibitemShut {NoStop}%
\bibitem [{\citenamefont {{Margalit}}\ and\ \citenamefont
  {{Metzger}}(2017)}]{MaMe2017}%
  \BibitemOpen
  \bibfield  {author} {\bibinfo {author} {\bibfnamefont {B.}~\bibnamefont
  {{Margalit}}}\ and\ \bibinfo {author} {\bibfnamefont {B.~D.}\ \bibnamefont
  {{Metzger}}},\ }\href {https://doi.org/10.3847/2041-8213/aa991c} {\bibfield
  {journal} {\bibinfo  {journal} {\apjl}\ }\textbf {\bibinfo {volume} {850}},\
  \bibinfo {eid} {L19} (\bibinfo {year} {2017})},\ \Eprint
  {https://arxiv.org/abs/1710.05938} {arXiv:1710.05938 [astro-ph.HE]}
  \BibitemShut {NoStop}%
\bibitem [{\citenamefont {{Coughlin}}\ \emph {et~al.}(2019)\citenamefont
  {{Coughlin}}, \citenamefont {{Dietrich}}, \citenamefont {{Margalit}} \emph
  {et~al.}}]{CoDi2018b}%
  \BibitemOpen
  \bibfield  {author} {\bibinfo {author} {\bibfnamefont {M.~W.}\ \bibnamefont
  {{Coughlin}}}, \bibinfo {author} {\bibfnamefont {T.}~\bibnamefont
  {{Dietrich}}}, \bibinfo {author} {\bibfnamefont {B.}~\bibnamefont
  {{Margalit}}}, \emph {et~al.},\ }\href
  {https://doi.org/10.1093/mnrasl/slz133} {\bibfield  {journal} {\bibinfo
  {journal} {\mnras}\ }\textbf {\bibinfo {volume} {489}},\ \bibinfo {pages}
  {L91} (\bibinfo {year} {2019})},\ \Eprint {https://arxiv.org/abs/1812.04803}
  {arXiv:1812.04803 [astro-ph.HE]} \BibitemShut {NoStop}%
\bibitem [{\citenamefont {{Coughlin}}\ \emph {et~al.}(2018)\citenamefont
  {{Coughlin}}, \citenamefont {{Dietrich}}, \citenamefont {{Doctor}} \emph
  {et~al.}}]{CoDi2018}%
  \BibitemOpen
  \bibfield  {author} {\bibinfo {author} {\bibfnamefont {M.~W.}\ \bibnamefont
  {{Coughlin}}}, \bibinfo {author} {\bibfnamefont {T.}~\bibnamefont
  {{Dietrich}}}, \bibinfo {author} {\bibfnamefont {Z.}~\bibnamefont
  {{Doctor}}}, \emph {et~al.},\ }\href {https://doi.org/10.1093/mnras/sty2174}
  {\bibfield  {journal} {\bibinfo  {journal} {\mnras}\ }\textbf {\bibinfo
  {volume} {480}},\ \bibinfo {pages} {3871} (\bibinfo {year} {2018})},\ \Eprint
  {https://arxiv.org/abs/1805.09371} {arXiv:1805.09371 [astro-ph.HE]}
  \BibitemShut {NoStop}%
\bibitem [{\citenamefont {{Coughlin}}\ \emph
  {et~al.}(2020{\natexlab{a}})\citenamefont {{Coughlin}}, \citenamefont
  {{Dietrich}}, \citenamefont {{Antier}} \emph {et~al.}}]{CoDi2019b}%
  \BibitemOpen
  \bibfield  {author} {\bibinfo {author} {\bibfnamefont {M.~W.}\ \bibnamefont
  {{Coughlin}}}, \bibinfo {author} {\bibfnamefont {T.}~\bibnamefont
  {{Dietrich}}}, \bibinfo {author} {\bibfnamefont {S.}~\bibnamefont
  {{Antier}}}, \emph {et~al.},\ }\href {https://doi.org/10.1093/mnras/stz3457}
  {\bibfield  {journal} {\bibinfo  {journal} {\mnras}\ }\textbf {\bibinfo
  {volume} {492}},\ \bibinfo {pages} {863} (\bibinfo {year}
  {2020}{\natexlab{a}})},\ \Eprint {https://arxiv.org/abs/1910.11246}
  {arXiv:1910.11246 [astro-ph.HE]} \BibitemShut {NoStop}%
\bibitem [{\citenamefont {{Annala}}\ \emph {et~al.}(2018)\citenamefont
  {{Annala}}, \citenamefont {{Gorda}}, \citenamefont {{Kurkela}} \emph
  {et~al.}}]{AnEe2018}%
  \BibitemOpen
  \bibfield  {author} {\bibinfo {author} {\bibfnamefont {E.}~\bibnamefont
  {{Annala}}}, \bibinfo {author} {\bibfnamefont {T.}~\bibnamefont {{Gorda}}},
  \bibinfo {author} {\bibfnamefont {A.}~\bibnamefont {{Kurkela}}}, \emph
  {et~al.},\ }\href {https://doi.org/10.1103/PhysRevLett.120.172703} {\bibfield
   {journal} {\bibinfo  {journal} {\prl}\ }\textbf {\bibinfo {volume} {120}},\
  \bibinfo {eid} {172703} (\bibinfo {year} {2018})},\ \Eprint
  {https://arxiv.org/abs/1711.02644} {arXiv:1711.02644 [astro-ph.HE]}
  \BibitemShut {NoStop}%
\bibitem [{\citenamefont {Most}\ \emph {et~al.}(2018)\citenamefont {Most},
  \citenamefont {Weih}, \citenamefont {Rezzolla} \emph {et~al.}}]{MoWe2018}%
  \BibitemOpen
  \bibfield  {author} {\bibinfo {author} {\bibfnamefont {E.~R.}\ \bibnamefont
  {Most}}, \bibinfo {author} {\bibfnamefont {L.~R.}\ \bibnamefont {Weih}},
  \bibinfo {author} {\bibfnamefont {L.}~\bibnamefont {Rezzolla}}, \emph
  {et~al.},\ }\href {https://doi.org/10.1103/PhysRevLett.120.261103} {\bibfield
   {journal} {\bibinfo  {journal} {Phys. Rev. Lett.}\ }\textbf {\bibinfo
  {volume} {120}},\ \bibinfo {pages} {261103} (\bibinfo {year}
  {2018})}\BibitemShut {NoStop}%
\bibitem [{\citenamefont {{Radice}}\ \emph {et~al.}(2018)\citenamefont
  {{Radice}}, \citenamefont {{Perego}}, \citenamefont {{Zappa}} \emph
  {et~al.}}]{RaPe2018}%
  \BibitemOpen
  \bibfield  {author} {\bibinfo {author} {\bibfnamefont {D.}~\bibnamefont
  {{Radice}}}, \bibinfo {author} {\bibfnamefont {A.}~\bibnamefont {{Perego}}},
  \bibinfo {author} {\bibfnamefont {F.}~\bibnamefont {{Zappa}}}, \emph
  {et~al.},\ }\href {https://doi.org/10.3847/2041-8213/aaa402} {\bibfield
  {journal} {\bibinfo  {journal} {\apjl}\ }\textbf {\bibinfo {volume} {852}},\
  \bibinfo {eid} {L29} (\bibinfo {year} {2018})},\ \Eprint
  {https://arxiv.org/abs/1711.03647} {arXiv:1711.03647 [astro-ph.HE]}
  \BibitemShut {NoStop}%
\bibitem [{\citenamefont {{Lai}}\ \emph {et~al.}(2019)\citenamefont {{Lai}},
  \citenamefont {{Zhou}},\ and\ \citenamefont {{Xu}}}]{Lai2019}%
  \BibitemOpen
  \bibfield  {author} {\bibinfo {author} {\bibfnamefont {X.}~\bibnamefont
  {{Lai}}}, \bibinfo {author} {\bibfnamefont {E.}~\bibnamefont {{Zhou}}},\ and\
  \bibinfo {author} {\bibfnamefont {R.}~\bibnamefont {{Xu}}},\ }\href
  {https://doi.org/10.1140/epja/i2019-12720-8} {\bibfield  {journal} {\bibinfo
  {journal} {European Physical Journal A}\ }\textbf {\bibinfo {volume} {55}},\
  \bibinfo {eid} {60} (\bibinfo {year} {2019})},\ \Eprint
  {https://arxiv.org/abs/1811.00193} {arXiv:1811.00193 [astro-ph.HE]}
  \BibitemShut {NoStop}%
\bibitem [{\citenamefont {{Dietrich}}\ \emph {et~al.}(2020)\citenamefont
  {{Dietrich}}, \citenamefont {{Coughlin}}, \citenamefont {{Pang}} \emph
  {et~al.}}]{DiCo2020}%
  \BibitemOpen
  \bibfield  {author} {\bibinfo {author} {\bibfnamefont {T.}~\bibnamefont
  {{Dietrich}}}, \bibinfo {author} {\bibfnamefont {M.~W.}\ \bibnamefont
  {{Coughlin}}}, \bibinfo {author} {\bibfnamefont {P.~T.~H.}\ \bibnamefont
  {{Pang}}}, \emph {et~al.},\ }\href {https://doi.org/10.1126/science.abb4317}
  {\bibfield  {journal} {\bibinfo  {journal} {Science}\ }\textbf {\bibinfo
  {volume} {370}},\ \bibinfo {pages} {1450} (\bibinfo {year} {2020})},\ \Eprint
  {https://arxiv.org/abs/2002.11355} {arXiv:2002.11355 [astro-ph.HE]}
  \BibitemShut {NoStop}%
\bibitem [{\citenamefont {{Huth}}\ \emph {et~al.}(2022)\citenamefont {{Huth}},
  \citenamefont {{Pang}}, \citenamefont {{Tews}} \emph
  {et~al.}}]{Huth:2021bsp}%
  \BibitemOpen
  \bibfield  {author} {\bibinfo {author} {\bibfnamefont {S.}~\bibnamefont
  {{Huth}}}, \bibinfo {author} {\bibfnamefont {P.~T.~H.}\ \bibnamefont
  {{Pang}}}, \bibinfo {author} {\bibfnamefont {I.}~\bibnamefont {{Tews}}},
  \emph {et~al.},\ }\href {https://doi.org/10.1038/s41586-022-04750-w}
  {\bibfield  {journal} {\bibinfo  {journal} {\nat}\ }\textbf {\bibinfo
  {volume} {606}},\ \bibinfo {pages} {276} (\bibinfo {year} {2022})},\ \Eprint
  {https://arxiv.org/abs/2107.06229} {arXiv:2107.06229 [nucl-th]} \BibitemShut
  {NoStop}%
\bibitem [{\citenamefont {{Coughlin}}\ \emph
  {et~al.}(2020{\natexlab{b}})\citenamefont {{Coughlin}}, \citenamefont
  {{Dietrich}}, \citenamefont {{Heinzel}} \emph {et~al.}}]{CoDi2019}%
  \BibitemOpen
  \bibfield  {author} {\bibinfo {author} {\bibfnamefont {M.~W.}\ \bibnamefont
  {{Coughlin}}}, \bibinfo {author} {\bibfnamefont {T.}~\bibnamefont
  {{Dietrich}}}, \bibinfo {author} {\bibfnamefont {J.}~\bibnamefont
  {{Heinzel}}}, \emph {et~al.},\ }\href
  {https://doi.org/10.1103/PhysRevResearch.2.022006} {\bibfield  {journal}
  {\bibinfo  {journal} {Physical Review Research}\ }\textbf {\bibinfo {volume}
  {2}},\ \bibinfo {eid} {022006} (\bibinfo {year} {2020}{\natexlab{b}})},\
  \Eprint {https://arxiv.org/abs/1908.00889} {arXiv:1908.00889 [astro-ph.HE]}
  \BibitemShut {NoStop}%
\bibitem [{\citenamefont {{Coughlin}}\ \emph
  {et~al.}(2020{\natexlab{c}})\citenamefont {{Coughlin}}, \citenamefont
  {{Antier}}, \citenamefont {{Dietrich}} \emph {et~al.}}]{CoAn2020}%
  \BibitemOpen
  \bibfield  {author} {\bibinfo {author} {\bibfnamefont {M.~W.}\ \bibnamefont
  {{Coughlin}}}, \bibinfo {author} {\bibfnamefont {S.}~\bibnamefont
  {{Antier}}}, \bibinfo {author} {\bibfnamefont {T.}~\bibnamefont
  {{Dietrich}}}, \emph {et~al.},\ }\href
  {https://doi.org/10.1038/s41467-020-17998-5} {\bibfield  {journal} {\bibinfo
  {journal} {Nature Communications}\ }\textbf {\bibinfo {volume} {11}},\
  \bibinfo {eid} {4129} (\bibinfo {year} {2020}{\natexlab{c}})},\ \Eprint
  {https://arxiv.org/abs/2008.07420} {arXiv:2008.07420 [astro-ph.HE]}
  \BibitemShut {NoStop}%
\bibitem [{\citenamefont {{Abbott}}\ \emph
  {et~al.}(2017{\natexlab{d}})\citenamefont {{Abbott}}, \citenamefont
  {{Abbott}}, \citenamefont {{Abbott}} \emph {et~al.}}]{2017Natur.551...85A}%
  \BibitemOpen
  \bibfield  {author} {\bibinfo {author} {\bibfnamefont {B.~P.}\ \bibnamefont
  {{Abbott}}}, \bibinfo {author} {\bibfnamefont {R.}~\bibnamefont {{Abbott}}},
  \bibinfo {author} {\bibfnamefont {T.~D.}\ \bibnamefont {{Abbott}}}, \emph
  {et~al.},\ }\href {https://doi.org/10.1038/nature24471} {\bibfield  {journal}
  {\bibinfo  {journal} {\nat}\ }\textbf {\bibinfo {volume} {551}},\ \bibinfo
  {pages} {85} (\bibinfo {year} {2017}{\natexlab{d}})},\ \Eprint
  {https://arxiv.org/abs/1710.05835} {arXiv:1710.05835} \BibitemShut {NoStop}%
\bibitem [{\citenamefont {{Hotokezaka}}\ \emph {et~al.}(2019)\citenamefont
  {{Hotokezaka}}, \citenamefont {{Nakar}}, \citenamefont {{Gottlieb}} \emph
  {et~al.}}]{HoNa2018}%
  \BibitemOpen
  \bibfield  {author} {\bibinfo {author} {\bibfnamefont {K.}~\bibnamefont
  {{Hotokezaka}}}, \bibinfo {author} {\bibfnamefont {E.}~\bibnamefont
  {{Nakar}}}, \bibinfo {author} {\bibfnamefont {O.}~\bibnamefont {{Gottlieb}}},
  \emph {et~al.},\ }\href {https://doi.org/10.1038/s41550-019-0820-1}
  {\bibfield  {journal} {\bibinfo  {journal} {Nature Astronomy}\ }\textbf
  {\bibinfo {volume} {3}},\ \bibinfo {pages} {940} (\bibinfo {year} {2019})},\
  \Eprint {https://arxiv.org/abs/1806.10596} {arXiv:1806.10596 [astro-ph.CO]}
  \BibitemShut {NoStop}%
\bibitem [{\citenamefont {{Chornock}}\ \emph {et~al.}(2017)\citenamefont
  {{Chornock}}, \citenamefont {{Berger}}, \citenamefont {{Kasen}} \emph
  {et~al.}}]{ChBe2017}%
  \BibitemOpen
  \bibfield  {author} {\bibinfo {author} {\bibfnamefont {R.}~\bibnamefont
  {{Chornock}}}, \bibinfo {author} {\bibfnamefont {E.}~\bibnamefont
  {{Berger}}}, \bibinfo {author} {\bibfnamefont {D.}~\bibnamefont {{Kasen}}},
  \emph {et~al.},\ }\href {https://doi.org/10.3847/2041-8213/aa905c} {\bibfield
   {journal} {\bibinfo  {journal} {\apjl}\ }\textbf {\bibinfo {volume} {848}},\
  \bibinfo {eid} {L19} (\bibinfo {year} {2017})},\ \Eprint
  {https://arxiv.org/abs/1710.05454} {arXiv:1710.05454 [astro-ph.HE]}
  \BibitemShut {NoStop}%
\bibitem [{\citenamefont {{Cowperthwaite}}\ \emph {et~al.}(2017)\citenamefont
  {{Cowperthwaite}}, \citenamefont {{Berger}}, \citenamefont {{Villar}} \emph
  {et~al.}}]{CoBe2017}%
  \BibitemOpen
  \bibfield  {author} {\bibinfo {author} {\bibfnamefont {P.~S.}\ \bibnamefont
  {{Cowperthwaite}}}, \bibinfo {author} {\bibfnamefont {E.}~\bibnamefont
  {{Berger}}}, \bibinfo {author} {\bibfnamefont {V.~A.}\ \bibnamefont
  {{Villar}}}, \emph {et~al.},\ }\href
  {https://doi.org/10.3847/2041-8213/aa8fc7} {\bibfield  {journal} {\bibinfo
  {journal} {\apjl}\ }\textbf {\bibinfo {volume} {848}},\ \bibinfo {eid} {L17}
  (\bibinfo {year} {2017})},\ \Eprint {https://arxiv.org/abs/1710.05840}
  {arXiv:1710.05840 [astro-ph.HE]} \BibitemShut {NoStop}%
\bibitem [{\citenamefont {{Rosswog}}\ \emph {et~al.}(2017)\citenamefont
  {{Rosswog}}, \citenamefont {{Feindt}}, \citenamefont {{Korobkin}} \emph
  {et~al.}}]{RoFe2017}%
  \BibitemOpen
  \bibfield  {author} {\bibinfo {author} {\bibfnamefont {S.}~\bibnamefont
  {{Rosswog}}}, \bibinfo {author} {\bibfnamefont {U.}~\bibnamefont {{Feindt}}},
  \bibinfo {author} {\bibfnamefont {O.}~\bibnamefont {{Korobkin}}}, \emph
  {et~al.},\ }\href {https://doi.org/10.1088/1361-6382/aa68a9} {\bibfield
  {journal} {\bibinfo  {journal} {Classical and Quantum Gravity}\ }\textbf
  {\bibinfo {volume} {34}},\ \bibinfo {eid} {104001} (\bibinfo {year}
  {2017})},\ \Eprint {https://arxiv.org/abs/1611.09822} {arXiv:1611.09822
  [astro-ph.HE]} \BibitemShut {NoStop}%
\bibitem [{\citenamefont {{Watson}}\ \emph {et~al.}(2019)\citenamefont
  {{Watson}}, \citenamefont {{Hansen}}, \citenamefont {{Selsing}} \emph
  {et~al.}}]{WaHa2019}%
  \BibitemOpen
  \bibfield  {author} {\bibinfo {author} {\bibfnamefont {D.}~\bibnamefont
  {{Watson}}}, \bibinfo {author} {\bibfnamefont {C.~J.}\ \bibnamefont
  {{Hansen}}}, \bibinfo {author} {\bibfnamefont {J.}~\bibnamefont {{Selsing}}},
  \emph {et~al.},\ }\href {https://doi.org/10.1038/s41586-019-1676-3}
  {\bibfield  {journal} {\bibinfo  {journal} {\nat}\ }\textbf {\bibinfo
  {volume} {574}},\ \bibinfo {pages} {497} (\bibinfo {year} {2019})},\ \Eprint
  {https://arxiv.org/abs/1910.10510} {arXiv:1910.10510 [astro-ph.HE]}
  \BibitemShut {NoStop}%
\bibitem [{\citenamefont {{Kasliwal}}\ \emph {et~al.}(2022)\citenamefont
  {{Kasliwal}}, \citenamefont {{Kasen}}, \citenamefont {{Lau}} \emph
  {et~al.}}]{KaKa2019}%
  \BibitemOpen
  \bibfield  {author} {\bibinfo {author} {\bibfnamefont {M.~M.}\ \bibnamefont
  {{Kasliwal}}}, \bibinfo {author} {\bibfnamefont {D.}~\bibnamefont {{Kasen}}},
  \bibinfo {author} {\bibfnamefont {R.~M.}\ \bibnamefont {{Lau}}}, \emph
  {et~al.},\ }\href {https://doi.org/10.1093/mnrasl/slz007} {\bibfield
  {journal} {\bibinfo  {journal} {\mnras}\ }\textbf {\bibinfo {volume} {510}},\
  \bibinfo {pages} {L7} (\bibinfo {year} {2022})},\ \Eprint
  {https://arxiv.org/abs/1812.08708} {arXiv:1812.08708 [astro-ph.HE]}
  \BibitemShut {NoStop}%
\bibitem [{\citenamefont {{The LVK
  Collaboration}}(2025)}]{2025arXiv250818080T}%
  \BibitemOpen
  \bibfield  {author} {\bibinfo {author} {\bibnamefont {{The LVK
  Collaboration}}},\ }\href@noop {} {\bibfield  {journal} {\bibinfo  {journal}
  {\apjl}\ }\textbf {\bibinfo {volume} {995}},\ \bibinfo {pages} {L18}
  (\bibinfo {year} {2025})},\ \Eprint {https://arxiv.org/abs/2508.18080}
  {arXiv:2508.18080 [gr-qc]} \BibitemShut {NoStop}%
\bibitem [{\citenamefont {{The LVK
  Collaboration}}(2026{\natexlab{b}})}]{2025arXiv250818081T}%
  \BibitemOpen
  \bibfield  {author} {\bibinfo {author} {\bibnamefont {{The LVK
  Collaboration}}},\ }\href@noop {} {\bibfield  {journal} {\bibinfo  {journal}
  {\apjl}\ }\textbf {\bibinfo {volume} {1004}},\ \bibinfo {pages} {L21}
  (\bibinfo {year} {2026}{\natexlab{b}})},\ \Eprint
  {https://arxiv.org/abs/2508.18081} {arXiv:2508.18081 [gr-qc]} \BibitemShut
  {NoStop}%
\bibitem [{\citenamefont {{Adams}}\ \emph {et~al.}(2016)\citenamefont
  {{Adams}}, \citenamefont {{Buskulic}}, \citenamefont {{Germain}} \emph
  {et~al.}}]{Adams:2015ulm}%
  \BibitemOpen
  \bibfield  {author} {\bibinfo {author} {\bibfnamefont {T.}~\bibnamefont
  {{Adams}}}, \bibinfo {author} {\bibfnamefont {D.}~\bibnamefont {{Buskulic}}},
  \bibinfo {author} {\bibfnamefont {V.}~\bibnamefont {{Germain}}}, \emph
  {et~al.},\ }\href {https://doi.org/10.1088/0264-9381/33/17/175012} {\bibfield
   {journal} {\bibinfo  {journal} {Classical and Quantum Gravity}\ }\textbf
  {\bibinfo {volume} {33}},\ \bibinfo {eid} {175012} (\bibinfo {year}
  {2016})},\ \Eprint {https://arxiv.org/abs/1512.02864} {arXiv:1512.02864
  [gr-qc]} \BibitemShut {NoStop}%
\bibitem [{\citenamefont {{Aubin}}\ \emph {et~al.}(2021)\citenamefont
  {{Aubin}}, \citenamefont {{Brighenti}}, \citenamefont {{Chierici}} \emph
  {et~al.}}]{Aubin:2020goo}%
  \BibitemOpen
  \bibfield  {author} {\bibinfo {author} {\bibfnamefont {F.}~\bibnamefont
  {{Aubin}}}, \bibinfo {author} {\bibfnamefont {F.}~\bibnamefont
  {{Brighenti}}}, \bibinfo {author} {\bibfnamefont {R.}~\bibnamefont
  {{Chierici}}}, \emph {et~al.},\ }\href
  {https://doi.org/10.1088/1361-6382/abe913} {\bibfield  {journal} {\bibinfo
  {journal} {Classical and Quantum Gravity}\ }\textbf {\bibinfo {volume}
  {38}},\ \bibinfo {eid} {095004} (\bibinfo {year} {2021})},\ \Eprint
  {https://arxiv.org/abs/2012.11512} {arXiv:2012.11512 [gr-qc]} \BibitemShut
  {NoStop}%
\bibitem [{\citenamefont {{All{\'e}n{\'e}}}\ \emph {et~al.}(2025)\citenamefont
  {{All{\'e}n{\'e}}}, \citenamefont {{Aubin}}, \citenamefont {{Bentara}} \emph
  {et~al.}}]{Allene:2025saz}%
  \BibitemOpen
  \bibfield  {author} {\bibinfo {author} {\bibfnamefont {C.}~\bibnamefont
  {{All{\'e}n{\'e}}}}, \bibinfo {author} {\bibfnamefont {F.}~\bibnamefont
  {{Aubin}}}, \bibinfo {author} {\bibfnamefont {I.}~\bibnamefont {{Bentara}}},
  \emph {et~al.},\ }\href {https://doi.org/10.1088/1361-6382/add234} {\bibfield
   {journal} {\bibinfo  {journal} {Classical and Quantum Gravity}\ }\textbf
  {\bibinfo {volume} {42}},\ \bibinfo {eid} {105009} (\bibinfo {year}
  {2025})},\ \Eprint {https://arxiv.org/abs/2501.04598} {arXiv:2501.04598
  [gr-qc]} \BibitemShut {NoStop}%
\bibitem [{\citenamefont {{Messick}}\ \emph {et~al.}(2017)\citenamefont
  {{Messick}}, \citenamefont {{Blackburn}}, \citenamefont {{Brady}} \emph
  {et~al.}}]{Messick:2016aqy}%
  \BibitemOpen
  \bibfield  {author} {\bibinfo {author} {\bibfnamefont {C.}~\bibnamefont
  {{Messick}}}, \bibinfo {author} {\bibfnamefont {K.}~\bibnamefont
  {{Blackburn}}}, \bibinfo {author} {\bibfnamefont {P.}~\bibnamefont
  {{Brady}}}, \emph {et~al.},\ }\href
  {https://doi.org/10.1103/PhysRevD.95.042001} {\bibfield  {journal} {\bibinfo
  {journal} {\prd}\ }\textbf {\bibinfo {volume} {95}},\ \bibinfo {eid} {042001}
  (\bibinfo {year} {2017})},\ \Eprint {https://arxiv.org/abs/1604.04324}
  {arXiv:1604.04324 [astro-ph.IM]} \BibitemShut {NoStop}%
\bibitem [{\citenamefont {{Sachdev}}\ \emph {et~al.}(2019)\citenamefont
  {{Sachdev}}, \citenamefont {{Caudill}}, \citenamefont {{Fong}} \emph
  {et~al.}}]{Sachdev:2019vvd}%
  \BibitemOpen
  \bibfield  {author} {\bibinfo {author} {\bibfnamefont {S.}~\bibnamefont
  {{Sachdev}}}, \bibinfo {author} {\bibfnamefont {S.}~\bibnamefont
  {{Caudill}}}, \bibinfo {author} {\bibfnamefont {H.}~\bibnamefont {{Fong}}},
  \emph {et~al.},\ }\href {https://doi.org/10.48550/arXiv.1901.08580}
  {\bibfield  {journal} {\bibinfo  {journal} {arXiv e-prints}\ ,\ \bibinfo
  {eid} {arXiv:1901.08580}} (\bibinfo {year} {2019})},\ \Eprint
  {https://arxiv.org/abs/1901.08580} {arXiv:1901.08580 [gr-qc]} \BibitemShut
  {NoStop}%
\bibitem [{\citenamefont {{Hanna}}\ \emph {et~al.}(2020)\citenamefont
  {{Hanna}}, \citenamefont {{Caudill}}, \citenamefont {{Messick}} \emph
  {et~al.}}]{Hanna:2019ezx}%
  \BibitemOpen
  \bibfield  {author} {\bibinfo {author} {\bibfnamefont {C.}~\bibnamefont
  {{Hanna}}}, \bibinfo {author} {\bibfnamefont {S.}~\bibnamefont {{Caudill}}},
  \bibinfo {author} {\bibfnamefont {C.}~\bibnamefont {{Messick}}}, \emph
  {et~al.},\ }\href {https://doi.org/10.1103/PhysRevD.101.022003} {\bibfield
  {journal} {\bibinfo  {journal} {\prd}\ }\textbf {\bibinfo {volume} {101}},\
  \bibinfo {eid} {022003} (\bibinfo {year} {2020})},\ \Eprint
  {https://arxiv.org/abs/1901.02227} {arXiv:1901.02227 [gr-qc]} \BibitemShut
  {NoStop}%
\bibitem [{\citenamefont {{Cannon}}\ \emph {et~al.}(2021)\citenamefont
  {{Cannon}}, \citenamefont {{Caudill}}, \citenamefont {{Chan}} \emph
  {et~al.}}]{Cannon:2020qnf}%
  \BibitemOpen
  \bibfield  {author} {\bibinfo {author} {\bibfnamefont {K.}~\bibnamefont
  {{Cannon}}}, \bibinfo {author} {\bibfnamefont {S.}~\bibnamefont {{Caudill}}},
  \bibinfo {author} {\bibfnamefont {C.}~\bibnamefont {{Chan}}}, \emph
  {et~al.},\ }\href {https://doi.org/10.1016/j.softx.2021.100680} {\bibfield
  {journal} {\bibinfo  {journal} {SoftwareX}\ }\textbf {\bibinfo {volume}
  {14}},\ \bibinfo {eid} {100680} (\bibinfo {year} {2021})},\ \Eprint
  {https://arxiv.org/abs/2010.05082} {arXiv:2010.05082 [astro-ph.IM]}
  \BibitemShut {NoStop}%
\bibitem [{\citenamefont {{Sakon}}\ \emph {et~al.}(2024)\citenamefont
  {{Sakon}}, \citenamefont {{Tsukada}}, \citenamefont {{Fong}}, \citenamefont
  {{Kennington}},\ and\ \citenamefont {et~al.}}]{2024PhRvD.109d4066S}%
  \BibitemOpen
  \bibfield  {author} {\bibinfo {author} {\bibfnamefont {S.}~\bibnamefont
  {{Sakon}}}, \bibinfo {author} {\bibfnamefont {L.}~\bibnamefont {{Tsukada}}},
  \bibinfo {author} {\bibfnamefont {H.}~\bibnamefont {{Fong}}}, \bibinfo
  {author} {\bibfnamefont {J.}~\bibnamefont {{Kennington}}},\ and\ \bibinfo
  {author} {\bibnamefont {et~al.}},\ }\href
  {https://doi.org/10.1103/PhysRevD.109.044066} {\bibfield  {journal} {\bibinfo
   {journal} {\prd}\ }\textbf {\bibinfo {volume} {109}},\ \bibinfo {eid}
  {044066} (\bibinfo {year} {2024})},\ \Eprint
  {https://arxiv.org/abs/2211.16674} {arXiv:2211.16674 [gr-qc]} \BibitemShut
  {NoStop}%
\bibitem [{\citenamefont {{Ewing}}\ \emph {et~al.}(2024)\citenamefont
  {{Ewing}}, \citenamefont {{Huxford}}, \citenamefont {{Singh}}, \citenamefont
  {{Tsukada}} \emph {et~al.}}]{Ewing:2023qqe}%
  \BibitemOpen
  \bibfield  {author} {\bibinfo {author} {\bibfnamefont {B.}~\bibnamefont
  {{Ewing}}}, \bibinfo {author} {\bibfnamefont {R.}~\bibnamefont {{Huxford}}},
  \bibinfo {author} {\bibfnamefont {D.}~\bibnamefont {{Singh}}}, \bibinfo
  {author} {\bibfnamefont {L.}~\bibnamefont {{Tsukada}}}, \emph {et~al.},\
  }\href {https://doi.org/10.1103/PhysRevD.109.042008} {\bibfield  {journal}
  {\bibinfo  {journal} {\prd}\ }\textbf {\bibinfo {volume} {109}},\ \bibinfo
  {eid} {042008} (\bibinfo {year} {2024})},\ \Eprint
  {https://arxiv.org/abs/2305.05625} {arXiv:2305.05625 [gr-qc]} \BibitemShut
  {NoStop}%
\bibitem [{\citenamefont {{Tsukada}}\ \emph {et~al.}(2023)\citenamefont
  {{Tsukada}}, \citenamefont {{Joshi}}, \citenamefont {{Adhicary}} \emph
  {et~al.}}]{Tsukada:2023edh}%
  \BibitemOpen
  \bibfield  {author} {\bibinfo {author} {\bibfnamefont {L.}~\bibnamefont
  {{Tsukada}}}, \bibinfo {author} {\bibfnamefont {P.}~\bibnamefont {{Joshi}}},
  \bibinfo {author} {\bibfnamefont {S.}~\bibnamefont {{Adhicary}}}, \emph
  {et~al.},\ }\href {https://doi.org/10.1103/PhysRevD.108.043004} {\bibfield
  {journal} {\bibinfo  {journal} {\prd}\ }\textbf {\bibinfo {volume} {108}},\
  \bibinfo {eid} {043004} (\bibinfo {year} {2023})},\ \Eprint
  {https://arxiv.org/abs/2305.06286} {arXiv:2305.06286 [astro-ph.IM]}
  \BibitemShut {NoStop}%
\bibitem [{\citenamefont {{Ray}}\ \emph {et~al.}(2023)\citenamefont {{Ray}},
  \citenamefont {{Niu}}, \citenamefont {{Sakon}}, \citenamefont {{Ewing}} \emph
  {et~al.}}]{Ray:2023nhx}%
  \BibitemOpen
  \bibfield  {author} {\bibinfo {author} {\bibfnamefont {A.}~\bibnamefont
  {{Ray}}}, \bibinfo {author} {\bibfnamefont {W.}~\bibnamefont {{Niu}}},
  \bibinfo {author} {\bibfnamefont {S.}~\bibnamefont {{Sakon}}}, \bibinfo
  {author} {\bibfnamefont {B.}~\bibnamefont {{Ewing}}}, \emph {et~al.},\ }\href
  {https://doi.org/10.48550/arXiv.2306.07190} {\bibfield  {journal} {\bibinfo
  {journal} {arXiv e-prints}\ ,\ \bibinfo {eid} {arXiv:2306.07190}} (\bibinfo
  {year} {2023})},\ \Eprint {https://arxiv.org/abs/2306.07190}
  {arXiv:2306.07190 [gr-qc]} \BibitemShut {NoStop}%
\bibitem [{\citenamefont {{Joshi}}\ \emph
  {et~al.}(2025{\natexlab{a}})\citenamefont {{Joshi}}, \citenamefont
  {{Tsukada}}, \citenamefont {{Hanna}} \emph {et~al.}}]{Joshi:2025nty}%
  \BibitemOpen
  \bibfield  {author} {\bibinfo {author} {\bibfnamefont {P.}~\bibnamefont
  {{Joshi}}}, \bibinfo {author} {\bibfnamefont {L.}~\bibnamefont {{Tsukada}}},
  \bibinfo {author} {\bibfnamefont {C.}~\bibnamefont {{Hanna}}}, \emph
  {et~al.},\ }\href {https://doi.org/10.48550/arXiv.2506.06497} {\bibfield
  {journal} {\bibinfo  {journal} {arXiv e-prints}\ ,\ \bibinfo {eid}
  {arXiv:2506.06497}} (\bibinfo {year} {2025}{\natexlab{a}})},\ \Eprint
  {https://arxiv.org/abs/2506.06497} {arXiv:2506.06497 [gr-qc]} \BibitemShut
  {NoStop}%
\bibitem [{\citenamefont {{Joshi}}\ \emph
  {et~al.}(2025{\natexlab{b}})\citenamefont {{Joshi}}, \citenamefont {{Niu}},
  \citenamefont {{Hanna}} \emph {et~al.}}]{Joshi:2025zdu}%
  \BibitemOpen
  \bibfield  {author} {\bibinfo {author} {\bibfnamefont {P.}~\bibnamefont
  {{Joshi}}}, \bibinfo {author} {\bibfnamefont {W.}~\bibnamefont {{Niu}}},
  \bibinfo {author} {\bibfnamefont {C.}~\bibnamefont {{Hanna}}}, \emph
  {et~al.},\ }\href {https://doi.org/10.48550/arXiv.2505.23959} {\bibfield
  {journal} {\bibinfo  {journal} {arXiv e-prints}\ ,\ \bibinfo {eid}
  {arXiv:2505.23959}} (\bibinfo {year} {2025}{\natexlab{b}})},\ \Eprint
  {https://arxiv.org/abs/2505.23959} {arXiv:2505.23959 [gr-qc]} \BibitemShut
  {NoStop}%
\bibitem [{\citenamefont {{Allen}}(2005)}]{Allen:2004gu}%
  \BibitemOpen
  \bibfield  {author} {\bibinfo {author} {\bibfnamefont {B.}~\bibnamefont
  {{Allen}}},\ }\href {https://doi.org/10.1103/PhysRevD.71.062001} {\bibfield
  {journal} {\bibinfo  {journal} {\prd}\ }\textbf {\bibinfo {volume} {71}},\
  \bibinfo {eid} {062001} (\bibinfo {year} {2005})},\ \Eprint
  {https://arxiv.org/abs/gr-qc/0405045} {arXiv:gr-qc/0405045 [gr-qc]}
  \BibitemShut {NoStop}%
\bibitem [{\citenamefont {{Usman}}\ \emph {et~al.}(2016)\citenamefont
  {{Usman}}, \citenamefont {{Nitz}}, \citenamefont {{Harry}} \emph
  {et~al.}}]{Usman_2016}%
  \BibitemOpen
  \bibfield  {author} {\bibinfo {author} {\bibfnamefont {S.~A.}\ \bibnamefont
  {{Usman}}}, \bibinfo {author} {\bibfnamefont {A.~H.}\ \bibnamefont {{Nitz}}},
  \bibinfo {author} {\bibfnamefont {I.~W.}\ \bibnamefont {{Harry}}}, \emph
  {et~al.},\ }\href {https://doi.org/10.1088/0264-9381/33/21/215004} {\bibfield
   {journal} {\bibinfo  {journal} {Classical and Quantum Gravity}\ }\textbf
  {\bibinfo {volume} {33}},\ \bibinfo {eid} {215004} (\bibinfo {year}
  {2016})},\ \Eprint {https://arxiv.org/abs/1508.02357} {arXiv:1508.02357
  [gr-qc]} \BibitemShut {NoStop}%
\bibitem [{\citenamefont {{Nitz}}(2018)}]{Nitz:2017lco}%
  \BibitemOpen
  \bibfield  {author} {\bibinfo {author} {\bibfnamefont {A.~H.}\ \bibnamefont
  {{Nitz}}},\ }\href {https://doi.org/10.1088/1361-6382/aaa13d} {\bibfield
  {journal} {\bibinfo  {journal} {Classical and Quantum Gravity}\ }\textbf
  {\bibinfo {volume} {35}},\ \bibinfo {eid} {035016} (\bibinfo {year}
  {2018})},\ \Eprint {https://arxiv.org/abs/1709.08974} {arXiv:1709.08974
  [gr-qc]} \BibitemShut {NoStop}%
\bibitem [{\citenamefont {{Dal Canton}}\ \emph {et~al.}(2021)\citenamefont
  {{Dal Canton}}, \citenamefont {{Nitz}}, \citenamefont {{Gadre}} \emph
  {et~al.}}]{DalCanton:2020vpm}%
  \BibitemOpen
  \bibfield  {author} {\bibinfo {author} {\bibfnamefont {T.}~\bibnamefont {{Dal
  Canton}}}, \bibinfo {author} {\bibfnamefont {A.~H.}\ \bibnamefont {{Nitz}}},
  \bibinfo {author} {\bibfnamefont {B.}~\bibnamefont {{Gadre}}}, \emph
  {et~al.},\ }\href {https://doi.org/10.3847/1538-4357/ac2f9a} {\bibfield
  {journal} {\bibinfo  {journal} {\apj}\ }\textbf {\bibinfo {volume} {923}},\
  \bibinfo {eid} {254} (\bibinfo {year} {2021})},\ \Eprint
  {https://arxiv.org/abs/2008.07494} {arXiv:2008.07494 [astro-ph.HE]}
  \BibitemShut {NoStop}%
\bibitem [{\citenamefont {{Davies}}\ \emph {et~al.}(2020)\citenamefont
  {{Davies}}, \citenamefont {{Dent}}, \citenamefont {{T{\'a}pai}} \emph
  {et~al.}}]{Davies:2020tsx}%
  \BibitemOpen
  \bibfield  {author} {\bibinfo {author} {\bibfnamefont {G.~S.}\ \bibnamefont
  {{Davies}}}, \bibinfo {author} {\bibfnamefont {T.}~\bibnamefont {{Dent}}},
  \bibinfo {author} {\bibfnamefont {M.}~\bibnamefont {{T{\'a}pai}}}, \emph
  {et~al.},\ }\href {https://doi.org/10.1103/PhysRevD.102.022004} {\bibfield
  {journal} {\bibinfo  {journal} {\prd}\ }\textbf {\bibinfo {volume} {102}},\
  \bibinfo {eid} {022004} (\bibinfo {year} {2020})},\ \Eprint
  {https://arxiv.org/abs/2002.08291} {arXiv:2002.08291 [astro-ph.HE]}
  \BibitemShut {NoStop}%
\bibitem [{\citenamefont {{Hooper}}\ \emph {et~al.}(2012)\citenamefont
  {{Hooper}}, \citenamefont {{Chung}}, \citenamefont {{Luan}}, \citenamefont
  {{Blair}}, \citenamefont {{Chen}},\ and\ \citenamefont
  {{Wen}}}]{s_hooper_2012}%
  \BibitemOpen
  \bibfield  {author} {\bibinfo {author} {\bibfnamefont {S.}~\bibnamefont
  {{Hooper}}}, \bibinfo {author} {\bibfnamefont {S.~K.}\ \bibnamefont
  {{Chung}}}, \bibinfo {author} {\bibfnamefont {J.}~\bibnamefont {{Luan}}},
  \bibinfo {author} {\bibfnamefont {D.}~\bibnamefont {{Blair}}}, \bibinfo
  {author} {\bibfnamefont {Y.}~\bibnamefont {{Chen}}},\ and\ \bibinfo {author}
  {\bibfnamefont {L.}~\bibnamefont {{Wen}}},\ }\href
  {https://doi.org/10.1103/PhysRevD.86.024012} {\bibfield  {journal} {\bibinfo
  {journal} {\prd}\ }\textbf {\bibinfo {volume} {86}},\ \bibinfo {eid} {024012}
  (\bibinfo {year} {2012})},\ \Eprint {https://arxiv.org/abs/1108.3186}
  {arXiv:1108.3186 [gr-qc]} \BibitemShut {NoStop}%
\bibitem [{\citenamefont {{Luan}}\ \emph {et~al.}(2012)\citenamefont {{Luan}},
  \citenamefont {{Hooper}}, \citenamefont {{Wen}} \emph
  {et~al.}}]{j_luan_2012}%
  \BibitemOpen
  \bibfield  {author} {\bibinfo {author} {\bibfnamefont {J.}~\bibnamefont
  {{Luan}}}, \bibinfo {author} {\bibfnamefont {S.}~\bibnamefont {{Hooper}}},
  \bibinfo {author} {\bibfnamefont {L.}~\bibnamefont {{Wen}}}, \emph {et~al.},\
  }\href {https://doi.org/10.1103/PhysRevD.85.102002} {\bibfield  {journal}
  {\bibinfo  {journal} {\prd}\ }\textbf {\bibinfo {volume} {85}},\ \bibinfo
  {eid} {102002} (\bibinfo {year} {2012})},\ \Eprint
  {https://arxiv.org/abs/1108.3174} {arXiv:1108.3174 [gr-qc]} \BibitemShut
  {NoStop}%
\bibitem [{\citenamefont {{Abbott}}\ \emph {et~al.}(2023)\citenamefont
  {{Abbott}}, \citenamefont {{Abbott}}, \citenamefont {{Acernese}} \emph
  {et~al.}}]{AbEA2021d}%
  \BibitemOpen
  \bibfield  {author} {\bibinfo {author} {\bibfnamefont {R.}~\bibnamefont
  {{Abbott}}}, \bibinfo {author} {\bibfnamefont {T.~D.}\ \bibnamefont
  {{Abbott}}}, \bibinfo {author} {\bibfnamefont {F.}~\bibnamefont
  {{Acernese}}}, \emph {et~al.},\ }\href
  {https://doi.org/10.1103/PhysRevX.13.041039} {\bibfield  {journal} {\bibinfo
  {journal} {Physical Review X}\ }\textbf {\bibinfo {volume} {13}},\ \bibinfo
  {eid} {041039} (\bibinfo {year} {2023})},\ \Eprint
  {https://arxiv.org/abs/2111.03606} {arXiv:2111.03606 [gr-qc]} \BibitemShut
  {NoStop}%
\bibitem [{\citenamefont {{The LVK Collaboration}}(2021)}]{AbEA2021b}%
  \BibitemOpen
  \bibfield  {author} {\bibinfo {author} {\bibnamefont {{The LVK
  Collaboration}}},\ }\href {https://doi.org/10.1103/PhysRevD.104.122004}
  {\bibfield  {journal} {\bibinfo  {journal} {\prd}\ }\textbf {\bibinfo
  {volume} {104}},\ \bibinfo {eid} {122004} (\bibinfo {year} {2021})},\ \Eprint
  {https://arxiv.org/abs/2107.03701} {arXiv:2107.03701 [gr-qc]} \BibitemShut
  {NoStop}%
\bibitem [{\citenamefont {{Klimenko}}\ \emph {et~al.}(2008)\citenamefont
  {{Klimenko}}, \citenamefont {{Yakushin}}, \citenamefont {{Mercer}} \emph
  {et~al.}}]{klimenko2008coherent}%
  \BibitemOpen
  \bibfield  {author} {\bibinfo {author} {\bibfnamefont {S.}~\bibnamefont
  {{Klimenko}}}, \bibinfo {author} {\bibfnamefont {I.}~\bibnamefont
  {{Yakushin}}}, \bibinfo {author} {\bibfnamefont {A.}~\bibnamefont
  {{Mercer}}}, \emph {et~al.},\ }\href
  {https://doi.org/10.1088/0264-9381/25/11/114029} {\bibfield  {journal}
  {\bibinfo  {journal} {Classical and Quantum Gravity}\ }\textbf {\bibinfo
  {volume} {25}},\ \bibinfo {eid} {114029} (\bibinfo {year} {2008})},\ \Eprint
  {https://arxiv.org/abs/0802.3232} {arXiv:0802.3232 [gr-qc]} \BibitemShut
  {NoStop}%
\bibitem [{\citenamefont {{Klimenko}}\ \emph {et~al.}(2016)\citenamefont
  {{Klimenko}}, \citenamefont {{Vedovato}}, \citenamefont {{Drago}} \emph
  {et~al.}}]{Klimenko:2015ypf}%
  \BibitemOpen
  \bibfield  {author} {\bibinfo {author} {\bibfnamefont {S.}~\bibnamefont
  {{Klimenko}}}, \bibinfo {author} {\bibfnamefont {G.}~\bibnamefont
  {{Vedovato}}}, \bibinfo {author} {\bibfnamefont {M.}~\bibnamefont {{Drago}}},
  \emph {et~al.},\ }\href {https://doi.org/10.1103/PhysRevD.93.042004}
  {\bibfield  {journal} {\bibinfo  {journal} {\prd}\ }\textbf {\bibinfo
  {volume} {93}},\ \bibinfo {eid} {042004} (\bibinfo {year} {2016})},\ \Eprint
  {https://arxiv.org/abs/1511.05999} {arXiv:1511.05999 [gr-qc]} \BibitemShut
  {NoStop}%
\bibitem [{\citenamefont {{Drago}}\ \emph {et~al.}(2021)\citenamefont
  {{Drago}}, \citenamefont {{Klimenko}}, \citenamefont {{Lazzaro}} \emph
  {et~al.}}]{drago2021coherent}%
  \BibitemOpen
  \bibfield  {author} {\bibinfo {author} {\bibfnamefont {M.}~\bibnamefont
  {{Drago}}}, \bibinfo {author} {\bibfnamefont {S.}~\bibnamefont {{Klimenko}}},
  \bibinfo {author} {\bibfnamefont {C.}~\bibnamefont {{Lazzaro}}}, \emph
  {et~al.},\ }\href {https://doi.org/10.1016/j.softx.2021.100678} {\bibfield
  {journal} {\bibinfo  {journal} {SoftwareX}\ }\textbf {\bibinfo {volume}
  {14}},\ \bibinfo {eid} {100678} (\bibinfo {year} {2021})},\ \Eprint
  {https://arxiv.org/abs/2006.12604} {arXiv:2006.12604 [gr-qc]} \BibitemShut
  {NoStop}%
\bibitem [{\citenamefont {{Lynch}}\ \emph {et~al.}(2017)\citenamefont
  {{Lynch}}, \citenamefont {{Vitale}}, \citenamefont {{Essick}}, \citenamefont
  {{Katsavounidis}},\ and\ \citenamefont {{Robinet}}}]{Lynch_2017}%
  \BibitemOpen
  \bibfield  {author} {\bibinfo {author} {\bibfnamefont {R.}~\bibnamefont
  {{Lynch}}}, \bibinfo {author} {\bibfnamefont {S.}~\bibnamefont {{Vitale}}},
  \bibinfo {author} {\bibfnamefont {R.}~\bibnamefont {{Essick}}}, \bibinfo
  {author} {\bibfnamefont {E.}~\bibnamefont {{Katsavounidis}}},\ and\ \bibinfo
  {author} {\bibfnamefont {F.}~\bibnamefont {{Robinet}}},\ }\href
  {https://doi.org/10.1103/PhysRevD.95.104046} {\bibfield  {journal} {\bibinfo
  {journal} {\prd}\ }\textbf {\bibinfo {volume} {95}},\ \bibinfo {eid} {104046}
  (\bibinfo {year} {2017})},\ \Eprint {https://arxiv.org/abs/1511.05955}
  {arXiv:1511.05955 [gr-qc]} \BibitemShut {NoStop}%
\bibitem [{\citenamefont {Piotrzkowski}(2022)}]{piotrzkowski2022searching}%
  \BibitemOpen
  \bibfield  {author} {\bibinfo {author} {\bibfnamefont {B.~J.}\ \bibnamefont
  {Piotrzkowski}},\ }\emph {\bibinfo {title} {Searching for Gravitational Wave
  Associations with High-Energy Astrophysical Transients}},\ \href@noop {}
  {Ph.D. thesis},\ \bibinfo  {school} {University of Wisconsin-Milwaukee},
  \bibinfo {address} {Milwaukee, WI} (\bibinfo {year} {2022}),\ \bibinfo {note}
  {available at \url{http://digital.library.wisc.edu/1793/93071}}\BibitemShut
  {NoStop}%
\bibitem [{Note2()}]{Note2}%
  \BibitemOpen
  \bibinfo {note} {\protect \url
  {https://git.ligo.org/emfollow/gwcelery}}\BibitemShut {NoStop}%
\bibitem [{Note3()}]{Note3}%
  \BibitemOpen
  \bibinfo {note} {\protect \url
  {https://rtd.igwn.org/projects/gwcelery/en/latest/}}\BibitemShut {NoStop}%
\bibitem [{Note4()}]{Note4}%
  \BibitemOpen
  \bibinfo {note} {\protect \url {https://gracedb.ligo.org/}}\BibitemShut
  {NoStop}%
\bibitem [{\citenamefont {{Chaudhary}}\ \emph {et~al.}(2024)\citenamefont
  {{Chaudhary}}, \citenamefont {{Toivonen}}, \citenamefont {{Waratkar}} \emph
  {et~al.}}]{doi:10.1073/pnas.2316474121}%
  \BibitemOpen
  \bibfield  {author} {\bibinfo {author} {\bibfnamefont {S.~S.}\ \bibnamefont
  {{Chaudhary}}}, \bibinfo {author} {\bibfnamefont {A.}~\bibnamefont
  {{Toivonen}}}, \bibinfo {author} {\bibfnamefont {G.}~\bibnamefont
  {{Waratkar}}}, \emph {et~al.},\ }\href
  {https://doi.org/10.1073/pnas.2316474121} {\bibfield  {journal} {\bibinfo
  {journal} {Proceedings of the National Academy of Science}\ }\textbf
  {\bibinfo {volume} {121}},\ \bibinfo {eid} {e2316474121} (\bibinfo {year}
  {2024})},\ \Eprint {https://arxiv.org/abs/2308.04545} {arXiv:2308.04545
  [astro-ph.HE]} \BibitemShut {NoStop}%
\bibitem [{Note5()}]{Note5}%
  \BibitemOpen
  \bibinfo {note} {\protect \url
  {https://emfollow.docs.ligo.org/userguide/}}\BibitemShut {NoStop}%
\bibitem [{\citenamefont {{Singer}}\ and\ \citenamefont
  {{Price}}(2016)}]{2016PhRvD..93b4013S}%
  \BibitemOpen
  \bibfield  {author} {\bibinfo {author} {\bibfnamefont {L.~P.}\ \bibnamefont
  {{Singer}}}\ and\ \bibinfo {author} {\bibfnamefont {L.~R.}\ \bibnamefont
  {{Price}}},\ }\href {https://doi.org/10.1103/PhysRevD.93.024013} {\bibfield
  {journal} {\bibinfo  {journal} {\prd}\ }\textbf {\bibinfo {volume} {93}},\
  \bibinfo {eid} {024013} (\bibinfo {year} {2016})},\ \Eprint
  {https://arxiv.org/abs/1508.03634} {arXiv:1508.03634 [gr-qc]} \BibitemShut
  {NoStop}%
\bibitem [{\citenamefont {{Creighton}}\ and\ \citenamefont
  {{Anderson}}(2011)}]{jolien}%
  \BibitemOpen
  \bibfield  {author} {\bibinfo {author} {\bibfnamefont {J.}~\bibnamefont
  {{Creighton}}}\ and\ \bibinfo {author} {\bibfnamefont {W.}~\bibnamefont
  {{Anderson}}},\ }\bibinfo {title} {Gravitational‐wave physics and
  astronomy: An introduction to theory, experiment and data analysis.}\
  (\bibinfo  {publisher} {John Wiley \& Sons, Ltd},\ \bibinfo {year} {2011})\
  Chap.~\bibinfo {chapter} {7}, pp.\ \bibinfo {pages} {269--347},\ \Eprint
  {https://arxiv.org/abs/https://onlinelibrary.wiley.com/doi/pdf/10.1002/9783527636037.ch7}
  {https://onlinelibrary.wiley.com/doi/pdf/10.1002/9783527636037.ch7}
  \BibitemShut {NoStop}%
\bibitem [{\citenamefont {{Dietrich}}\ \emph {et~al.}(2017)\citenamefont
  {{Dietrich}}, \citenamefont {{Bernuzzi}},\ and\ \citenamefont
  {{Tichy}}}]{NRPv2}%
  \BibitemOpen
  \bibfield  {author} {\bibinfo {author} {\bibfnamefont {T.}~\bibnamefont
  {{Dietrich}}}, \bibinfo {author} {\bibfnamefont {S.}~\bibnamefont
  {{Bernuzzi}}},\ and\ \bibinfo {author} {\bibfnamefont {W.}~\bibnamefont
  {{Tichy}}},\ }\href {https://doi.org/10.1103/PhysRevD.96.121501} {\bibfield
  {journal} {\bibinfo  {journal} {\prd}\ }\textbf {\bibinfo {volume} {96}},\
  \bibinfo {eid} {121501} (\bibinfo {year} {2017})},\ \Eprint
  {https://arxiv.org/abs/1706.02969} {arXiv:1706.02969 [gr-qc]} \BibitemShut
  {NoStop}%
\bibitem [{\citenamefont {{Chabanat}}\ \emph {et~al.}(1998)\citenamefont
  {{Chabanat}}, \citenamefont {{Bonche}}, \citenamefont {{Haensel}} \emph
  {et~al.}}]{ChBo1998}%
  \BibitemOpen
  \bibfield  {author} {\bibinfo {author} {\bibfnamefont {E.}~\bibnamefont
  {{Chabanat}}}, \bibinfo {author} {\bibfnamefont {P.}~\bibnamefont
  {{Bonche}}}, \bibinfo {author} {\bibfnamefont {P.}~\bibnamefont {{Haensel}}},
  \emph {et~al.},\ }\href {https://doi.org/10.1016/S0375-9474(98)00180-8}
  {\bibfield  {journal} {\bibinfo  {journal} {\nphysa}\ }\textbf {\bibinfo
  {volume} {635}},\ \bibinfo {pages} {231} (\bibinfo {year}
  {1998})}\BibitemShut {NoStop}%
\bibitem [{\citenamefont {{Planck Collaboration}}(2021)}]{Planck18}%
  \BibitemOpen
  \bibfield  {author} {\bibinfo {author} {\bibnamefont {{Planck
  Collaboration}}},\ }\href {https://doi.org/10.1051/0004-6361/201833910e}
  {\bibfield  {journal} {\bibinfo  {journal} {\aap}\ }\textbf {\bibinfo
  {volume} {652}},\ \bibinfo {eid} {C4} (\bibinfo {year} {2021})}\BibitemShut
  {NoStop}%
\bibitem [{\citenamefont {{Urban}}(2016)}]{urban2016}%
  \BibitemOpen
  \bibfield  {author} {\bibinfo {author} {\bibfnamefont {A.~L.}\ \bibnamefont
  {{Urban}}},\ }\emph {\bibinfo {title} {{Monsters in the Dark: High Energy
  Signatures of Black Hole Formation with Multimessenger Astronomy}}},\
  \href@noop {} {Ph.D. thesis},\ \bibinfo  {school} {University of Wisconsin,
  Milwaukee} (\bibinfo {year} {2016})\BibitemShut {NoStop}%
\bibitem [{\citenamefont {Cho}(2019)}]{cho2019low}%
  \BibitemOpen
  \bibfield  {author} {\bibinfo {author} {\bibfnamefont {M.-A.}\ \bibnamefont
  {Cho}},\ }\emph {\bibinfo {title} {Low-Latency Searches for Gravitational
  Waves and Their Electromagnetic Counterparts with Advanced LIGO and Virgo}},\
  \href@noop {} {Ph.D. thesis},\ \bibinfo  {school} {University of Maryland}
  (\bibinfo {year} {2019})\BibitemShut {NoStop}%
\bibitem [{Note6()}]{Note6}%
  \BibitemOpen
  \bibinfo {note} {\protect \url
  {https://emfollow.docs.ligo.org/userguide/}}\BibitemShut {NoStop}%
\bibitem [{\citenamefont {Goodfellow}\ \emph {et~al.}(2016)\citenamefont
  {Goodfellow}, \citenamefont {Bengio},\ and\ \citenamefont
  {Courville}}]{Goodfellow}%
  \BibitemOpen
  \bibfield  {author} {\bibinfo {author} {\bibfnamefont {I.}~\bibnamefont
  {Goodfellow}}, \bibinfo {author} {\bibfnamefont {Y.}~\bibnamefont {Bengio}},\
  and\ \bibinfo {author} {\bibfnamefont {A.}~\bibnamefont {Courville}},\
  }\href@noop {} {\emph {\bibinfo {title} {Deep Learning}}}\ (\bibinfo
  {publisher} {MIT Press},\ \bibinfo {year} {2016})\ \bibinfo {note}
  {\url{http://www.deeplearningbook.org}}\BibitemShut {NoStop}%
\bibitem [{\citenamefont {Ioffe}\ and\ \citenamefont
  {Szegedy}(2015)}]{ioffe2015batchnormalizationacceleratingdeep}%
  \BibitemOpen
  \bibfield  {author} {\bibinfo {author} {\bibfnamefont {S.}~\bibnamefont
  {Ioffe}}\ and\ \bibinfo {author} {\bibfnamefont {C.}~\bibnamefont
  {Szegedy}},\ }\href {https://arxiv.org/abs/1502.03167} {\bibinfo {title}
  {Batch normalization: Accelerating deep network training by reducing internal
  covariate shift}} (\bibinfo {year} {2015}),\ \Eprint
  {https://arxiv.org/abs/1502.03167} {arXiv:1502.03167 [cs.LG]} \BibitemShut
  {NoStop}%
\bibitem [{Note7()}]{Note7}%
  \BibitemOpen
  \bibinfo {note} {\protect \url
  {https://docs.pytorch.org/docs/stable/generated/torch.nn.CrossEntropyLoss.html}}\BibitemShut
  {NoStop}%
\bibitem [{\citenamefont {{Shlens}}(2014)}]{KLD}%
  \BibitemOpen
  \bibfield  {author} {\bibinfo {author} {\bibfnamefont {J.}~\bibnamefont
  {{Shlens}}},\ }\href {https://doi.org/10.48550/arXiv.1404.2000} {\bibfield
  {journal} {\bibinfo  {journal} {arXiv e-prints}\ ,\ \bibinfo {eid}
  {arXiv:1404.2000}} (\bibinfo {year} {2014})},\ \Eprint
  {https://arxiv.org/abs/1404.2000} {arXiv:1404.2000 [cs.IT]} \BibitemShut
  {NoStop}%
\bibitem [{\citenamefont {Hinton}\ \emph {et~al.}(2012)\citenamefont {Hinton},
  \citenamefont {Srivastava}, \citenamefont {Krizhevsky} \emph
  {et~al.}}]{hinton2012improvingneuralnetworkspreventing}%
  \BibitemOpen
  \bibfield  {author} {\bibinfo {author} {\bibfnamefont {G.~E.}\ \bibnamefont
  {Hinton}}, \bibinfo {author} {\bibfnamefont {N.}~\bibnamefont {Srivastava}},
  \bibinfo {author} {\bibfnamefont {A.}~\bibnamefont {Krizhevsky}}, \emph
  {et~al.},\ }\href {https://arxiv.org/abs/1207.0580} {\bibinfo {title}
  {Improving neural networks by preventing co-adaptation of feature detectors}}
  (\bibinfo {year} {2012}),\ \Eprint {https://arxiv.org/abs/1207.0580}
  {arXiv:1207.0580 [cs.NE]} \BibitemShut {NoStop}%
\bibitem [{\citenamefont {Glorot}\ and\ \citenamefont
  {Bengio}(2010)}]{pmlr-v9-glorot10a}%
  \BibitemOpen
  \bibfield  {author} {\bibinfo {author} {\bibfnamefont {X.}~\bibnamefont
  {Glorot}}\ and\ \bibinfo {author} {\bibfnamefont {Y.}~\bibnamefont
  {Bengio}},\ }in\ \href {https://proceedings.mlr.press/v9/glorot10a.html}
  {\emph {\bibinfo {booktitle} {Proceedings of the Thirteenth International
  Conference on Artificial Intelligence and Statistics}}},\ \bibinfo {series}
  {Proceedings of Machine Learning Research}, Vol.~\bibinfo {volume} {9},\
  \bibinfo {editor} {edited by\ \bibinfo {editor} {\bibfnamefont {Y.~W.}\
  \bibnamefont {Teh}}\ and\ \bibinfo {editor} {\bibfnamefont {M.}~\bibnamefont
  {Titterington}}}\ (\bibinfo  {publisher} {PMLR},\ \bibinfo {address} {Chia
  Laguna Resort, Sardinia, Italy},\ \bibinfo {year} {2010})\ pp.\ \bibinfo
  {pages} {249--256}\BibitemShut {NoStop}%
\bibitem [{\citenamefont {Kingma}\ and\ \citenamefont
  {Ba}(2017)}]{kingma2017adammethodstochasticoptimization}%
  \BibitemOpen
  \bibfield  {author} {\bibinfo {author} {\bibfnamefont {D.~P.}\ \bibnamefont
  {Kingma}}\ and\ \bibinfo {author} {\bibfnamefont {J.}~\bibnamefont {Ba}},\
  }\href {https://arxiv.org/abs/1412.6980} {\bibinfo {title} {Adam: A method
  for stochastic optimization}} (\bibinfo {year} {2017}),\ \Eprint
  {https://arxiv.org/abs/1412.6980} {arXiv:1412.6980 [cs.LG]} \BibitemShut
  {NoStop}%
\bibitem [{\citenamefont {{Marx}}\ \emph {et~al.}(2025)\citenamefont {{Marx}},
  \citenamefont {{Benoit}}, \citenamefont {{Gunny}} \emph {et~al.}}]{Aframe}%
  \BibitemOpen
  \bibfield  {author} {\bibinfo {author} {\bibfnamefont {E.}~\bibnamefont
  {{Marx}}}, \bibinfo {author} {\bibfnamefont {W.}~\bibnamefont {{Benoit}}},
  \bibinfo {author} {\bibfnamefont {A.}~\bibnamefont {{Gunny}}}, \emph
  {et~al.},\ }\href {https://doi.org/10.1103/PhysRevD.111.042010} {\bibfield
  {journal} {\bibinfo  {journal} {\prd}\ }\textbf {\bibinfo {volume} {111}},\
  \bibinfo {eid} {042010} (\bibinfo {year} {2025})},\ \Eprint
  {https://arxiv.org/abs/2403.18661} {arXiv:2403.18661 [gr-qc]} \BibitemShut
  {NoStop}%
\bibitem [{\citenamefont {Baral}\ \emph {et~al.}()\citenamefont {Baral},
  \citenamefont {Messick},\ and\ \citenamefont {Brady}}]{preferred_event_ml}%
  \BibitemOpen
  \bibfield  {author} {\bibinfo {author} {\bibfnamefont {P.}~\bibnamefont
  {Baral}}, \bibinfo {author} {\bibfnamefont {C.}~\bibnamefont {Messick}},\
  and\ \bibinfo {author} {\bibfnamefont {P.}~\bibnamefont {Brady}},\
  }\href@noop {} {}\bibinfo {howpublished}
  {\url{https://git.ligo.org/pratyusava.baral/preferred_event_ml}}\BibitemShut
  {NoStop}%
\end{thebibliography}%
\end{document}